\pdfoutput=1

\documentclass[11pt,twoside,a4paper,cmspaper,final,collab]{cms-tdr}
\def\svnVersion{380078}\def\cmsCernNoTag{CERN-EP/2016-214}\def\cmsCernDate{\today}\def\cmsMessage{Published in Physical Review D as \href{http://dx.doi.org/10.1103/PhysRevD.95.012003}{\doi{10.1103/PhysRevD.95.012003.}}}

\begin{document}\cmsNoteHeader{SUS-15-004}

\hyphenation{had-ron-i-za-tion}
\hyphenation{cal-or-i-me-ter}
\hyphenation{de-vices}
\RCS$Revision: 379022 $
\RCS$HeadURL: svn+ssh://svn.cern.ch/reps/tdr2/papers/SUS-15-004/trunk/SUS-15-004.tex $
\RCS$Id: SUS-15-004.tex 379022 2016-12-26 13:09:30Z alverson $
\newlength\cmsFigWidth
\ifthenelse{\boolean{cms@external}}{\setlength\cmsFigWidth{0.85\columnwidth}}{\setlength\cmsFigWidth{0.4\textwidth}}
\ifthenelse{\boolean{cms@external}}{\providecommand{\cmsLeft}{top\xspace}}{\providecommand{\cmsLeft}{left\xspace}}
\ifthenelse{\boolean{cms@external}}{\providecommand{\cmsRight}{bottom\xspace}}{\providecommand{\cmsRight}{right\xspace}}
\ifthenelse{\boolean{cms@external}}{\providecommand{\CL}{C.L.\xspace}}{\providecommand{\CL}{CL\xspace}}
\ifthenelse{\boolean{cms@external}}{\providecommand{\cmsUpperLeft}{top}}{\providecommand{\cmsUpperLeft}{upper left}}
\ifthenelse{\boolean{cms@external}}{\providecommand{\cmsUpperRight}{middle}}{\providecommand{\cmsUpperRight}{upper right}}
\newcommand{\CLs}{\ensuremath{\mathrm{CL}_{\mathrm{s}}}\xspace}
\newcommand{\MR}{\ensuremath{M_\mathrm{R}}\xspace}
\newcommand{\dPhiR}{\ensuremath{\Delta\phi_\mathrm{R}}\xspace}
\newcommand{\MRz}{\ensuremath{M_\mathrm{R}^0}\xspace}
\newcommand{\Rtwo}{\ensuremath{\mathrm{R}^2}\xspace}
\newcommand{\Rtwoz}{\ensuremath{\mathrm{R}^2_0}\xspace}
\newcommand{\R}{\ensuremath{\mathrm{R}}\xspace}
\newcommand{\MRT}{\ensuremath{M_\mathrm{T}^\mathrm{R}}\xspace}
\providecommand{\cPV}{\ensuremath{\cmsSymbolFace{V}}\xspace}
\providecommand{\MATNLO}{{\MADGRAPH{}\_a\textsc{mc@nlo}}\xspace}
\newcommand{\x}{\ensuremath{\phantom{0}}}

\cmsNoteHeader{SUS-15-004}
\title{Inclusive search for supersymmetry using razor variables in \texorpdfstring{$\Pp\Pp$ collisions at $\sqrt{s}=13\TeV$}{pp collisions at sqrt(s) = 13 TeV}}

\date{\today}
\abstract{
An inclusive search for supersymmetry using razor variables
is performed in events with four or more jets and no more than one lepton.
The results are based on a sample of proton-proton collisions corresponding to an
integrated luminosity of 2.3\fbinv collected with the CMS experiment at a
center-of-mass energy of $\sqrt{s} = 13\TeV$. No significant excess
over the background prediction is observed in data, and 95\%
confidence level exclusion limits are placed on the masses of new
heavy particles in a variety of simplified models.
Assuming that pair-produced gluinos decay only via three-body
processes involving third-generation quarks plus a neutralino, and
that the neutralino is the lightest supersymmetric particle with a
mass of 200\GeV, gluino masses below 1.6\TeV are excluded for any
branching fractions for the individual gluino decay modes. For some
specific decay mode scenarios, gluino masses up to 1.65\TeV are
excluded. For decays to first- and second-generation quarks and a
neutralino with a mass of 200\GeV, gluinos with masses up to 1.4\TeV
are excluded. Pair production of top squarks decaying to a top quark
and a neutralino with a mass of 100\GeV is excluded for top squark masses
up to 750\GeV.}
\hypersetup{%
pdfauthor={CMS Collaboration},%
pdftitle={Inclusive search for supersymmetry using razor variables in pp collisions at sqrt(s) = 13 TeV},%
pdfsubject={CMS},%
pdfkeywords={CMS, physics, software, computing}}
\maketitle
\section{Introduction}
\label{sec:intro}
Supersymmetry (SUSY) is a proposed extended spacetime symmetry that
introduces a bosonic (fer\-mi\-onic) partner for every fermion (boson) in the
standard model
(SM)~\cite{Wess,Golfand,Volkov,Chamseddine,Kane,Fayet,Barbieri,Hall,Ramond}.
Supersymmetric extensions of the SM are particularly compelling
because they yield solutions to the gauge hierarchy problem without the
need for large fine tuning of fundamental parameters~\cite{Witten:1981nf,Dimopoulos:1981zb,Dine:1981za,Dimopoulos:1981au,Sakai:1981gr,Kaul:1981hi},
exhibit gauge coupling unification~\cite{Dimopoulos:1981yj,Marciano:1981un,Einhorn:1981sx,Ibanez:1981yh,Amaldi:1991cn,Langacker:1995fk},
and can provide weakly interacting particle candidates for dark matter~\cite{Ellis:1983ew,Jungman:1995df}.
For SUSY to provide a ``natural'' solution to the gauge hierarchy
problem, the three Higgsinos, two neutral and one charged, must
be light, and two top squarks, one bottom squark, and the gluino must have masses below a few
TeV, making them potentially accessible at the CERN LHC. Previous searches for SUSY by the
CMS~\cite{1LepMVA,SUS12024,Chatrchyan:2014lfa,Chatrchyan:2013iqa,Chatrchyan:2013fea,Chatrchyan:2013lya,MT2at8TeV}
and ATLAS
~\cite{Aad:2013wta,Aad:2014lra,Aad:2014pda,Aad:2014bva,Aad:2014qaa,Atlas3rdGen,Atlas8tevSummary} Collaborations
have probed SUSY particle masses near the TeV scale, and the increase in the center-of-mass
energy of the LHC from 8 to 13\TeV provides an opportunity to
significantly extend the sensitivity to higher SUSY particle
masses~\cite{Khachatryan:2016kdk,
Khachatryan:2016xvy, Khachatryan:2016uwr, Khachatryan:2016kod,
Khachatryan:2016fll, Aad:2016jxj, Aad:2016tuk, Aaboud:2016tnv, Aaboud:2016zdn,
Aad:2016qqk, Aad:2016eki, Aaboud:2016lwz, Aaboud:2016nwl,
ATLASCollaboration:2016wlb}.

In R-parity~\cite{Farrar:1978xj} conserving SUSY scenarios, the lightest SUSY particle (LSP) is stable and assumed
to be weakly interacting. For many of these models, the experimental signatures at the LHC
are characterized by an abundance of jets and a large transverse momentum imbalance,
but the exact form of the final state can vary significantly,
depending on the values of the unconstrained model parameters. To ensure sensitivity
to a broad range of SUSY parameter space, we adopt an inclusive search
strategy, categorizing events according to the number of identified leptons and \PQb-tagged jets. The razor kinematic variables $\MR$ and $\Rtwo$~\cite{razorPRL,razorPRD}
are used as search variables and are generically sensitive to
pair production of massive particles with subsequent direct or cascading
decays to weakly interacting stable particles. Searches for SUSY and
other beyond the SM phenomena using razor variables have been performed by both the
CMS~\cite{Chatrchyan:2011ek,razorPRL,razorPRD,razor8TeV,Khachatryan:2016zcu,Khachatryan:2016reg} and
ATLAS~\cite{Aad:2012naa,Aad:2015mia}
Collaborations in the past.

\begin{figure*}[!htb]
\centering
\includegraphics[width=0.25\textwidth]{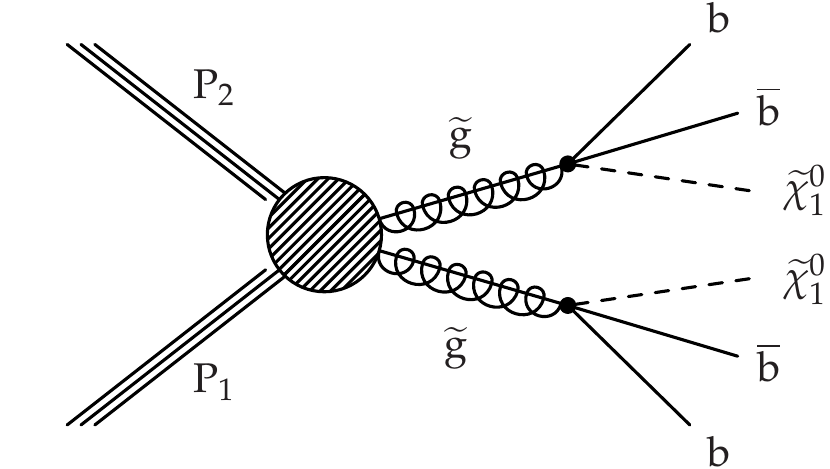}
\includegraphics[width=0.25\textwidth]{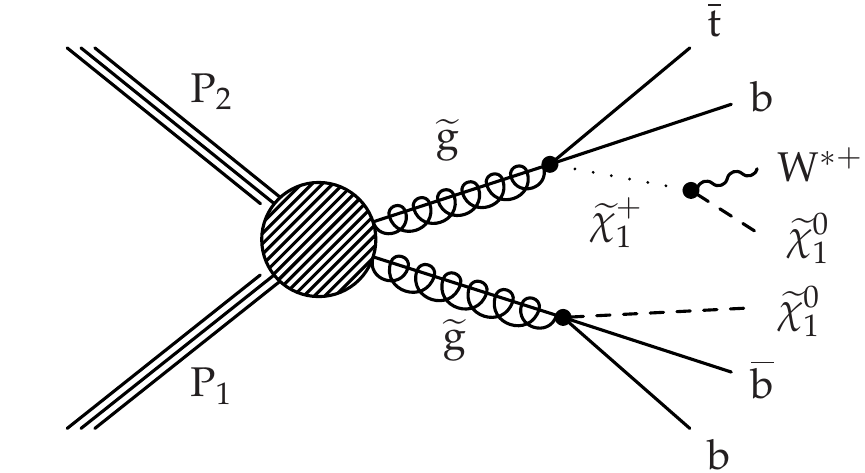}
\includegraphics[width=0.25\textwidth]{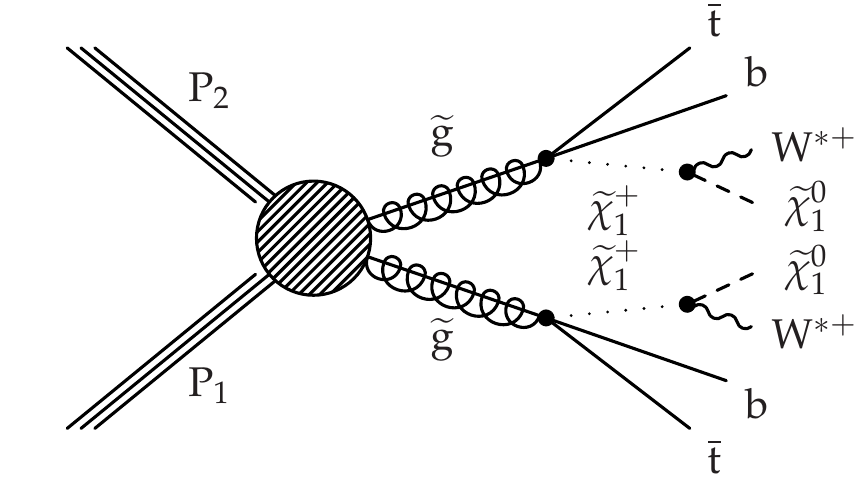} \\
\includegraphics[width=0.25\textwidth]{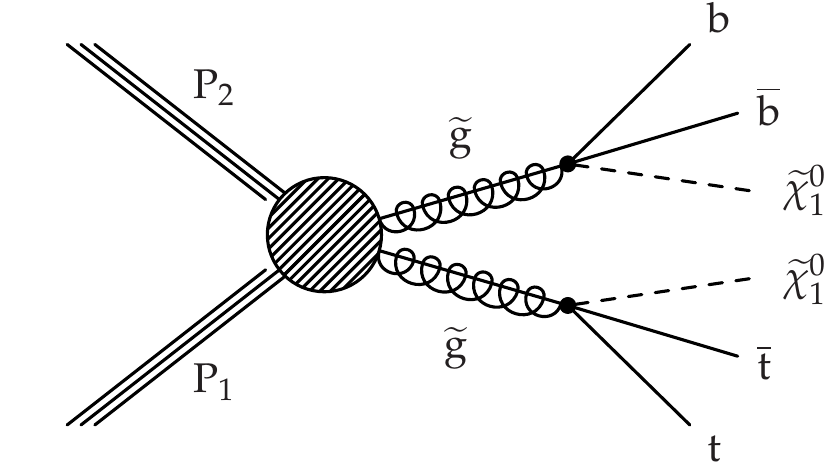}
\includegraphics[width=0.25\textwidth]{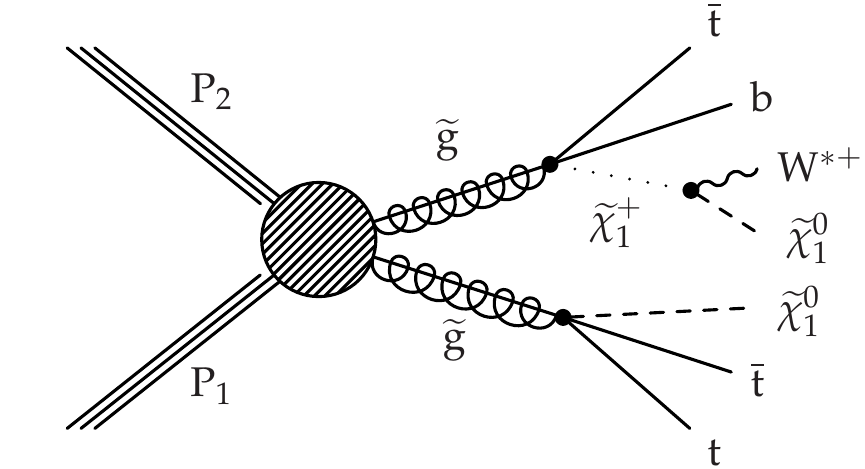}
\includegraphics[width=0.25\textwidth]{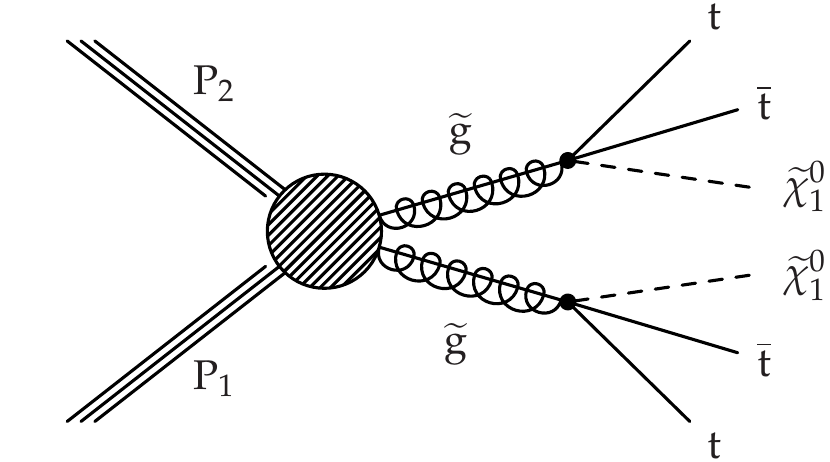} \\
\includegraphics[width=0.25\textwidth]{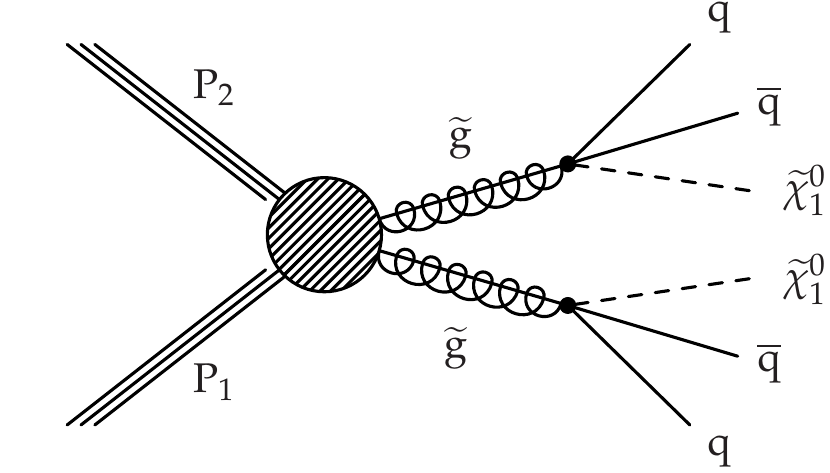}
\includegraphics[width=0.25\textwidth]{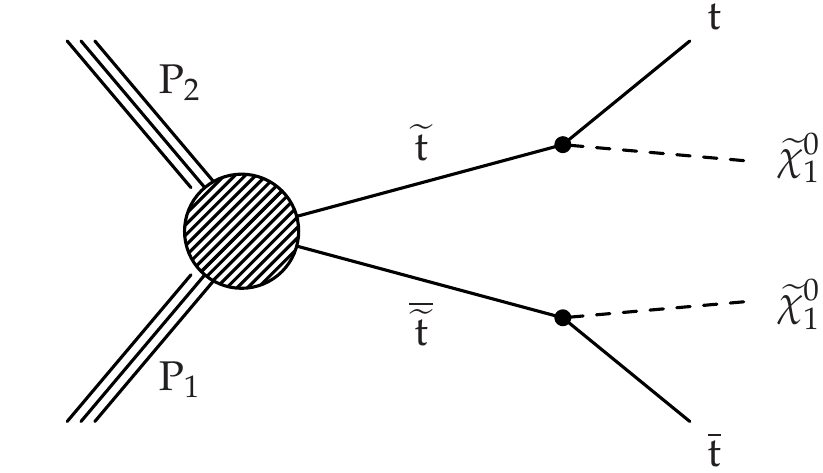}
\caption{Diagrams displaying the distinct event topologies of gluino (all but last) and top squark (last) pair production
considered in this paper. Diagrams corresponding to charge conjugate
decay modes are implied. The symbol $\PW^*$ is used to denote a
virtual $\PW$ boson. \label{fig:SMSDiagrams}}
\end{figure*}

We interpret the results of the inclusive search using
simplified SUSY scenarios for pair production of gluinos and top
squarks. First, we consider models in which the gluino undergoes
three-body decay, either to a
bottom or top quark-antiquark pair and the lightest neutralino $\chiz_{1}$, assumed to be the lightest SUSY
particle; or to a bottom quark (antiquark), a top antiquark (quark), and the
lightest chargino $\chipm_{1}$, assumed to be the next-to-lightest SUSY
particle (NLSP). The NLSP is assumed to have a mass that is 5\GeV larger
than the mass of the LSP, motivated by the fact that in many natural
SUSY scenarios the lightest chargino and the two lightest neutralinos
are Higgsino-like and quasi-degenerate~\cite{naturalSUSY}. The NLSP
decays to an LSP and an off-shell
$\PW$ boson, whose decay products mostly have too low momentum to be
identifiable. The specific choice of the NLSP-LSP mass splitting does not
have a large impact on the results of the interpretation.
The full range of branching fractions to the three possible decay modes ($\bbbar\chiz_{1}$,
$\PQb\cPaqt\chip_1$ or $\cPaqb\cPqt\chim_1$, and $\ttbar\chiz_{1}$) is considered, assuming that these sum to 100\%. We also consider
a model in which the gluino decays to a first- or second-generation quark-antiquark pair and the LSP.
Finally, we consider top squark pair production with the top squark decaying to
a top quark and the LSP. Diagrams of these simplified model processes are shown in Fig.~\ref{fig:SMSDiagrams}.

This paper is organized as follows. Section~\ref{sec:CMSDetector}
presents an overview of the CMS detector. A description of simulated
signal and background samples is given in
Section~\ref{sec:simulation}. Section~\ref{sec:Objects} describes
physics object reconstruction and the event
selection. Section~\ref{sec:StrategySelection} describes the analysis
strategy and razor variables, and the background estimation techniques
used in this analysis are described in
Section~\ref{sec:Background}. Section~\ref{sec:Systematics} covers the
systematic uncertainties. Finally, our results and their interpretation
are presented in Section~\ref{sec:Results}, followed by a summary in Section~\ref{sec:Summary}.
\section{The CMS detector}
\label{sec:CMSDetector}
The central feature of the CMS detector is a
superconducting solenoid of 6\unit{m} internal diameter, providing a
magnetic field of 3.8\unit{T}. Within the solenoid
volume are a silicon pixel and a silicon strip tracker, a
lead tungstate crystal electromagnetic calorimeter (ECAL), and a
brass and scintillator hadron calorimeter (HCAL), each comprising a barrel and
two endcap sections. Muons are measured in gas-ionization detectors
embedded in the magnet steel flux-return yoke outside the
solenoid. Extensive forward calorimetry complements the coverage
provided by the barrel and endcap detectors. Jets are
reconstructed within the pseudorapidity region $\abs{\eta}<5$
covered by the ECAL and HCAL,
where $\eta \equiv -\ln [\tan (\theta/2)]$ and $\theta$ is the
polar angle of the trajectory of the particle with respect to
the counterclockwise beam direction.
Electrons and muons are reconstructed in the region with
$\abs{\eta}<2.5$ and 2.4, respectively.
Events are selected by a two-level trigger system. The first
level is based on a hardware
trigger, followed by a software-based high level trigger. A more
detailed description of the CMS detector, together with a definition
of the coordinate system used and the relevant kinematic variables,
can be found in Ref.~\cite{Adolphi:2008zzk}.
\section{Simulated event samples}
\label{sec:simulation}

Simulated Monte Carlo (MC) samples are used for modeling of the SM backgrounds
in the search regions and for calculating the selection efficiencies for
SUSY signal models. The production of $\ttbar$+jets, $\PW$+jets, $\cPZ$+jets, $\cPgg$+jets,
and QCD multijet events, as well as production of gluino and top squark
pairs,  is simulated with the MC generator \MADGRAPH v5~\cite{Alwall:2011uj}. Single
top quark events are modeled at next-to-leading order (NLO) with \MATNLO~v2.2~\cite{Alwall:2014hca}
for the $s$-channel, and with \POWHEG~v2~\cite{Alioli:2009je, Re:2010bp}
for the $t$-channel and $\PW$-associated production. Contributions from
$\ttbar\PW$, $\ttbar\cPZ$ are also simulated with
\MATNLO~v2.2. Simulated events are interfaced with \PYTHIA
v8.2~\cite{Sjostrand2008852} for fragmentation and parton
showering.
The \textsc{NNPDF3.0LO} and \textsc{NNPDF3.0NLO}~\cite{Ball:2014uwa} parton distribution functions (PDF) are
used, respectively, with \MADGRAPH, and with \POWHEG and \MATNLO.

The SM background events are simulated using a \GEANTfour-based model~\cite{geant4} of the CMS detector.
The simulation of SUSY signal model events is performed using the CMS fast
simulation package~\cite{FastSim}. All simulated events include the
effects of pileup, i.e.~multiple $\Pp\Pp$ collisions within the same or
neighboring bunch crossings, and are processed with the same chain of
reconstruction programs as is used for collision data. Simulated events are weighted to
reproduce the observed distribution of pileup vertices in the data set, calculated based on the measured
instantaneous luminosity.

The SUSY signal production cross sections are calculated to next-to-leading
order (NLO) plus next-to-leading-logarithm (NLL)
accuracy~\cite{NLONLL1,NLONLL2,NLONLL3,NLONLL4,NLONLL5,Borschensky:2014cia}, assuming all
SUSY particles other than those in the relevant diagram to be too
heavy to participate in the interaction. The NLO+NLL cross sections and
their associated uncertainties~\cite{Borschensky:2014cia} are used to derive
the exclusion limits on the masses of the SUSY
particles. The hard scattering is generated using \MADGRAPH with up to
two extra partons to model initial-state radiation at the matrix element level, and
simulated events are interfaced to \PYTHIA v8.2 for the showering,
fragmentation and hadronization steps.
\section{Object reconstruction and selection}
\label{sec:Objects}
Physics objects are defined using the particle-flow (PF)
algorithm~\cite{CMS-PAS-PFT-09-001, CMS-PAS-PFT-10-001}. The PF
algorithm reconstructs and identifies each individual particle with an optimized
combination of information from the various elements of the CMS
detector. All reconstructed PF candidates are clustered into jets using the
anti-$\kt$ algorithm~\cite{Cacciari:2008gp, Cacciari:2011ma}
with a distance parameter
of 0.4. The jet momentum is determined as the vector sum of all particle momenta
in the jet, and jet-energy corrections are derived from simulation and
confirmed by in-situ measurements of the energy balance in dijet
and photon+jet events. Jets are required to pass loose identification criteria
on the jet composition designed to reject spurious signals arising from noise and
failures in the event reconstruction~\cite{CMS-PAS-JME-10-003, Khachatryan:2016kdb}.
For this search, we consider jets with transverse momentum $\pt>40\GeV$ and
$|\eta|<3.0$. The missing transverse momentum vector \ptvecmiss
is defined as the projection on the plane perpendicular to the beams of
the negative vector sum of the momenta of all reconstructed  PF
candidates in an event. Its magnitude is referred to as the missing
transverse energy \ETmiss.

Electrons are reconstructed by associating a cluster of
energy deposited in the ECAL with a reconstructed track~\cite{Khachatryan:2015hwa},
and are required to have $\pt > 5\GeV$ and $|\eta|<2.5$. A ``tight'' selection
used to identify prompt electrons with $\pt > 25\GeV$ is based on requirements
on the electromagnetic shower shape, the geometric matching of
the track to the calorimeter cluster, the track quality and impact
parameter, and isolation. The isolation of electrons and muons is
defined as the scalar sum of the transverse momenta of all neutral and
charged PF candidates within a cone $\Delta R = \sqrt{\smash[b]{(\Delta\eta)^2+(\Delta\phi)^2}}$ along the lepton
direction. The variable is corrected for the effects of pileup using an
effective area correction~\cite{CMS-PAS-JME-14-001}, and the cone size
$\Delta R$ shrinks with increasing lepton $\pt$  according to
\begin{eqnarray}
\label{eq:miniIsolation}
\Delta R=
\begin{cases}
0.2, & \pt \le 50\GeV\\
{10\GeV}/{\pt}, & 50 < \pt \le 200\GeV \\
0.05, & \pt > 200\GeV. \\
\end{cases}
\end{eqnarray}
The use of the lepton $\pt$ dependent isolation cone enhances the
efficiency for identifying leptons in events containing a large amount of hadronic
energy, such as those with $\ttbar$ production. For tight electrons, the isolation is required to be less than 10\% of
the electron $\pt$. The selection efficiency for tight electrons
increases from 60\% for $\pt$ around 20\GeV
to 70\% for $\pt$ around 40\GeV and to 80\% for $\pt$ above 50\GeV.

To improve the purity of all-hadronic signals in the zero-lepton event categories, a looser ``veto''
selection is also defined.  For this selection, electrons are required to have $\pt>5\GeV$.  The output of a boosted decision tree is used to identify electrons based on shower
shape and track information~\cite{Khachatryan:2015hwa}.
For electrons with $\pt>20\GeV$, the isolation is required to be less than 20\% of the
electron $\pt$.  For electrons with $\pt$ between 5 and 20\GeV, the value of the
isolation, computed by summing the $\pt$'s of all particle flow candidates within a
$\Delta R$ cone of 0.3, is required to be less than 5\GeV. For the
veto electron selection, the efficiency increases from 60\% for
$\pt$ around 5\GeV to 80\% for $\pt$ around 15\GeV and 90\% for $\pt$ above 20\GeV.

Muons are reconstructed by combining tracks found in the muon system with
corresponding tracks in the silicon detectors~\cite{Chatrchyan:2012xi},
and are required to have $\pt > 5\GeV$ and $|\eta|<2.4$. Muons are identified
based on the quality of the track fit, the number of detector hits used in the
tracking algorithm, and the compatibility between track
segments. The absolute value of the 3D impact
parameter significance of the muon track, which is defined as the ratio of the impact
parameter to its estimated uncertainty, is required to be less than
4. As for electrons, we define a ``tight'' selection for muons with $\pt > 20\GeV$
and a ``veto'' selection for muons with $\pt > 5\GeV$.
For both tight and veto muons with
$\pt > 20\GeV$ the isolation is required to be less than 20\%
of the muon $\pt$, while for veto muons with $\pt$ between 5 and 20\GeV
the isolation computed using a $\Delta R$ cone of 0.4
is required to be less than 10\GeV. For tight muons we require $d_0<0.2 \unit{cm}$, where $d_0$ is the transverse impact parameter of the muon
track, while this selection is not applied for veto muons.
The selection efficiency for tight muons increases from 65\% for
$\pt$ around 20\GeV to 75\% for $\pt$ around 40\GeV and to 80\% for $\pt$ above 50\GeV.
For the veto muon selection, the efficiency increases from 85\% for
$\pt$ around 5\GeV to 95\% for $\pt$ above 20\GeV.

We additionally reconstruct and identify hadronically decaying $\ensuremath{\tau}$
leptons ($\ensuremath{\tau_{\mathrm{h}}}$) to further enhance the all-hadronic purity
of the zero-lepton event categories, using the hadron-plus-strips algorithm~\cite{Khachatryan:2015dfa}, which
identifies $\ensuremath{\tau}$ decay modes
with one charged hadron and up to two neutral pions, or three charged hadrons.
The $\ensuremath{\tau_{\mathrm{h}}}$ candidate is required to have
$\pt>20\GeV$, and the isolation, defined as the $\pt$ sum of other nearby PF candidates, must be below a certain threshold.
The loose cutoff-based selection~\cite{Khachatryan:2015dfa} is used and results in an efficiency
of about 50\% for successfully reconstructed $\ensuremath{\tau_{\mathrm{h}}}$ decays.

To identify jets originating from \PQb-hadron decays, we use the
combined secondary vertex \PQb~jet tagger, which uses the inclusive
vertex finder to select \PQb~jets~\cite{CMS-PAS-BTV-15-001, Chatrchyan:2012jua}. The ``medium''
working point is used to define the event categories for the search signal regions.
For jets with $\pt$ between 40 and 200\GeV the \PQb~jet tagging efficiency
is approximately 70\% and the probability of misidentifying a light-flavor quark or gluon as
a \PQb~jet is 1.5\% in typical background events relevant for this search.

Photon candidates are reconstructed from clusters of energy deposits in
the ECAL. They are identified using
selections on the transverse shower width $\sigma_{\eta\eta}$ as defined
in~Ref.~\cite{Khachatryan:2015iwa}, and the hadronic to electromagnetic energy ratio ($H/E$).
Photon isolation, defined as the scalar $\pt$ sum of charged particles within a cone of
$\Delta R<0.3$, must be less than 2.5\GeV. Finally, photon candidates that share
the same energy cluster as an identified electron are vetoed.
\section{Analysis strategy and event selection}
\label{sec:StrategySelection}
We select events with four or more jets, using search categories
defined by the number of leptons and \PQb-tagged jets in the event.
The Multijet category consists of events with no electrons or muons passing the tight or veto selection, and no selected $\ensuremath{\tau_{\mathrm{h}}}$.
Events in the one electron (muon) category, denoted as the Electron Multijet (Muon Multijet) category,
are required to have one and only one electron (muon) passing the tight selection.
Within these three event classes, we divide the events further into categories depending on
whether the events have zero, one, two, or more than two \PQb-tagged jets.

Each event in the above categories is treated as a dijet-like event by grouping selected leptons
and jets in the event into two ``megajets'', whose four-momenta are
defined as the vector sum of the four-momenta of their constituent physics objects~\cite{Chatrchyan:2011ek}. The
clustering algorithm selects the grouping that minimizes the sum of the squares of the invariant masses
of the two megajets. We define the razor variables $\MR$ and $\MRT$ as
\begin{align}
\label{eq:MRstar}
\MR &\equiv
\sqrt{
(\abs{\vec{p}^{\,\mathrm{j}_{1}}}+\abs{\vec{p}^{\,\mathrm{j}_{2}}})^2 -({p}^{\,\mathrm{j}_1}_z+{p}^{\,\mathrm{j}_2}_z)^2},\\
\MRT &\equiv \sqrt{ \frac{\ETm(\pt^{\,\mathrm{j}_1}+\pt^{\,\mathrm{j}_2}) -
\ptvecmiss \cdot
(\ptvec^{\,\mathrm{j}_1}+\ptvec^{\,\mathrm{j}_2}) }{2}},
\end{align}
where $\vec{p}_{\mathrm{j}_i}$, $\ptvec^{\,\mathrm{j}_i}$, and
$p^{\,\mathrm{j}_i}_z$ are the momentum of the $i$th megajet and its transverse and longitudinal components with
respect to the beam axis, respectively.  The dimensionless variable $\R$ is defined as
\begin{equation}
\R \equiv \frac{\MRT}{\MR}.
\end{equation}

For a typical SUSY decay of a superpartner $\PSq$ decaying into an
invisible neutralino $\PSGczDo$ and the standard model partner $\Pq$,
the mass variable $\MR$ peaks at a characteristic mass scale~\cite{razorPRL,razorPRD}
${(m_{\PSq}^{2}-m_{\PSGczDo}^{2})/m_{\PSGczDo}}$. For
standard model background processes, the distribution of $\MR$ has an
exponentially falling shape. The variable $\Rtwo$ is
related to the missing transverse energy and is used to
suppress QCD multijet background.

The events of interest are triggered either by the presence of a high-$\pt$ electron or muon, or
through dedicated hadronic triggers requiring the presence of at least two highly energetic jets
and with loose thresholds on the razor variables $\MR$ and $\Rtwo$. The single-electron (single-muon) triggers require at least one isolated electron
(muon) with $\pt> $ 23 (20)\GeV. The isolation requirement is dropped for electrons (muons) with
$\pt> $ 105 (50)\GeV. The efficiencies for the single electron (muon) triggers
are above 70\% for $\pt$ around 25 (20)\GeV, and reach a
plateau above 97\% for $\pt> 40\GeV$. The efficiencies for the single electron trigger were measured in data and simulation and
found to be in good agreement, as were the corresponding efficiencies for muons.
The hadronic razor trigger requires at least two jets with $\pt > 80\GeV$ or at least
four jets with $\pt > 40\GeV$. The events are also required to pass selections on the
razor variables $\MR>200\GeV$ and $\Rtwo>0.09$ and on the product
$(\MR + 300\GeV)\times(\Rtwo + 0.25)>240\GeV$.
The efficiency of the hadronic razor trigger for events passing the baseline
$\MR$ and $\Rtwo$ selections described below is 97\% and is consistent
with the prediction from MC simulation.

For events in the Electron or Muon Multijet categories, the search region
is defined by the selections $\MR > 400\GeV$ and $\Rtwo > 0.15$.
The $\pt$ of the electron (muon)
is required to be larger than 25 (20)\GeV. To suppress backgrounds from the $\PW(\ell\nu)$+jets
and $\ttbar$ processes, we require that the transverse mass
$M_{\mathrm{T}}$ formed by the lepton momentum
and \ptvecmiss be larger than 120\GeV.

For events in the Multijet category, the search uses a region defined by the
selections $\MR > 500\GeV$ and $\Rtwo > 0.25$ and requires the presence of at least
two jets with $\pt >80\GeV$ within $|\eta|<3.0$, for compatibility with the requirements
imposed by the hadronic razor triggers. For QCD multijet background events, the
\ETmiss arises mainly from mismeasurement of
the energy of one of the leading jets.  In such cases, the two razor
megajets tend to lie in a back-to-back configuration. Therefore, to suppress the QCD multijet
background we require that the azimuthal angle $\dPhiR$ between the two razor
megajets be less than 2.8 radians.

Finally, events containing signatures consistent with beam-induced background or anomalous noise
in the calorimeters are rejected using dedicated
filters~\cite{Chatrchyan:2011tn,Khachatryan:2014gga}.
\section{Background modeling}
\label{sec:Background}
The main background processes in the search regions considered are
$\PW(\ell\nu)$+jets (with $\ell=\Pe$, $\Pgm$, $\ensuremath{\tau}$), $\cPZ(\nu\PAGn)$+jets, $\ttbar$, and QCD multijet production. For event categories with
zero \PQb-tagged jets, the background is primarily composed of the $\PW(\ell\nu)$+jets and $\cPZ(\nu\PAGn)$+jets
processes, while for categories with two or more \PQb-tagged jets it is
dominated by the $\ttbar$ process. There are also very small contributions from
the production of two or three electroweak bosons and from the production of $\ttbar$ in
association with a $\PW$ or $\cPZ$ boson. These
contributions are summed and labeled ``Other'' in Fig.~\ref{fig:TTBarWJetsCR_MR}-\ref{fig:Znn_PhotonJets}.

We model the background using two independent methods based on control samples in data with entirely
independent sets of systematic assumptions. The first method (A) is based on the use of
dedicated control regions that isolate specific background processes in order
to control and correct the predictions of the MC simulation.
The second method (B) is based on a fit to an assumed functional
form for the shape of the observed data distribution in the two-dimensional $\MR$-$\Rtwo$ plane.
These two background predictions are compared and cross-checked against each other in order
to significantly enhance the robustness of the background estimate.
\subsection{Method A: simulation-assisted background prediction from data}
\label{sec:MADD}
The simulation-assisted method defines dedicated control regions that isolate
each of the main background processes. Data in these control regions are used
to control and correct the accuracy of the MC prediction for each of the
background processes. Corrections for the jet energy response and lepton momentum response
are applied to the MC, as are corrections for the trigger
efficiency and the selection efficiency of electrons, muons, and \PQb-tagged jets. Any
disagreement observed in these control regions is then interpreted as an inaccuracy of the
MC in predicting the hadronic recoil spectrum and jet multiplicity.
Two alternative formulations of the method are typically used in searches for
new physics~\cite{SUS12024,MT2at8TeV,Aad:2013wta}.
In the first formulation, the data control region yields are extrapolated to the search regions
via translation factors derived from simulation.  In the second formulation, simulation to data
correction factors are derived in bins of the razor variables $\MR$ and $\Rtwo$
and are then applied to the simulation prediction of the search region yields.
The two formulations are identical and the choice of which formulation is used
depends primarily on the convenience of the given data processing sequence.
In both cases, the contributions from background processes other than the one
under study are subtracted using the MC prediction.
We employ the first formulation of the method for the estimate of the QCD background,
while the second formulation is used for modeling all other major backgrounds.
Details of the control regions used for each of
the dominant background processes are described in the subsections below.

Finally, the small contribution from rare background processes such as $\ttbar\cPZ$ is
modeled using simulation. Systematic uncertainties on the cross sections of these processes
are propagated to the final result.
\subsubsection{The \texorpdfstring{$\ttbar$ and $\PW(\ell\nu)$+jets}{ttbar and W(lnu)} background}
\label{sec:TTBarWJetsCR}

The control region to isolate the $\ttbar$ and $\PW(\ell\nu)$+jets processes is defined by requiring
at least one tight electron or muon. To suppress QCD multijet
background, the quantities \ETmiss and $M_{\mathrm{T}}$ are both required to be larger than 30\GeV. To minimize
contamination from potential SUSY processes and to explicitly separate the control region
from the search regions, we require $M_{\mathrm{T}} < 100\GeV$. The $\ttbar$ enhanced control region is defined by requiring that there be at
least one \PQb-tagged jet, and the $\PW(\ell\nu)$+jets enhanced control region is defined by
requiring no such \PQb-tagged jets. Other than these  \PQb-tagged jet
requirements, we place no explicit requirement on the number of jets in
the event, in order to benefit from significantly larger control
samples.

\begin{figure}[!ptb] \centering
\includegraphics[width=0.45\textwidth]{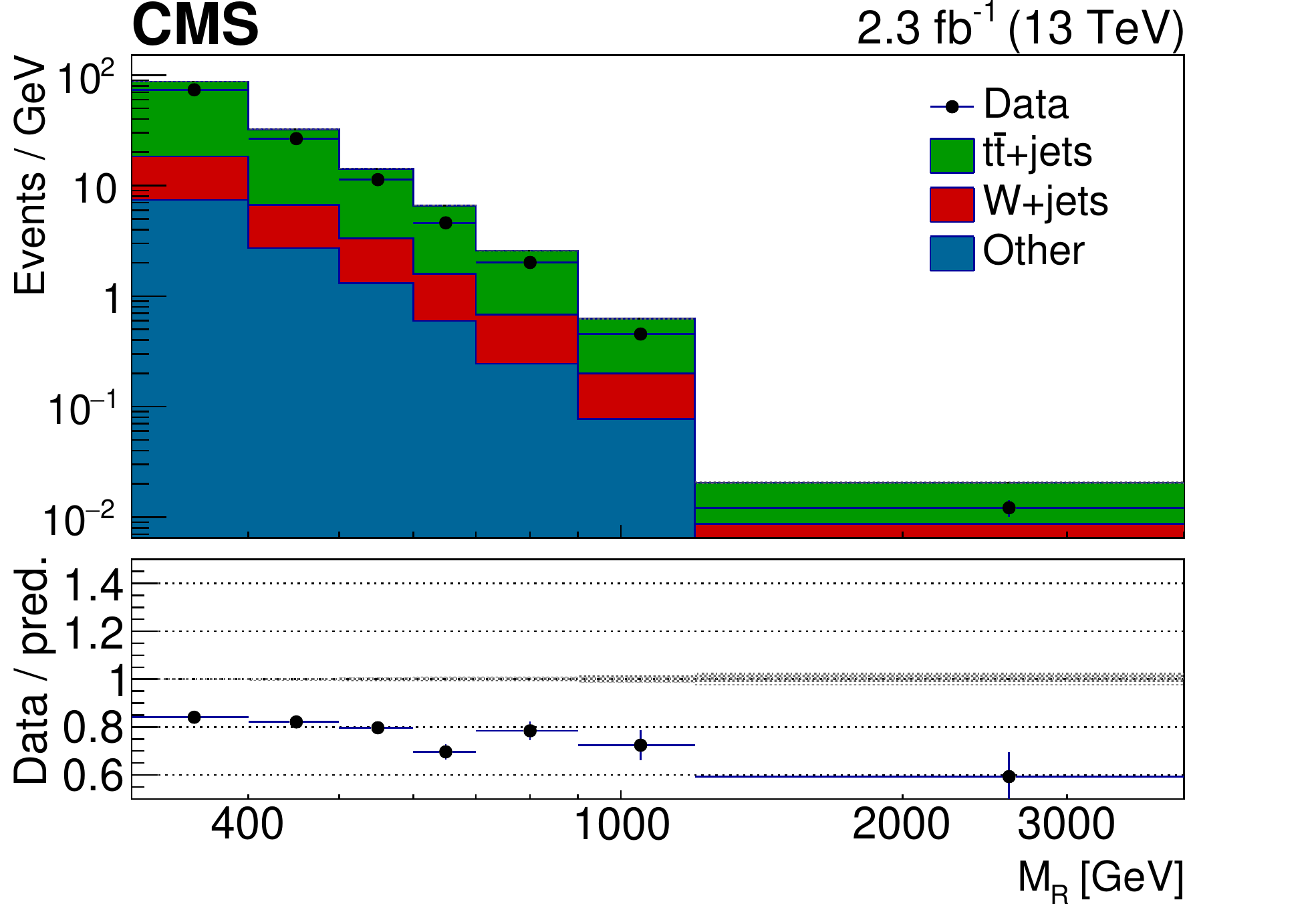}
\includegraphics[width=0.45\textwidth]{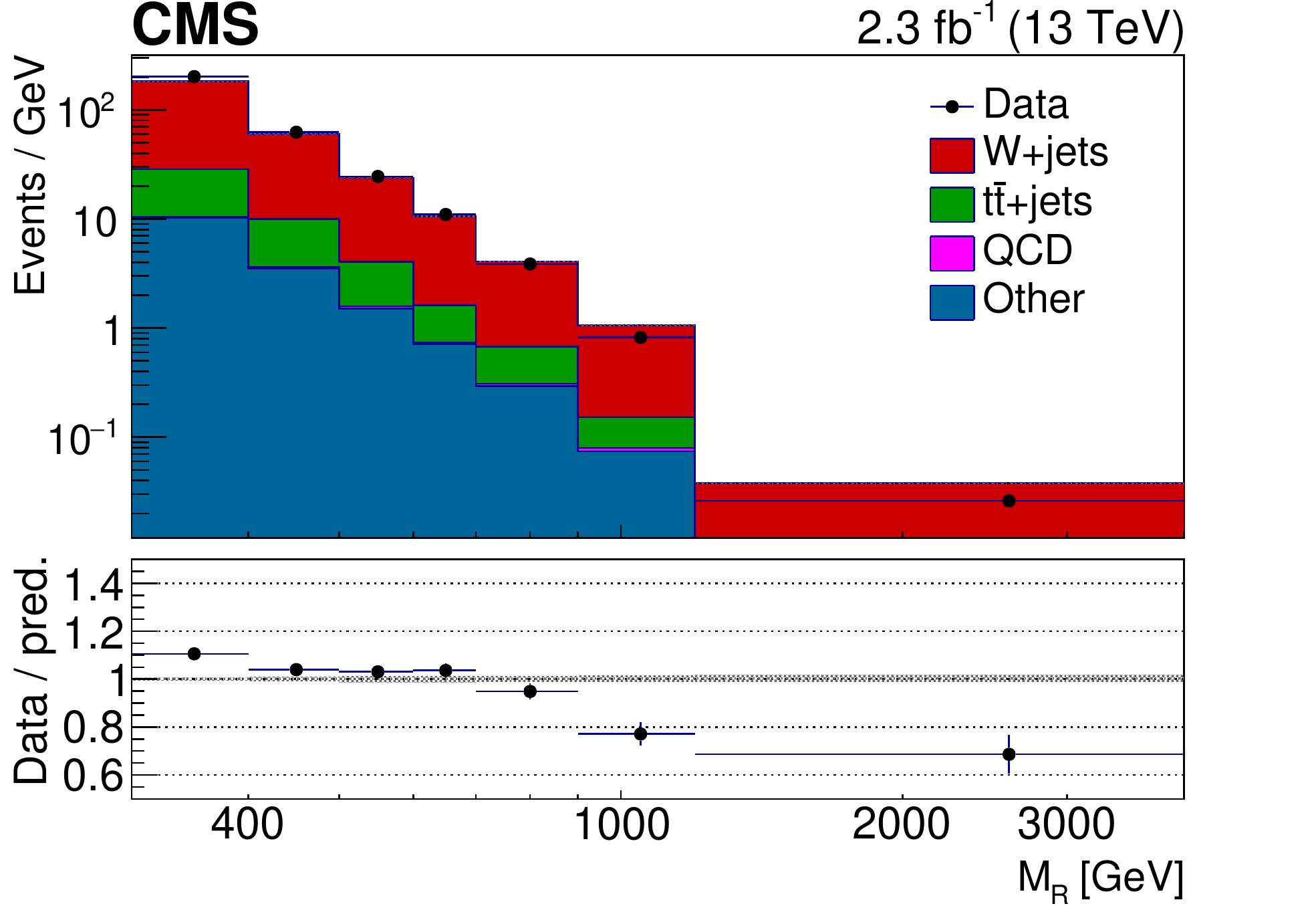}
\caption{ The $\MR$ distributions for events in the $\ttbar$ (\cmsLeft) and $\PW(\ell\nu)$+jets (\cmsRight)
control regions are shown, comparing data with the MC prediction.
The ratio of data to the background prediction is shown on the bottom panel,
with the statistical uncertainty expressed through the data point error bars and the
systematic uncertainty of the background prediction represented by the shaded region.
In the right-hand plot, the $\ttbar$ MC events have been reweighted
according to the corrections derived in the $\ttbar$-enhanced control
region.
}
\label{fig:TTBarWJetsCR_MR}
\end{figure}

\begin{figure}[!ptb] \centering
\includegraphics[width=0.50\textwidth]{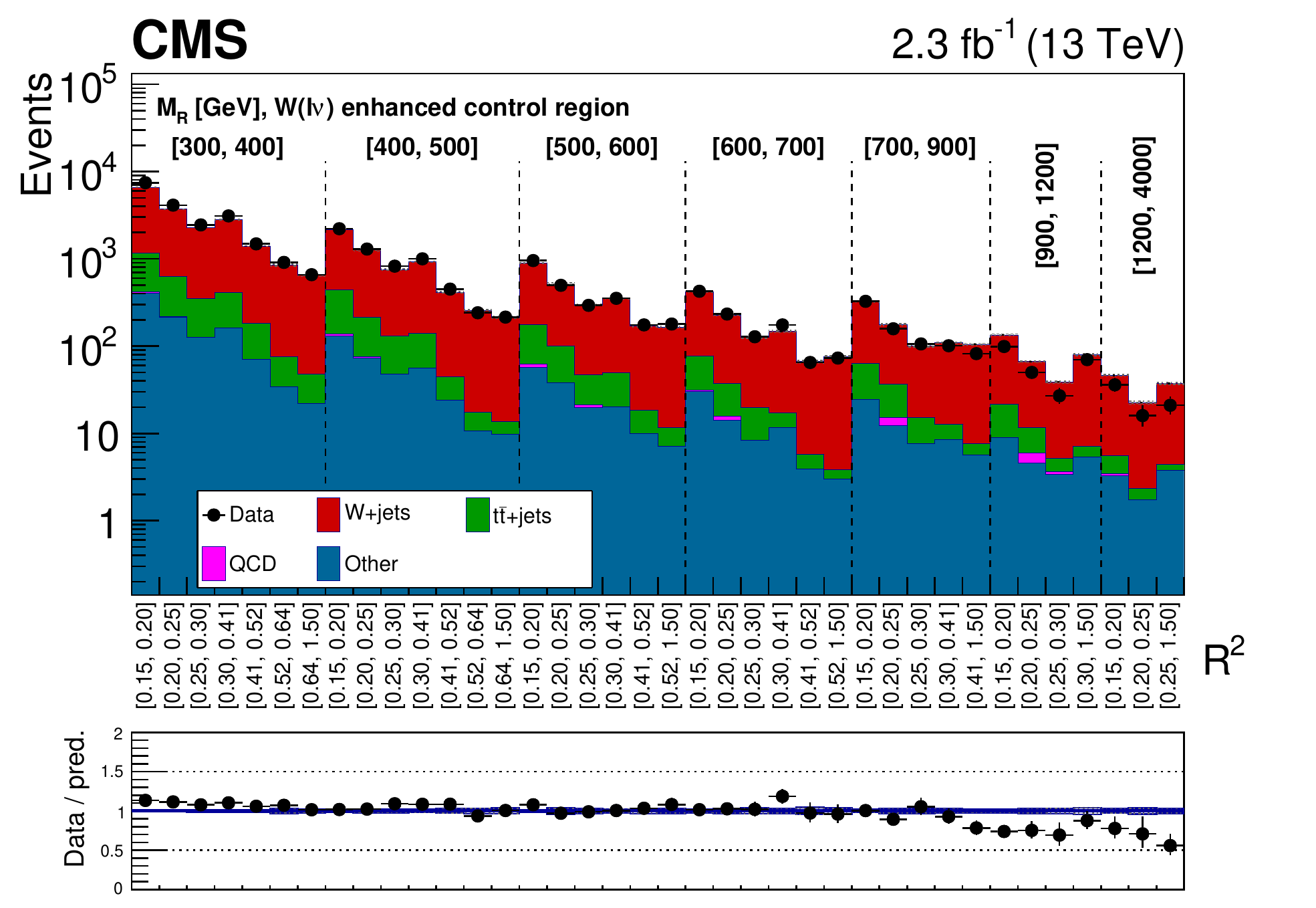}
\includegraphics[width=0.50\textwidth]{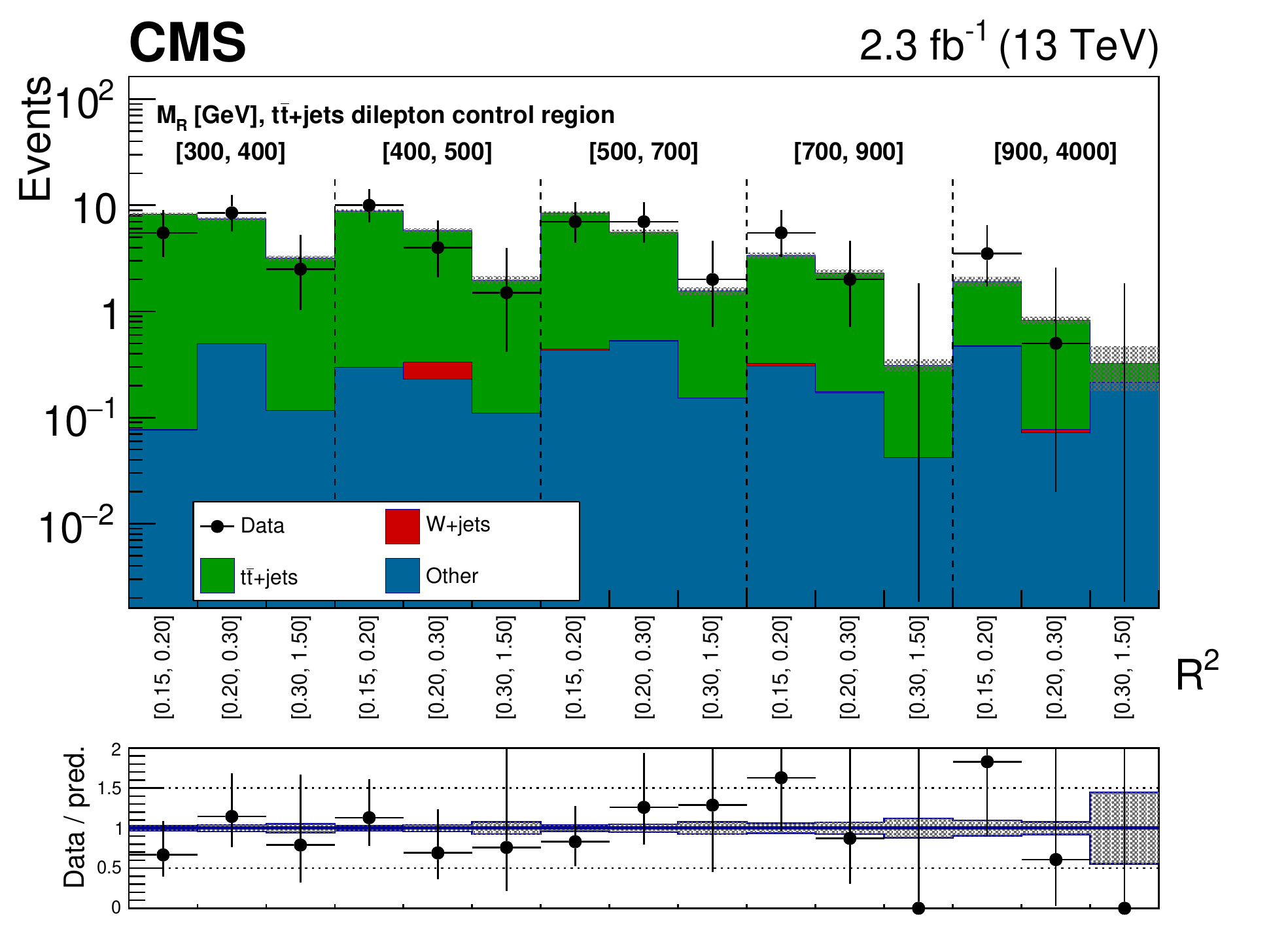}
\caption{The two-dimensional $\MR$-$\Rtwo$ distribution for the
$\PW(\ell\nu)$+jets enhanced (upper) and the $\ttbar$ dilepton (lower)
control regions are shown, comparing data with the MC prediction. The $\ttbar$ MC events have been reweighted according to the correction factors
derived in the $\ttbar$-enhanced control region.  The two-dimensional $\MR$-$\Rtwo$ distribution is shown
in a one dimensional representation, with each $\MR$ bin marked by the dashed lines and labeled near the top
, and each $\Rtwo$ bin labeled below.
The bottom panel shows the ratio of data to the background prediction, with uncertainties displayed as in Fig.~\ref{fig:TTBarWJetsCR_MR}.
}
\label{fig:WJets_TTBarDileptonCR_MRRsqUnrolled}
\end{figure}

We first derive corrections for the $\ttbar$ background, and then measure
corrections for the $W(\ell\nu)$+jets process after first applying the corrections already obtained
for the $\ttbar$ background in the $W(\ell\nu)$+jets control region.
As discussed above, the corrections to the MC prediction are derived in two-dimensional bins of the
$\MR$-$\Rtwo$ plane. We observe that the $\MR$ spectrum predicted by the simulation
falls off less steeply than the control region data for both the $\ttbar$ and $\PW(\ell\nu)$+jets
processes, as shown in Fig.~\ref{fig:TTBarWJetsCR_MR}.
In Fig.~\ref{fig:WJets_TTBarDileptonCR_MRRsqUnrolled}, we show the two dimensional $\MR$-$\Rtwo$ distributions
for data and simulation in the $\PW(\ell\nu)$+jets control region. The statistical uncertainties in the correction factors
due to limited event yields in the control region bins are propagated and dominate the total uncertainty
of the background prediction. For bins at large $\MR$ (near 1000\GeV), the statistical uncertainties
range between 15\% and 50\%.

Corrections to the MC simulation are first measured and applied as a function of $\MR$ and $\Rtwo$, inclusively in the
number of selected jets. As our search region requires a higher multiplicity of jets, an additional correction factor
is required to accurately model the jet multiplicity. We measure this additional
correction factor to be $0.90 \pm 0.03$ by comparing the data and the MC prediction in the $\PW(\ell\nu)$+jets and $\ttbar$
control region for events with four or more jets.
To control for possible simulation mismodeling that is correlated between the number of jets and the razor
variables, we perform additional cross-checks of the $\MR$ and $\Rtwo$ distributions in bins of
the number of \PQb-tagged jets in the $\ttbar$ and $\PW(\ell\nu)$+jets
control regions for events with four or more jets. For bins that show statistically significant disagreement,
the size of the disagreement is propagated as a systematic uncertainty. The typical range of these additional
systematic uncertainties is between 10\% and 30\%.

\begin{figure}[!ptb] \centering
\includegraphics[width=0.45\textwidth]{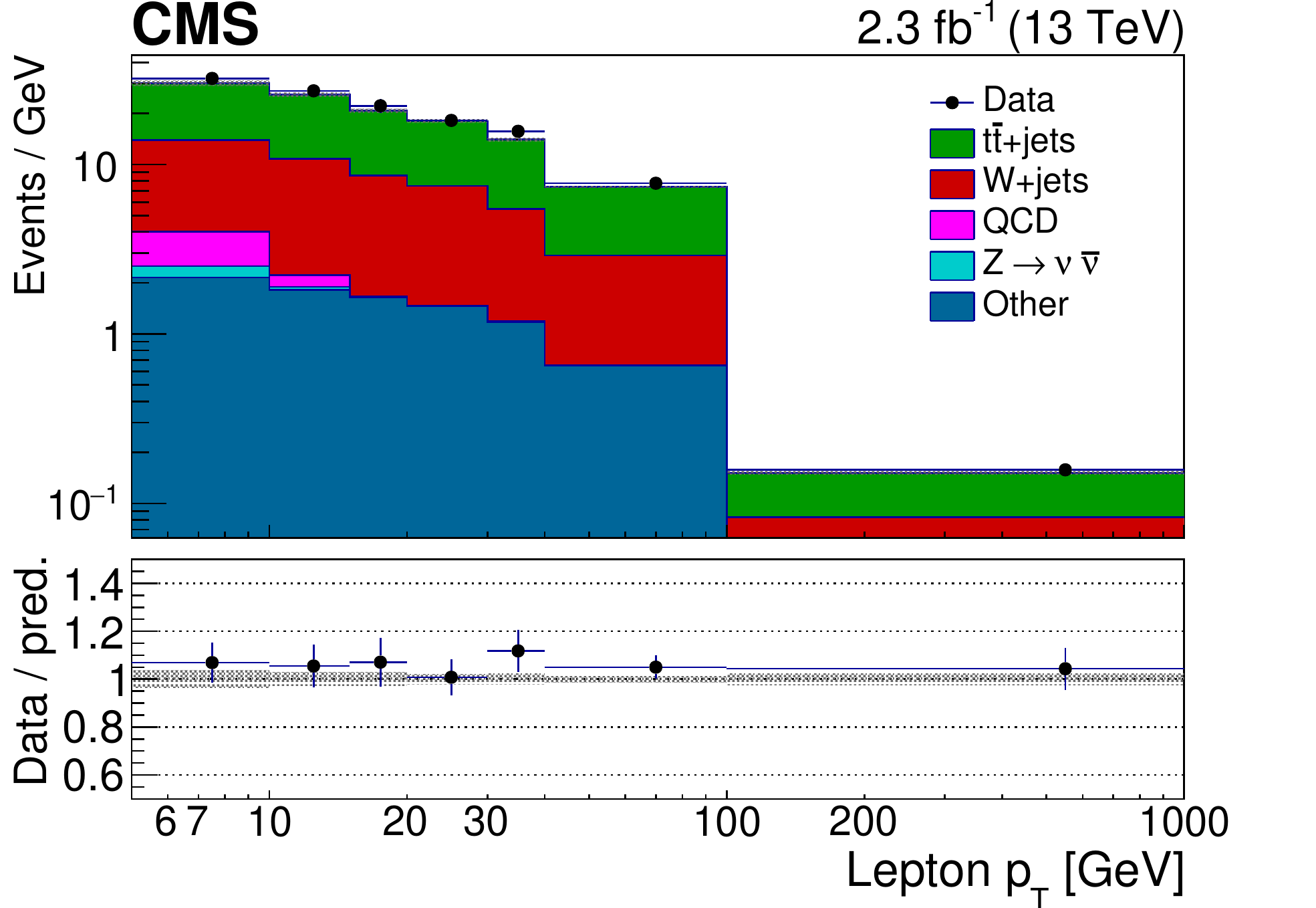}
\includegraphics[width=0.45\textwidth]{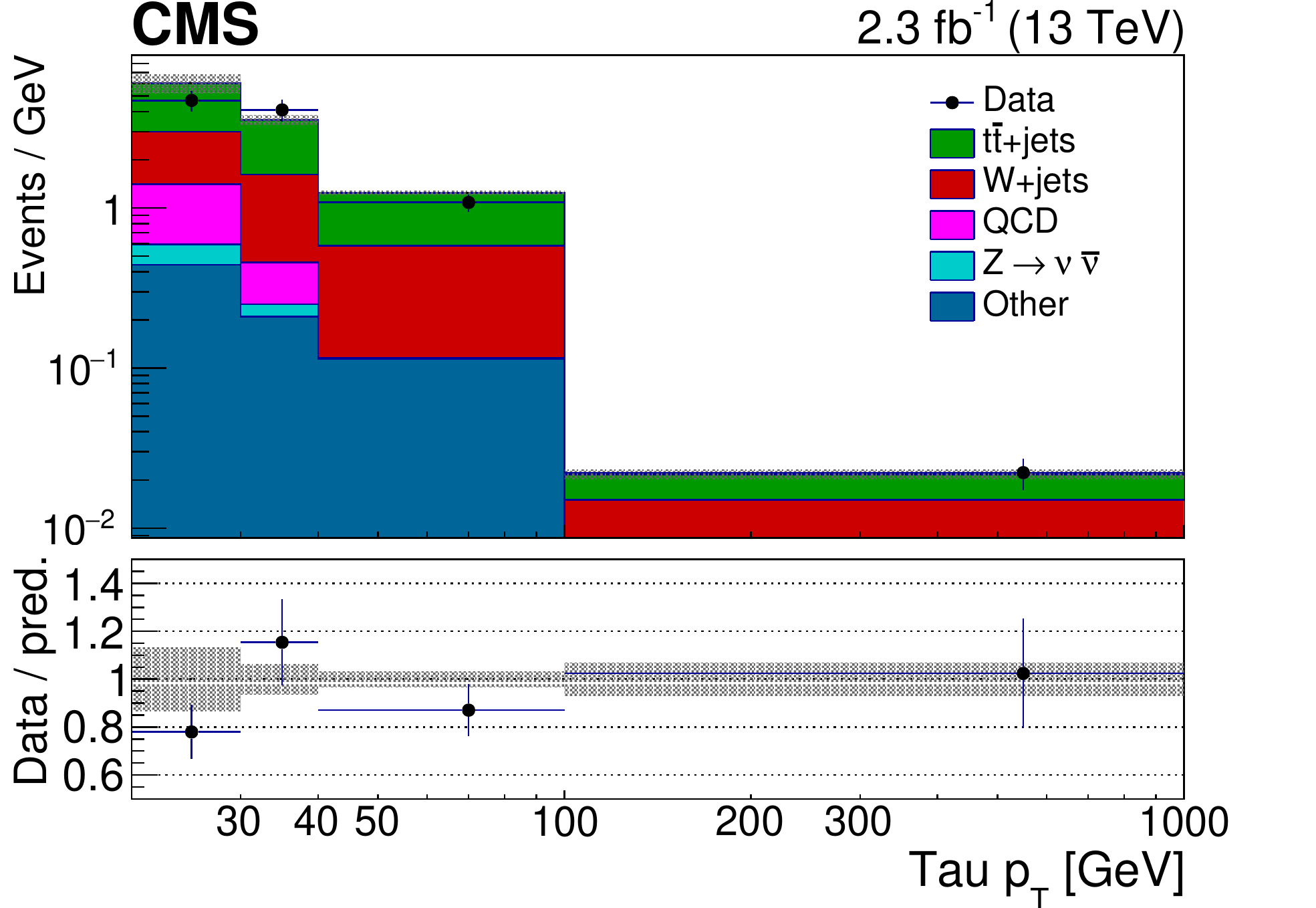}
\caption{ The $\pt$ distributions of the veto electron or muon (\cmsLeft) and the veto $\ensuremath{\tau_{\mathrm{h}}}$ (\cmsRight)
is shown for events in the veto lepton control regions, comparing data with the MC prediction. The $\ttbar$ and
$\PW(\ell\nu)$+jets MC events have been reweighted according to the correction factors derived
in the $\ttbar$ enhanced and $\PW(\ell\nu)$+jets enhanced control regions, respectively.
The bottom panel shows the ratio of data to the background prediction, with uncertainties displayed as in Fig.~\ref{fig:TTBarWJetsCR_MR}.
}
\label{fig:VetoLeptonCR_LepPt}
\end{figure}

The $\ttbar$ and $\PW(\ell\nu)$+jets backgrounds in the zero-lepton Multijet event category are
composed of lost lepton events with at least one lepton in the final state, which is either out of
acceptance or fails the veto electron, veto muon, or
$\ensuremath{\tau_{\mathrm{h}}}$ selection. To ensure a good
understanding of the rate of lost lepton events in data and the MC simulation, two
additional control regions are defined to evaluate accuracy of the
modeling of the acceptance and efficiency for selecting veto
electrons, veto muons, or $\ensuremath{\tau_{\mathrm{h}}}$.
We require events in the veto lepton ($\ensuremath{\tau_{\mathrm{h}}}$ candidate) control region to have at least one veto electron or muon
($\ensuremath{\tau_{\mathrm{h}}}$ candidate) selected. The $M_{\mathrm{T}}$ is required to be between 30 and 100\GeV in order to
suppress QCD multijet background and contamination from potential new physics processes. At least two jets
with $\pt>80$~GeV and at least four jets with $\pt>40\GeV$ are required,
consistent with the search region requirements. Finally, we consider events with
$\MR > 400$~GeV and $\Rtwo>0.25$. The distribution of the veto lepton $\pt$ for events in the veto
lepton and veto $\ensuremath{\tau_{\mathrm{h}}}$ control regions are shown in Fig.~\ref{fig:VetoLeptonCR_LepPt},
and demonstrate that the MC models describe well the observed data.
The observed discrepancies in any bin are propagated as systematic uncertainties in the
prediction of the $\ttbar$ and $\PW(\ell\nu)$+jets backgrounds in the Multijet category search region.

The $\ttbar$ background in the Electron and Muon Multijet categories is primarily from
the dilepton decay mode as the $M_{\mathrm{T}}$ requirement highly suppresses the semi-leptonic decay
mode. Corrections to the MC simulation derived from the $\ttbar$
control region primarily arise from semi-leptonic decays.
We define an additional control region enhanced in dilepton $\ttbar$ decays
to confirm that the MC corrections derived from a region dominated by
semi-leptonic decays also apply to dilepton decays. We select events with two tight leptons,
both with $\pt>30\GeV$, $\ETmiss>40\GeV$, and
dilepton mass larger than 20\GeV. For events with two leptons of the same flavor, we additionally
veto events with a dilepton mass between 76 and 106\GeV in order to suppress background from $\cPZ$ boson
decays. At least one \PQb-tagged jet is required to enhance the purity for the $\ttbar$
process. Finally, we mimic the phase space region similar to our search region in the Electron and
Muon Multijet categories by treating one lepton as having failed the identification criteria
and applying the $M_{\mathrm{T}}$ requirement using the other lepton. The correction factors measured in the
$\ttbar$ control region are applied to the MC prediction of the dilepton
$\ttbar$ cross-check region in bins of $\MR$ and $\Rtwo$.
In Fig.~\ref{fig:WJets_TTBarDileptonCR_MRRsqUnrolled}
we show the $\MR$-$\Rtwo$ distribution for the dilepton $\ttbar$ cross-check region
in events with four or more jets, and we observe no significant
mismodeling by the simulation, indicating that the measured corrections are accurate.
\subsubsection{The \texorpdfstring{$\cPZ\to\nu\PAGn$}{Z to nu nubar} background}
\label{sec:ZInvCR}

\begin{figure}[!ptb] \centering
\includegraphics[width=0.50\textwidth]{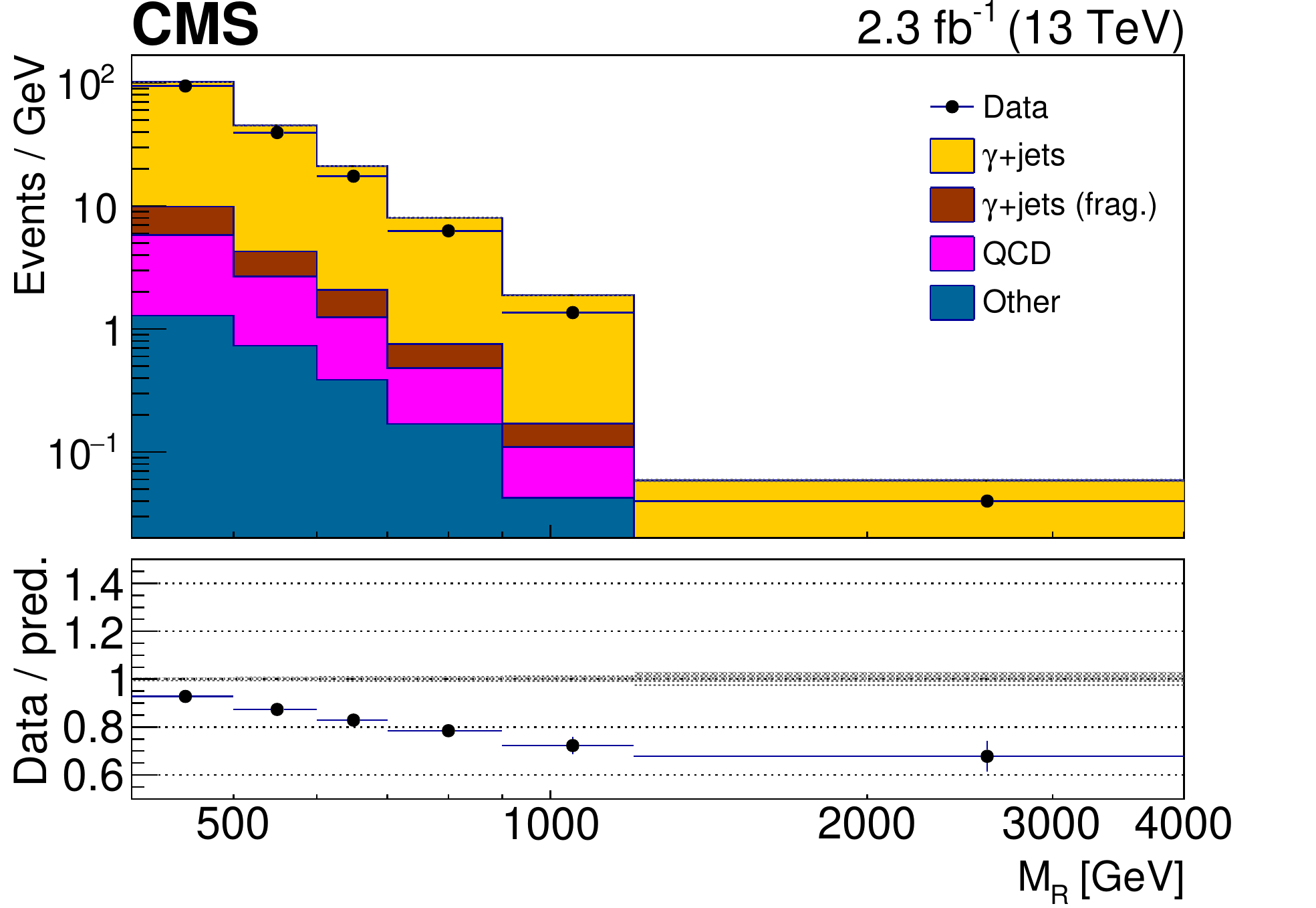} \\
\includegraphics[width=0.50\textwidth]{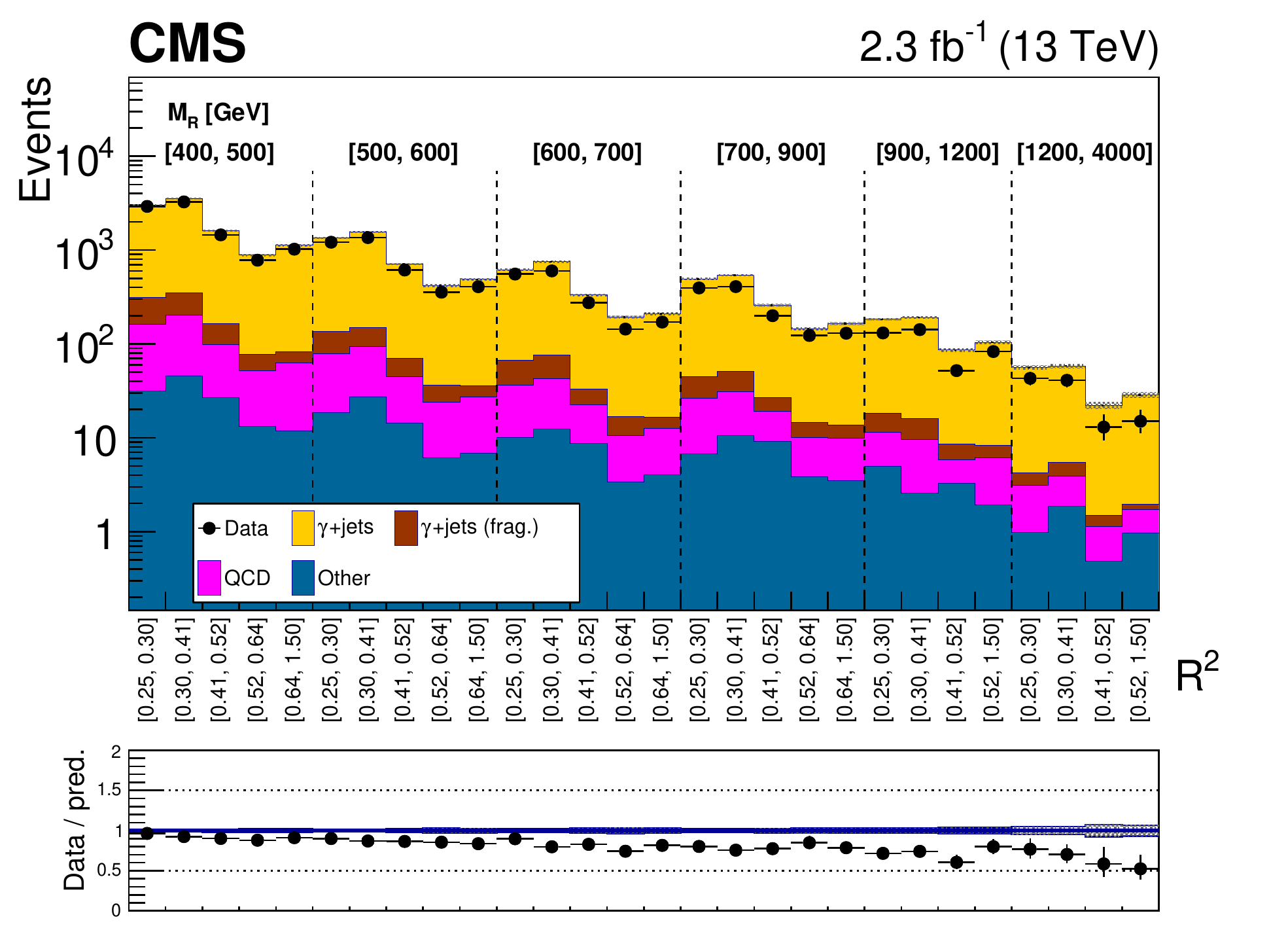}
\caption{The one-dimensional distribution of $\MR$ in the $\cPgg$+jets control region
(above) and the two-dimensional $\MR$-$\Rtwo$ distribution in
the $\cPgg$+jets control region (below) are shown. The two-dimensional $\MR$-$\Rtwo$ distribution
is shown in a one-dimensional representation as in Fig.~\ref{fig:WJets_TTBarDileptonCR_MRRsqUnrolled}.
The bottom panel shows the ratio of data to the background prediction, with uncertainties displayed as in Fig.~\ref{fig:TTBarWJetsCR_MR}.
}
\label{fig:Znn_PhotonJets}
\end{figure}

Three independent control regions are used to predict the $\cPZ(\nu\PAGn)$+jets background,
relying on the assumption that Monte Carlo simulation mismodeling of the hadronic recoil spectrum and the
jet multiplicity distribution of the $\cPZ(\nu\PAGn)$+jets process are similar to those
of the $\PW(\ell\nu)$+jets and $\cPgg$+jets processes. The primary and most populated
control region is the $\cPgg$+jets control region,
defined by selecting events with at least one photon passing loose identification and
isolation requirements. The events are triggered using single-photon triggers, and
the photon is required to have $\pt>50\GeV$. The momentum of the photon candidate
in the transverse plane is added vectorially to $\ptvecmiss$
in order to simulate an invisible particle, as one would have in the case of a
$\cPZ\to\nu\PAGn$ decay, and the $\MR$ and $\Rtwo$ variables are computed according to
this invisible decay scenario.
A template fit to the distribution of $\sigma_{\eta\eta}$ is
performed to determine the contribution from misidentified photons to the $\cPgg$+jets
control region and this is found to be about 5\%, independent of $\MR$ and $\Rtwo$.
Events from the $\cPgg$+jets process where the photon is produced within the cone of a jet
(labeled as $\cPgg$+jets fragmentation) are considered to be background and subtracted
using the MC prediction. Backgrounds from rarer processes such as $\PW\cPgg$, $\cPZ\cPgg$,
and $\ttbar\cPgg$ are also subtracted.
In Fig.~\ref{fig:Znn_PhotonJets}, we show the $\MR$ distribution as well as the two-dimensional $\MR$-$\Rtwo$ distribution for the $\cPgg$+jets control region, where we again
observe a steeper $\MR$ falloff in the data compared to the simulation. Correction factors
are derived in bins of $\MR$ and $\Rtwo$ and applied to the MC prediction for the
$\cPZ\to\nu\PAGn$ background in the search region. The statistical uncertainties for the
correction factors range between 10\% and 30\% and are among the dominant uncertainties
for the $\cPZ\to\nu\PAGn$ background prediction.
Analogously to the procedure for the $\ttbar$ and $\PW(\ell\nu)$+jets control region, we derive an additional correction
factor of $0.87 \pm 0.05$ to accurately describe the yield in events with four or more jets. Additional
cross-checks are performed in bins of the number of b-tagged jets and systematic uncertainties ranging
from 4\% for events with zero b-tagged jets to 58\% for events with three or more b-tagged jets are
derived.

The second control region, enhanced in the $\PW(\ell\nu)$+jets process, is defined
identically to the $\PW(\ell\nu)$+jets control region described in Section~\ref{sec:TTBarWJetsCR}, except that the lepton is treated as invisible
by adding its momentum vectorially to $\ptvecmiss$, and the $\MR$ and $\Rtwo$
variables are computed accordingly. Correction factors computed using events from this control region
are compared to those computed from the $\cPgg$+jets control region and exhibit differences ranging
between 10\% and 40\% depending on the $\MR$-$\Rtwo$ bin. These differences are
propagated as a systematic uncertainty.

The third control region, enhanced in $\cPZ\to\ell^+\ell^-$ decays,
is defined by selecting events with two tight electrons or two tight muons, and requiring that the dilepton mass is
between 76 and 106\GeV. Events are required to have no \PQb-tagged jets
in order to suppress $\ttbar$ background. The two leptons are treated as invisible by adding their
momenta vectorially to $\ptvecmiss$. We apply the correction factors obtained from the
$\cPgg$+jet control region to the $\cPZ\to\ell^+\ell^-$ MC prediction and perform a cross-check against data
in this control region. No significant discrepancy between the data and the prediction is observed.
\subsubsection{The QCD Multijet background}
\label{sec:QCDCR}

The QCD multijet processes contribute about 10\% of the total background in the zero-lepton Multijet
event category for bins with zero or one \PQb-tagged jets. Such events enter the search regions
in the tails of the \MET distribution when the energy of
one of the jets in the event is significantly under- or over-measured.
In most such situations, the \ptvecmiss points either toward
or away from the leading jets and therefore the two megajets tend to
be in a back-to-back configuration. The search region is defined by requiring that
the azimuthal angle between the two megajets $\Delta\phi_R$ be less than
2.8, which was found to be an optimal selection based on studies
of QCD multijet and signal simulated samples. We define the control region for the QCD background process to be events
with $\dPhiR>2.8$, keeping all other selection requirements identical to those for
the search region. The purity of the QCD multijet process in the control region
is more than 70\%.

\begin{figure}[!ptb]
\begin{center}
\includegraphics[width=0.45\textwidth]{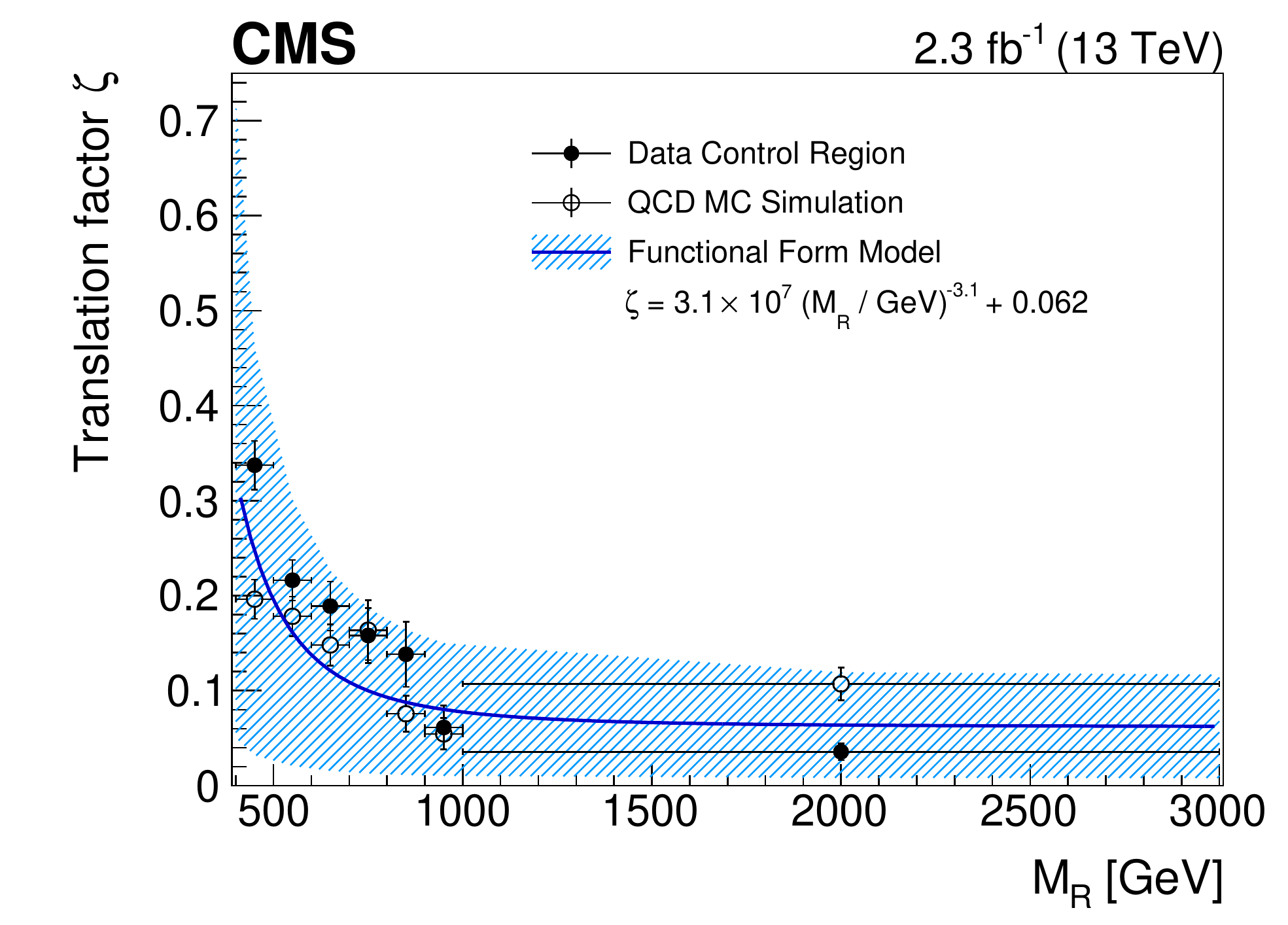}
\caption{\label{fig:QCDTranslationFactor}
The translation factor $\zeta$ is shown as a function of $\MR$. The curve shows the
functional form used to model the $\MR$ dependence, and the open circle
and black dot data points are  the values of $\zeta$ measured in the low-$\Rtwo$ data
control region and the QCD MC simulation, respectively. The hashed region indicates the size of the systematic uncertainty in
$\zeta$.
}
\end{center}
\end{figure}

After subtracting the non-QCD background,
we project the observed data yield in the control region to the search region using
the translation factor $\zeta$:
\begin{equation}
\zeta = \frac{N(|\dPhiR|<2.8)}{N(|\dPhiR|>2.8)},
\end{equation}
where the numerator and denominator are the number of events
passing and failing the selection on $|\dPhiR|<2.8$, respectively. We
find that the translation factor calculated from the MC simulation
decreases as a function of $\MR$ and is, to a large degree, constant as a function of $\Rtwo$.
Using data events in the low $\Rtwo$ region (0.15 to 0.25), dominated
by QCD multijet background, we measure the translation factor $\zeta$ as a function of
$\MR$ to cross-check the values obtained from the simulation.
The $\MR$ dependence of $\zeta$ is modeled as the sum of a power law
and a constant. This functional shape is fitted to the values of $\zeta$ calculated from the MC.
A systematic uncertainty of 87\% is propagated, covering both the
spread around the fitted model as a function of $\MR$ and $\Rtwo$ in
simulation, and the difference between the values measured in simulation
and data. The function used
for $\zeta$ and the values measured in data and simulation are
shown in Fig.~\ref{fig:QCDTranslationFactor}.

We perform two additional cross-checks on the accuracy of the MC prediction for
$\zeta$ in control regions dominated by processes similar to the QCD multijet
background with no invisible neutrinos in the final state. The first
cross-check is performed on a dimuon control region enhanced in $\cPZ\to\Pgm^+\Pgm^-$ decays,
and the second cross-check is performed on a dijet control region enhanced in QCD dijet events.
In both cases, the events at large $\Rtwo$ result from cases similar to our search region
where the energy of a leading jet is severely mismeasured. We compare the values of
$\zeta$ measured in these data control regions to the values predicted
by the simulation and observe agreement within 20\%, well within the
systematic uncertainty of 87\% assigned to the QCD background estimate.
\subsection{Method B: fit-based background prediction}
\label{sec:FitBkg}
The second background prediction method is based on a fit to the data with an
assumed functional form for the shape of the background distribution in the $\MR$-$\Rtwo$ plane.
Based on past studies~\cite{razorPRD,razor8TeV}, the shape of the background in
the $\MR$ and $\Rtwo$ variables is found to be well described by the following functional form:
\ifthenelse{\boolean{cms@external}}{
\begin{multline}
f_{\mathrm{SM}}(\MR,\Rtwo) =  \bigl[b(\MR-{\MRz})^{1/n}(\Rtwo-{\Rtwoz})
^{1/n}-1\bigr]\times\\
\re^{-bn(\MR-{\MRz})^{1/n}(\Rtwo-{\Rtwoz})
^{1/n}} ,
\label{eq:razFunction}
\end{multline}
}{
\begin{equation}
f_{\mathrm{SM}}(\MR,\Rtwo) =  \bigl[b(\MR-{\MRz})^{1/n}(\Rtwo-{\Rtwoz})
^{1/n}-1\bigr]\re^{-bn(\MR-{\MRz})^{1/n}(\Rtwo-{\Rtwoz})
^{1/n}} ,
\label{eq:razFunction}
\end{equation}
}
where $\MRz$, $\Rtwoz$, $b$, and $n$ are free parameters.
In the original study~\cite{razorPRD}, this function with $n$ fixed to
1 was used to model the data in each category. The function choice
was motivated by the observation that for $n=1$, the
function projects to an exponential both on $\Rtwo$ and $\MR$, and $b$
is proportional to the exponential rate parameter in each
one-dimensional projection. The generalized function
in Eq.~(\ref{eq:razFunction}) was found to be in better agreement with the SM
backgrounds over a larger range of $\Rtwo$ and $\MR$~\cite{razor8TeV}
in comparison to the choice with $n$ fixed to 1. The two
parameters $b$ and $n$ determine the tail of the distribution in the
two-dimensional plane, while the $\MRz$ ($\Rtwoz$) parameter affects the tail of the
one-dimensional projection on $\Rtwo$ ($\MR$).

The background estimation is performed using an extended, binned, maximum likelihood fit to the $\MR$ and $\Rtwo$
distribution in one of two ways:
\begin{itemize}
\item A fit to the data in the sideband regions in $\MR$ and
$\Rtwo$, defined more precisely below, as a model-independent way to look for excesses or
discrepancies. The fit is performed using only the data in the
sideband, and the functional form is extrapolated to the full $\MR$ and $\Rtwo$ plane.
\item A fit to the data in the full search region in $\MR$ and $\Rtwo$ under
background-only and signal-plus-background hypotheses, following
a modified frequentist approach (LHC $\CLs$)~\cite{Junk:1999kv, Read:2002hq,Read:2000ru,Cowan:2010js,ATLAS:2011tau}
to interpret the data in the context of particular SUSY simplified models.
\end{itemize}
The sideband region is defined to be 100\GeV in width in $\MR$
and 0.05 in $\Rtwo$. Explicitly, for the
Multijet event category, it comprises the region $500\GeV < \MR < 600\GeV$ and $\Rtwo > 0.3$, plus the region $\MR > 500\GeV$
and $0.25 < \Rtwo < 0.3$.  For the Muon and Electron Multijet
event categories, it comprises the region $400\GeV < \MR < 500\GeV$
and $\Rtwo > 0.2$, plus the region $\MR > 400\GeV$ and
$0.15 < \Rtwo < 0.2$.

For each event category, we fit the two-dimensional distribution of
$\MR$ and $\Rtwo$ in the sideband region using the
above functional form, separately for events with zero, one, two, and three or more \PQb-tagged jets. The
normalization in each event category and each \PQb-tagged jet bin is
independently varied in the fit. Due to the lack of data events in the category
with three or more \PQb-tagged jets, we constrain
the shape in this category to be related to the shape for events with two
\PQb-tagged jets as follows:
\begin{equation}
f^{{\geq}3\PQb}_{\mathrm{SM}}(\MR, \Rtwo)  = (1+m_{\MR}(\MR - M^{\mathrm{offset}}_R))f^{2\PQb}_{\mathrm{SM}}(\MR, \Rtwo),
\label{eq:3btagFunction}
\end{equation}
where $f^{2\PQb}_{\mathrm{SM}}(\MR, \Rtwo)$ and $f^{{\geq}3\PQb}_{\mathrm{SM}}(\MR, \Rtwo)$ are the
probability density functions for events with two and with three or more \PQb-tagged jets,
respectively; $\MR^{\mathrm{offset}}$ is the lowest $\MR$ value in a particular
event category; and $m_{\MR}$ is a floating parameter constrained by a Gaussian distribution
centered at the value measured using the simulation and with a
100\% uncertainty. The above form for the shape of the background events
with three or more \PQb-tagged jets is verified in simulation.

Numerous tests are performed to establish the robustness of the fit
model in adequately describing the underlying distributions. To
demonstrate that the background model gives an accurate description of the
background distributions, we construct a representative
data set using MC samples, and perform the background fit using
the form given by Eq.~(\ref{eq:razFunction}). Goodness of fit is
evaluated by comparing the background prediction from the fit with the
prediction from the simulation. This procedure is performed
separately for each of the search categories and we find
that the fit function yields an accurate representation of the
background predicted by the simulation.

We also observe that the accuracy of the fit model is insensitive to variations of the background
composition predicted by the simulation in each event category by altering
relative contributions of the dominant backgrounds, performing a
new fit with the alternative background composition, and comparing
the new fit results to the nominal fit result. The contributions
of the main $\ttbar$, $\PW(\ell\nu)$+jets, and $\cPZ(\nu\PAGn)$ backgrounds
are varied by 30\%, and the rare backgrounds from QCD multijet and
$\mathrm{\ttbar}\cPZ$  processes are varied by 100\%. For the Muon and Electron Multijet event categories,
we also vary the contributions from the dileptonic and semi-leptonic decays
of the $\ttbar$ background separately by 30\%. In each of these tests, we
observe that the chosen functional form can adequately describe the shapes of
the $\MR$ and $\Rtwo$ distributions as predicted by the modified MC simulation.

Additional pseudo-experiment studies are performed comparing the background prediction
from the sideband fit and the full region fit to evaluate the average
deviation between the two fit predictions. We observe that
the sideband fit and the full region fit predictions in the
signal-sensitive region differ by up to 15\% and we propagate an
additional systematic uncertainty to the sideband fit background
prediction to cover this average difference.
\begin{figure}[!ptb] \centering
\includegraphics[width=0.50\textwidth]{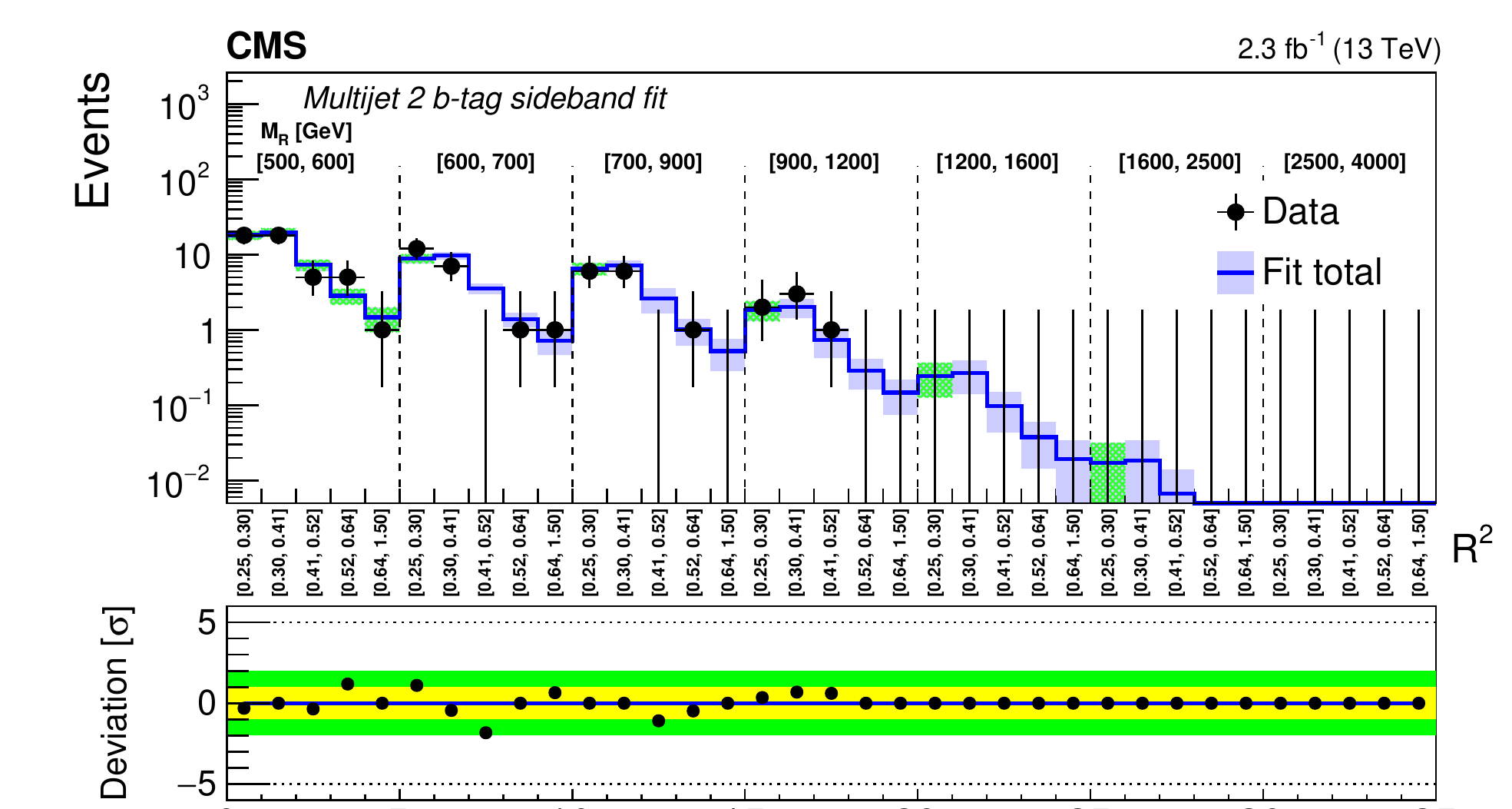}\\
\includegraphics[width=0.50\textwidth]{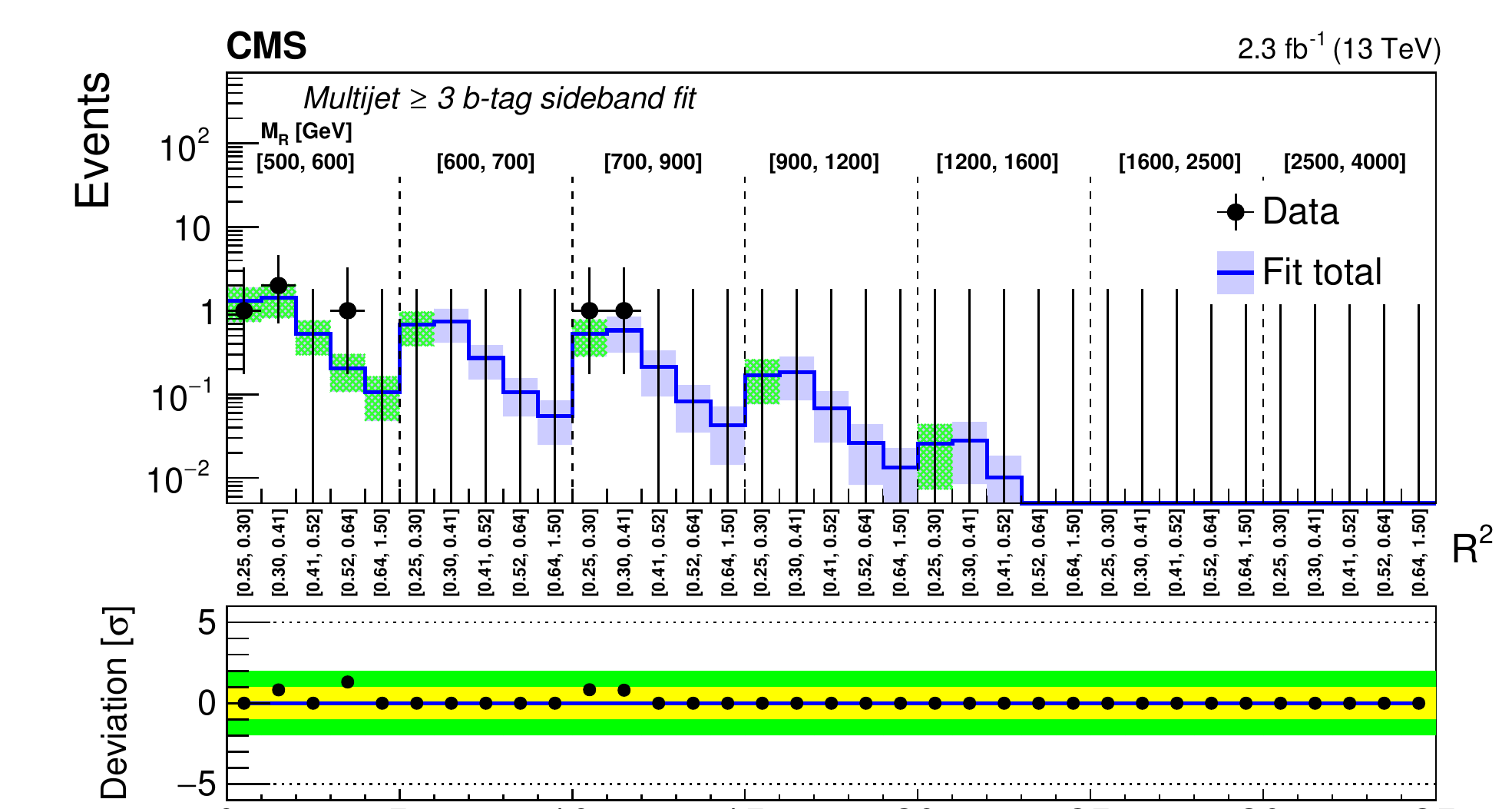}
\caption{Comparison of the sideband fit background prediction with the observed data
in bins of $\MR$ and $\Rtwo$ variables in the Multijet category for
the 2 \PQb-tag (upper) and ${\geq}3$ \PQb-tag (lower) bins. Vertical dashed lines denote the
boundaries of different $\MR$ bins. On the
upper panels, the colored bands represent the
systematic uncertainties in the background prediction, and the uncertainty bands for the
sideband bins are shown in green. On the bottom panels, the deviations between the observed
data and the background prediction are plotted in units of standard
deviation ($\sigma$), taking into
account both statistical and systematic uncertainties. The green
and yellow horizontal bands show the boundaries of 1 and $2\sigma$. }
\label{fig:results_Multijet2btag3btag}
\end{figure}
\begin{figure}[!ptb] \centering
\includegraphics[width=0.50\textwidth]{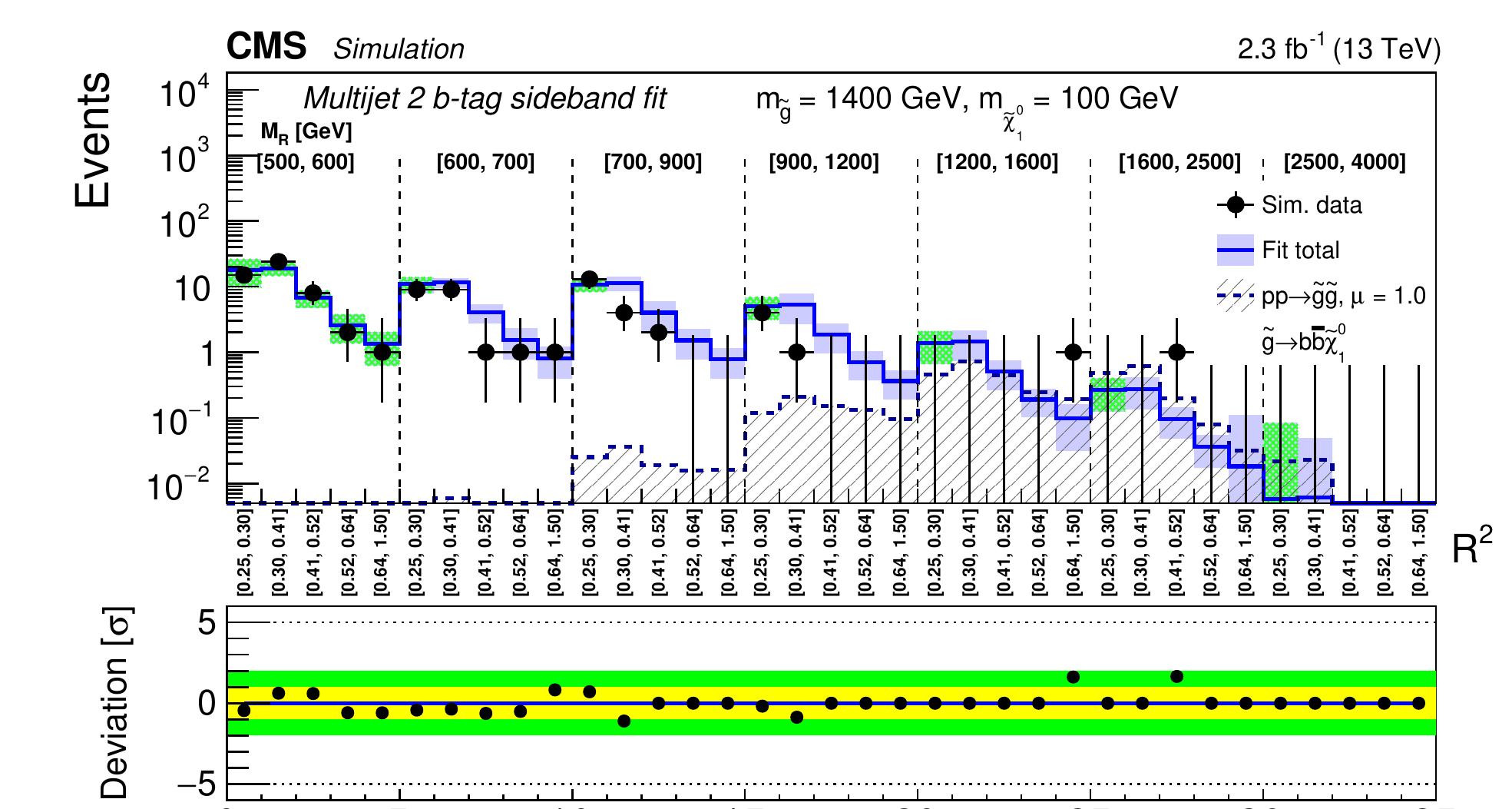}\\
\includegraphics[width=0.50\textwidth]{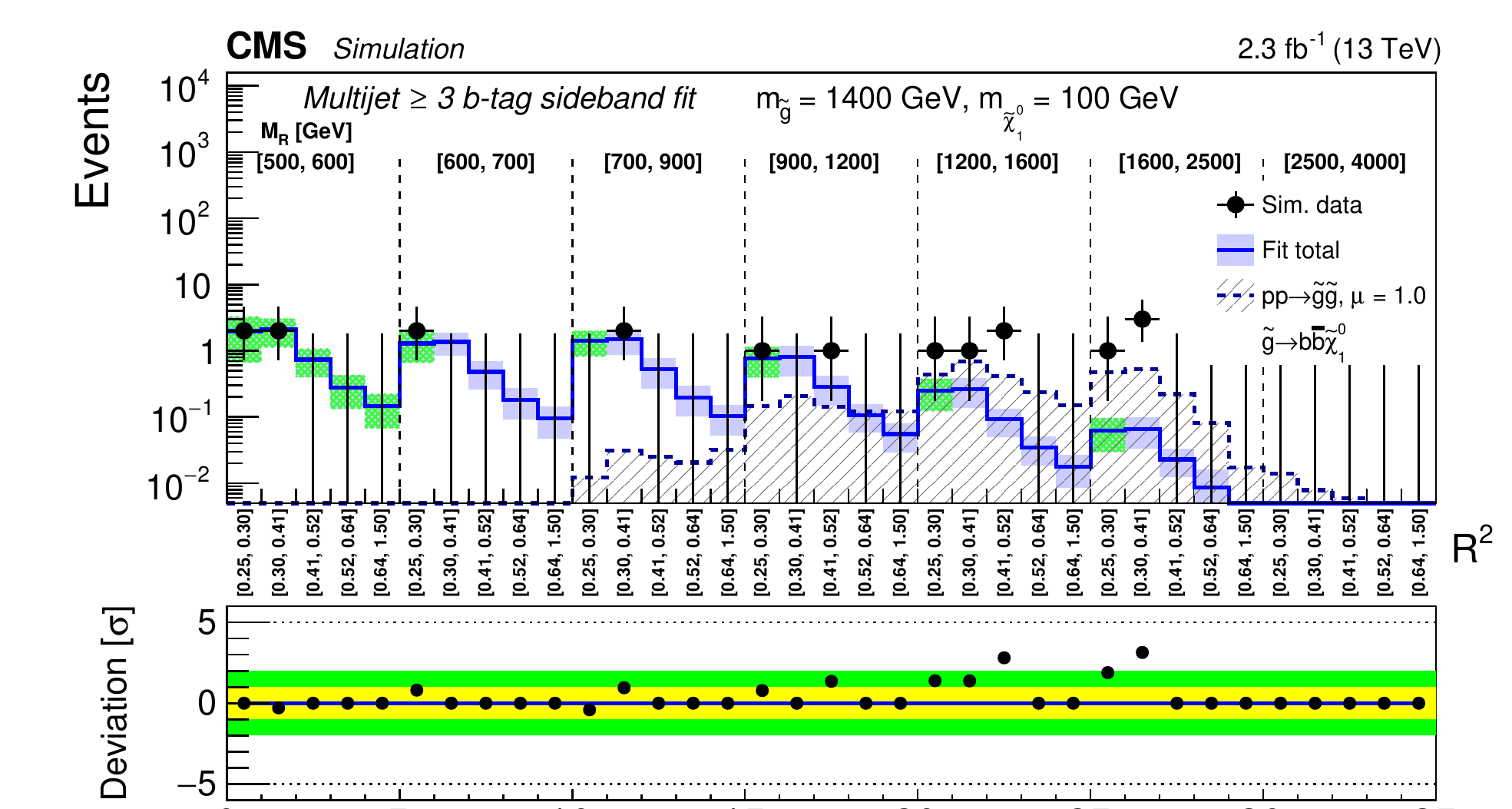}
\caption{The result of the background-only fit performed in the
sideband of the 2 \PQb-tag (upper) and ${\geq}3$ \PQb-tag (lower) bins of the
Multijet category on a signal-plus-background pseudo-data set assuming a gluino pair production simplified
model signal, where gluinos decay with a 100\% branching fraction to a $\bbbar$ pair and the
LSP, with $m_{\PSg} = 1.4\TeV$ and  $m_{\PSGczDo}=100\GeV$, at
nominal signal strength. A detailed explanation of the figure format is given in the caption of
Fig.~\ref{fig:results_Multijet2btag3btag}.}
\label{fig:signal_Multijet2btag3btag}
\end{figure}

To illustrate method B, we present the data and fit-based background predictions
in Fig.~\ref{fig:results_Multijet2btag3btag}, for events in the 2 \PQb-tag and ${\geq}3$ \PQb-tag
Multijet categories. The number of events observed in data is compared to the
prediction from the sideband fit in the $\MR$ and $\Rtwo$ bins. To
quantify the agreement between the background model and the observation, we generate
alternative sets of background shape parameters from the covariance matrix calculated
by the fit. An ensemble of pseudo-experiment data sets is created, generating
random ($\MR$, $\Rtwo$) pairs distributed according to each of these alternative shapes.
For each $\MR$-$\Rtwo$ bin, the distribution of the predicted yields from the
ensemble of pseudo-experiments is compared to the observed yield in data.
The agreement between the predicted and the observed yields is described as a two-sided
\textit{p}-value and translated into the corresponding number of standard deviations for a normal
distribution. Positive (negative) significance indicates the observed
yield is larger (smaller) than the predicted one. We find that the pattern of
differences between data and background predictions in the different
bins considered is consistent with statistical fluctuations.

To demonstrate that the model-independent sideband fit procedure
used in the analysis would be sensitive to the presence of a
signal, we perform a signal injection test. We sample a signal-plus-background
pseudo-data set and perform a background-only fit in the sideband. We show one illustrative
example of such a test in Fig.~\ref{fig:signal_Multijet2btag3btag}, where we inject a signal
corresponding to gluino pair production, in which each gluino decays to a neutralino and
a $\bbbar$ pair with $m_{\PSg} = 1.4\TeV$ and $m_{\PSGczDo}=100\GeV$. The
deviations with respect to the fit predictions are shown for the 2
\PQb-tag and ${\geq}3$ \PQb-tag Multijet categories. We observe characteristic patterns
of excesses in two adjacent groups of bins neighboring in $\MR$.
\subsection{Comparison of two methods}
\begin{figure}[!ptb] \centering
\includegraphics[width=0.50\textwidth]{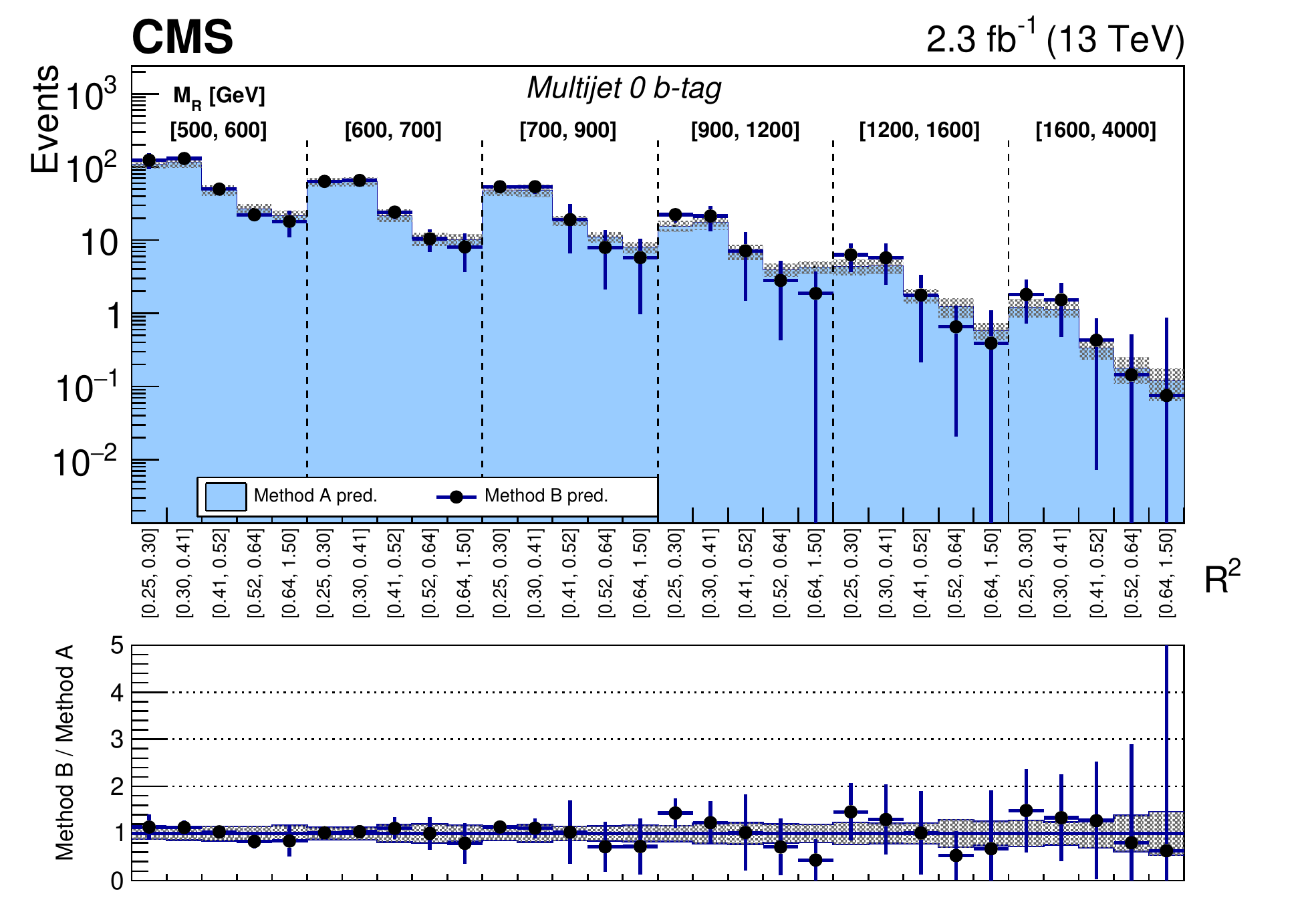}
\includegraphics[width=0.50\textwidth]{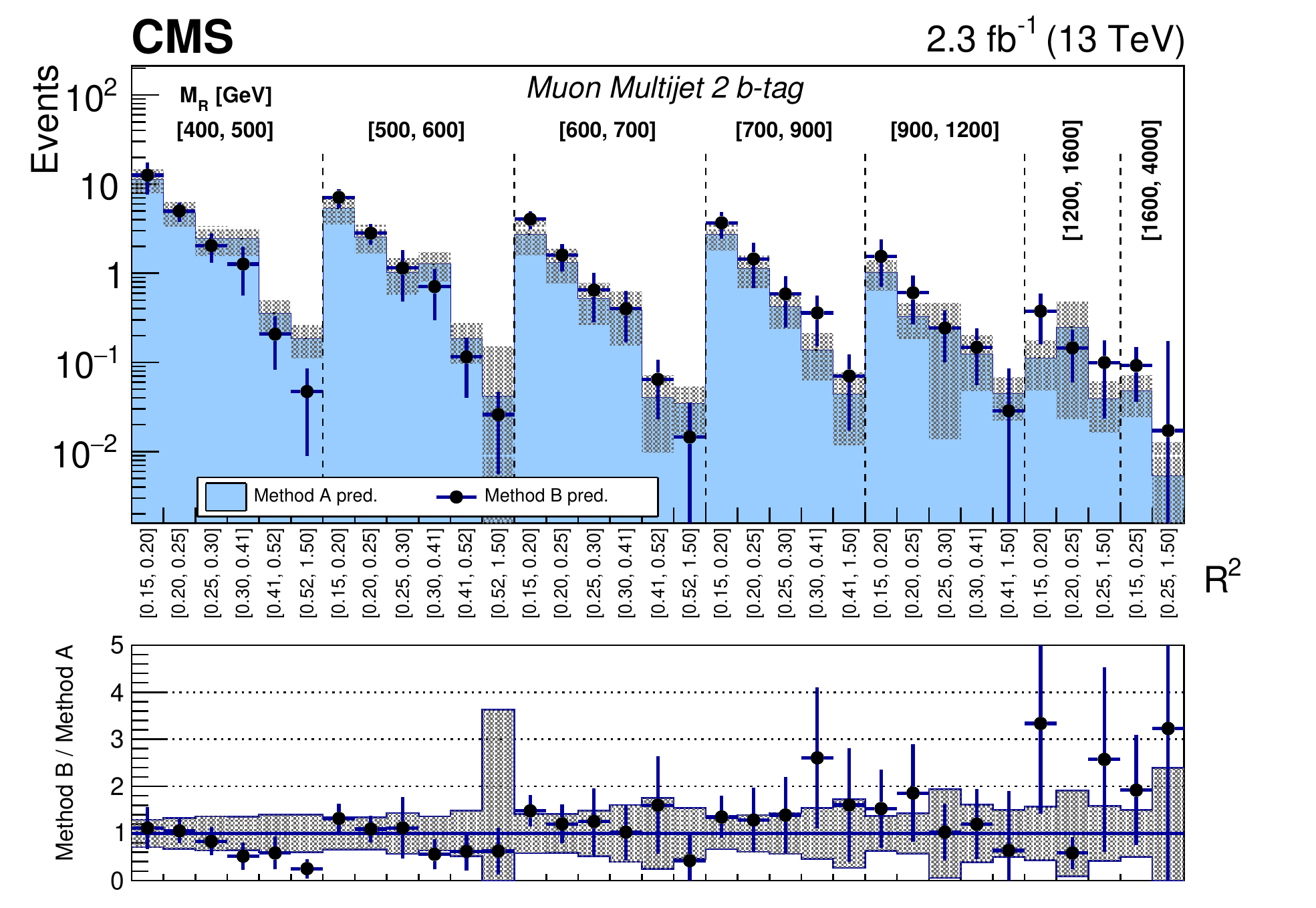}
\caption{Comparisons of the two alternative background predictions for the $\MR$-$\Rtwo$ distribution
for the 0 \PQb-tag bin of the Multijet category (upper) and the 2 \PQb-tag bin of the Muon Multijet
category (lower). The two-dimensional $\MR$-$\Rtwo$ distribution is shown
in a one dimensional representation, with each $\MR$ bin marked by the dashed lines and labeled near the top
and each $\Rtwo$ bin labeled below. The ratios of the method B fit-based predictions to the method A simulation-assisted predictions are shown
on the bottom panels. The method B uncertainty is represented by the error bars on the data points and the
method A uncertainty is represented by the shaded region.
}
\label{fig:FitVsMADD}
\end{figure}

The background predictions obtained from methods A and B are systematically compared
in all of the search region categories. For method B, the model-independent
fit to the sideband is used for this comparison. In Fig.~\ref{fig:FitVsMADD},
we show the comparison of the two background predictions for two example event categories.
The predictions from the two methods agree within the uncertainties of each method.
The uncertainty from the fit-based method tends to be slightly larger at high
$\MR$ and $\Rtwo$ due to the additional uncertainty in the exact shape of
the tail of the distribution, as the $n$ and $b$ parameters are not strongly
constrained by the sideband data.

The two background predictions use methods based on data that make
very different systematic assumptions. Method A assumes that corrections
to the simulation prediction measured in control regions apply also to the
signal regions, while method B assumes that the shape of the background distribution
in $\MR$ and $\Rtwo$ is well described by a particular exponentially falling functional
form. The agreement observed between predictions obtained using these two very different
methods significantly enhances the confidence of the background modeling, and
also validates the respective assumptions.

\section{Systematic uncertainties}
\label{sec:Systematics}

Various systematic uncertainties are considered in the evaluation of the
signal and background predictions. Different types of systematic
uncertainties are considered for the two different background models.

For method A, the largest uncertainties arise from the precision with
which the MC corrections are measured. The dominant uncertainties
in the correction factors result from statistical uncertainties due to
the limited size of the control region event sample. We also propagate systematic
uncertainties in the theoretical cross-section for the small residual backgrounds
present in the control regions, and they contribute $2-5\%$ to the
correction factor uncertainty.
Additional systematic uncertainties are computed from the procedure that
tests that the accuracy of the MC corrections as a function of
($\MR$, $\Rtwo$), and the number of \PQb-tagged jets in events with four or more jets.
The total uncertainty from this procedure ranges from 10\% for the most populated bins to
50\% and 100\% for the least populated bins. For the $\cPZ\to\nu\PAGn$ process, we
also propagate the difference in the correction factors measured in the three alternative
control regions as a systematic uncertainty, intended to estimate the possible differences in
the simulation mismodeling of the hadronic recoil for the $\cPgg$+jets process and
the $\cPZ(\nu\PAGn)$+jets process. These systematic uncertainties
range from 10 to 40\%. For the QCD multijet background prediction the statistical uncertainty
due to limited event counts in the $\dPhiR>2.8$ control regions and the systematic
uncertainty of 87\% in the translation factor $\zeta$ are propagated.

For method B, the systematic uncertainties in the background are propagated as part of
the maximum likelihood fit procedure. For each event category, the background shape in
$\MR$ and $\Rtwo$ is described by four independent parameters: two
that control the exponential fall off and two that control the behavior of the
nonexponential tail. Systematic uncertainties in the background are propagated
through the freedom of these unconstrained shape parameters in the fit model. For more populated bins, such as
the 0 \PQb-tag and 1 \PQb-tag bins in the Multijet category, the systematic uncertainties range from
about 30\% at low $\MR$ and $\Rtwo$ to about 70\% at high $\MR$ and $\Rtwo$.
For sparsely populated bins such as the 3-or-more \PQb-tag bin in the Muon Multijet or Electron
Multijet categories, the systematic uncertainties range from
about 60\% at low $\MR$ and $\Rtwo$ to more than 200\% at high $\MR$ and $\Rtwo$.
\begin{table}[!tpb]
\centering
\renewcommand{\arraystretch}{1.2}
\topcaption{Summary of the main instrumental and theoretical systematic
uncertainties. The systematic uncertainty associated to the modeling
of the initial-state radition is only applied for events with recoil above 400\GeV.}
\label{tab:BackgroundSystematics}
\resizebox{\columnwidth}{!}{
\begin{scotch}{lcc}
\multirow{2}{*}{Source}      & On signal    &  Typical values  \\
    &  and/or bkg    &   [\%]  \\
\hline
Jet energy scale                   & Both                   & \x2--15 \\
Electron energy scale              & Both                   & 7--9 \\
Muon momentum scale                & Both                   & 7--9 \\
Muon efficiency                    & Both                   & 7--8 \\
Electron efficiency                & Both                   & 7--8 \\
Trigger efficiency                 & Both                   & 3 \\
\PQb-tagging  efficiency           & Both                   & \x6--15 \\
\PQb mistagging  efficiency        & Both                   & 4--7 \\
Missing higher orders            & Both                   & 10--25 \\
Integrated luminosity                         & Both                   & 2.7 \\
Fast simulation corrections        & Signal only            & \x0--10 \\
Initial-state radiation            & Signal only            & 15--30 \\
\end{scotch}
}
\end{table}

Systematic uncertainties due to instrumental and theoretical effects are propagated as shape
uncertainties in the signal predictions for methods A and B, and on the background
predictions for method A. The background prediction from method B is not affected
by these uncertainties as the shape and normalization are measured from data.
Uncertainties in the trigger and lepton selection efficiency, and the
integrated luminosity~\cite{CMS-PAS-LUM-15-001} primarily affect the total normalization. Uncertainties in
the \PQb-tagging efficiency affect the relative yields between different \PQb-tag categories.
The uncertainties from missing higher-order corrections and the uncertainties in the jet
energy and lepton momentum scale affect the shapes of the $\MR$ and $\Rtwo$ distributions.

For the signal predictions, we also propagate systematic uncertainties due to
possible inaccuracies of the fast simulation in modeling the lepton selection and
\PQb tagging efficiencies. These uncertainties were evaluated by comparing
the $\ttbar$ and signal $\GEANT$ based MC samples with those
that used fast simulation. Finally, we propagate an uncertainty in the modeling of initial-state radiation for signal predictions, that ranges from 15\% for signal events with
recoil between 400 and 600\GeV to 30\% for events with recoil above 600\GeV.
The systematic uncertainties and their typical impact on the background and signal
predictions are summarized in Table~\ref{tab:BackgroundSystematics}.
\section{Results and interpretations}
\label{sec:Results}
\begin{figure}[!ptb] \centering
\includegraphics[width=0.50\textwidth]{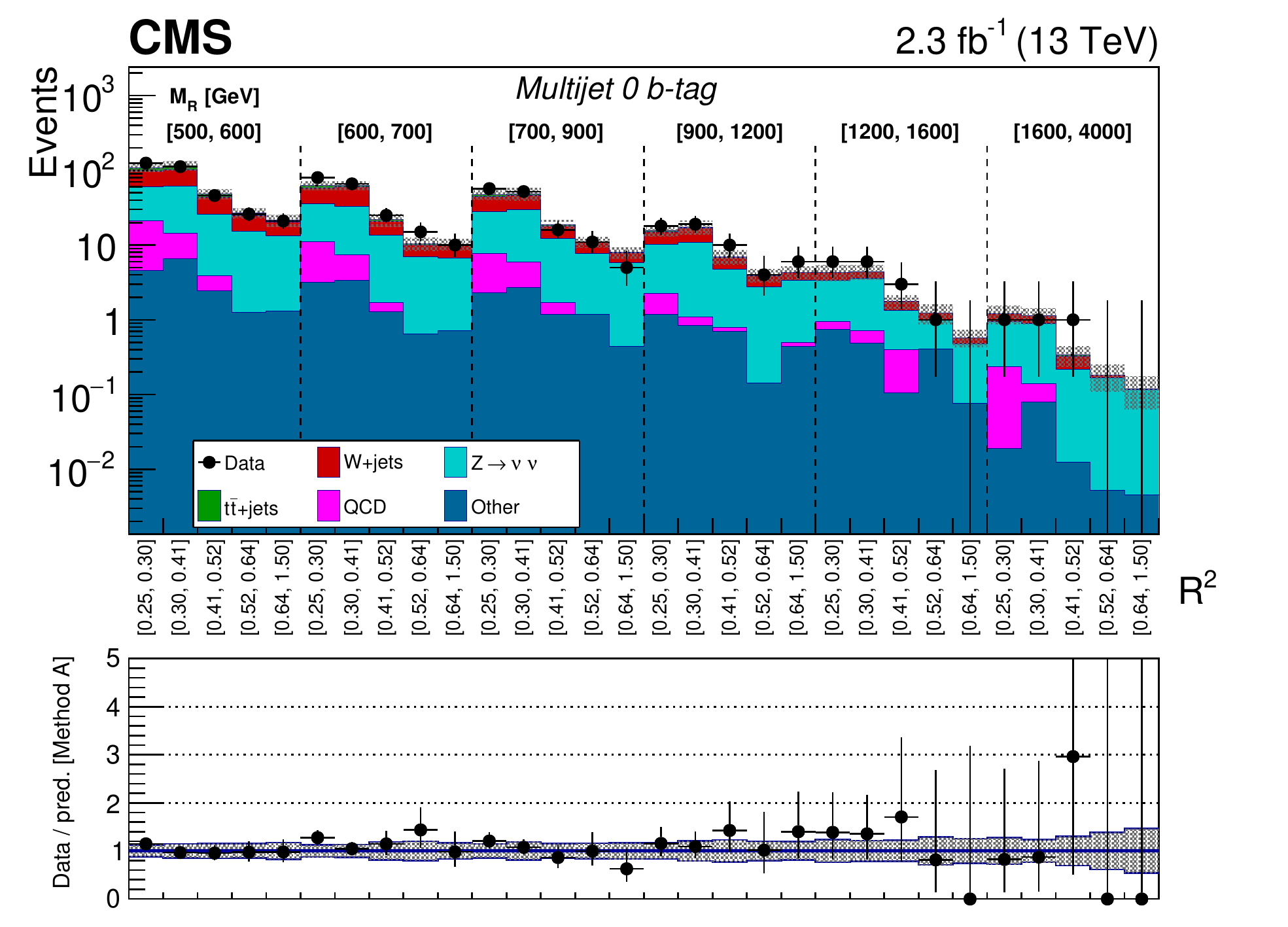}\\
\includegraphics[width=0.50\textwidth]{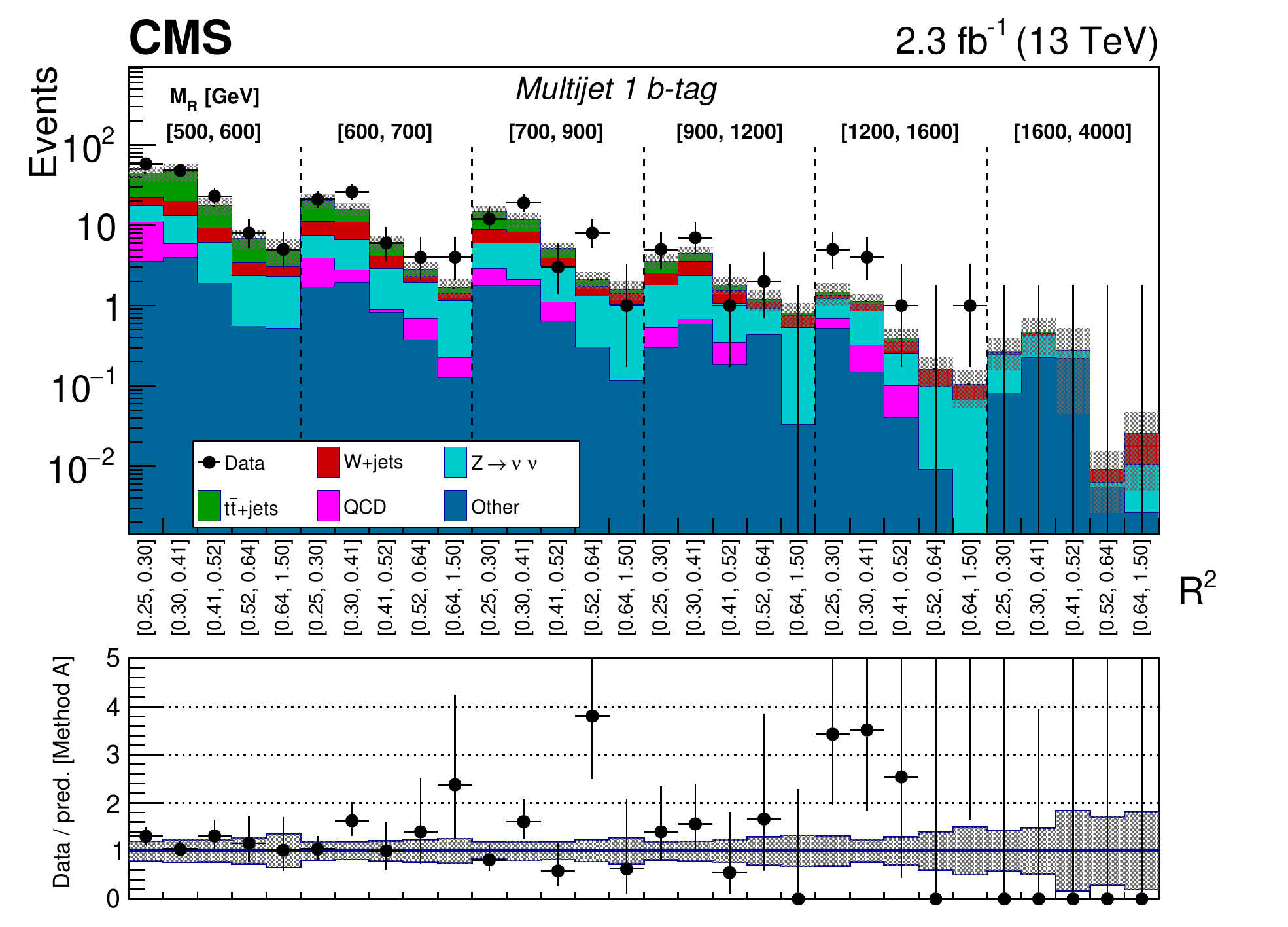}
\caption{ The $\MR$-$\Rtwo$ distribution observed in data is shown along with the background prediction
obtained from method A for the Multijet event category in the 0
\PQb-tag (upper) and 1 \PQb-tag (lower) bins. The two-dimensional $\MR$-$\Rtwo$ distribution is shown
in a one-dimensional representation, with each $\MR$ bin marked by the dashed lines and labeled near the top,
and each $\Rtwo$ bin labeled below. The ratio of data to the background
prediction is shown on the bottom panels, with
the statistical uncertainty expressed through the data point error bars and the systematic uncertainty of the
background prediction represented by the shaded region.
}
\label{fig:ResultsMultiJet0btag1btag}
\end{figure}

\begin{figure}[!ptb] \centering
\includegraphics[width=0.50\textwidth]{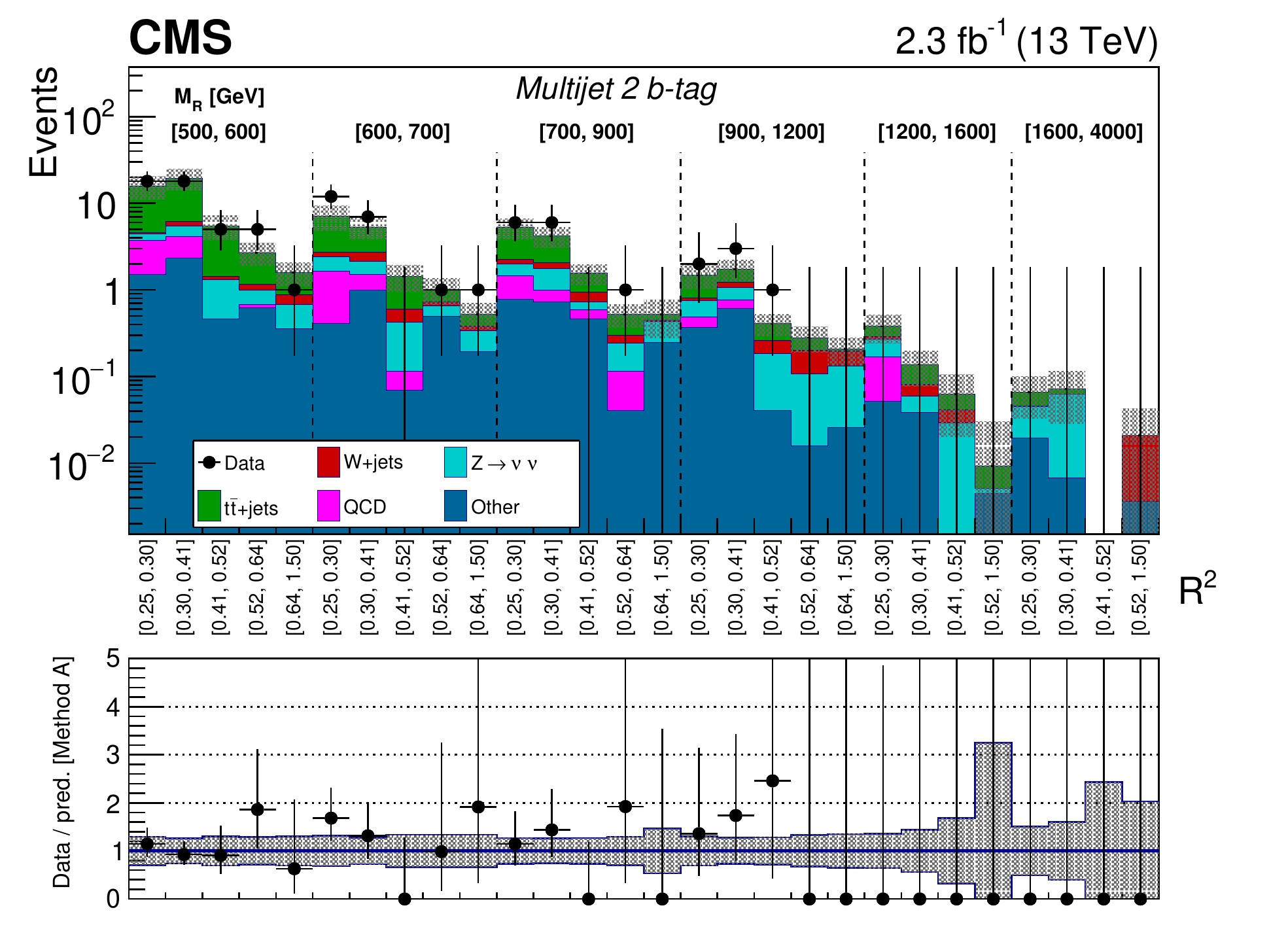}\\
\includegraphics[width=0.50\textwidth]{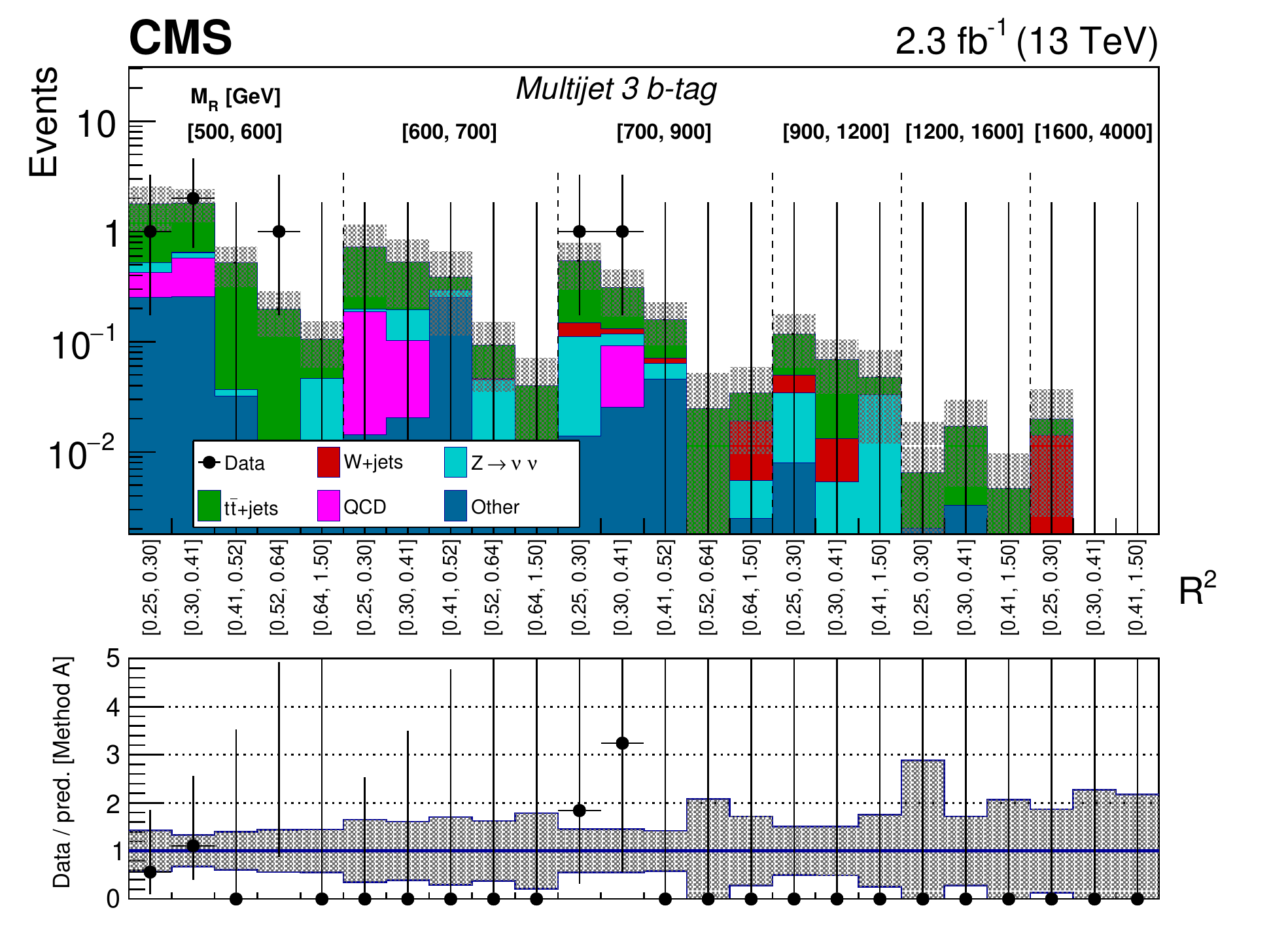}
\caption{ The $\MR$-$\Rtwo$ distribution observed in data is shown along with the background prediction
obtained from method A for the Multijet event category in the 2 \PQb-tag (upper) and ${\geq}3$ \PQb-tag (lower) bins.
A detailed explanation of the panels is given in the caption of   Fig.~\ref{fig:ResultsMultiJet0btag1btag}.
}
\label{fig:ResultsMultiJet2btag3btag}
\end{figure}

\begin{figure}[!ptb] \centering
\includegraphics[width=0.50\textwidth]{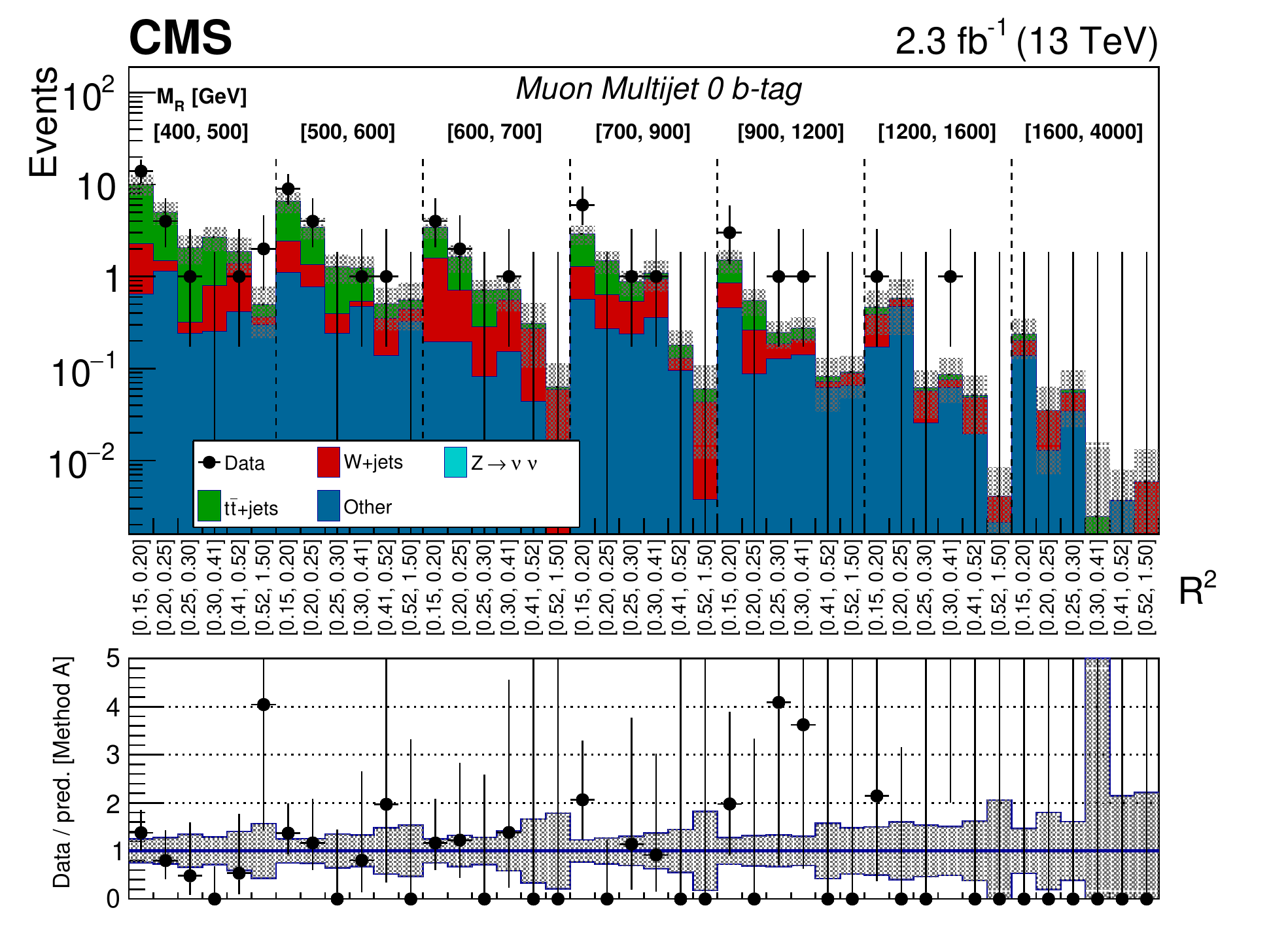}\\
\includegraphics[width=0.50\textwidth]{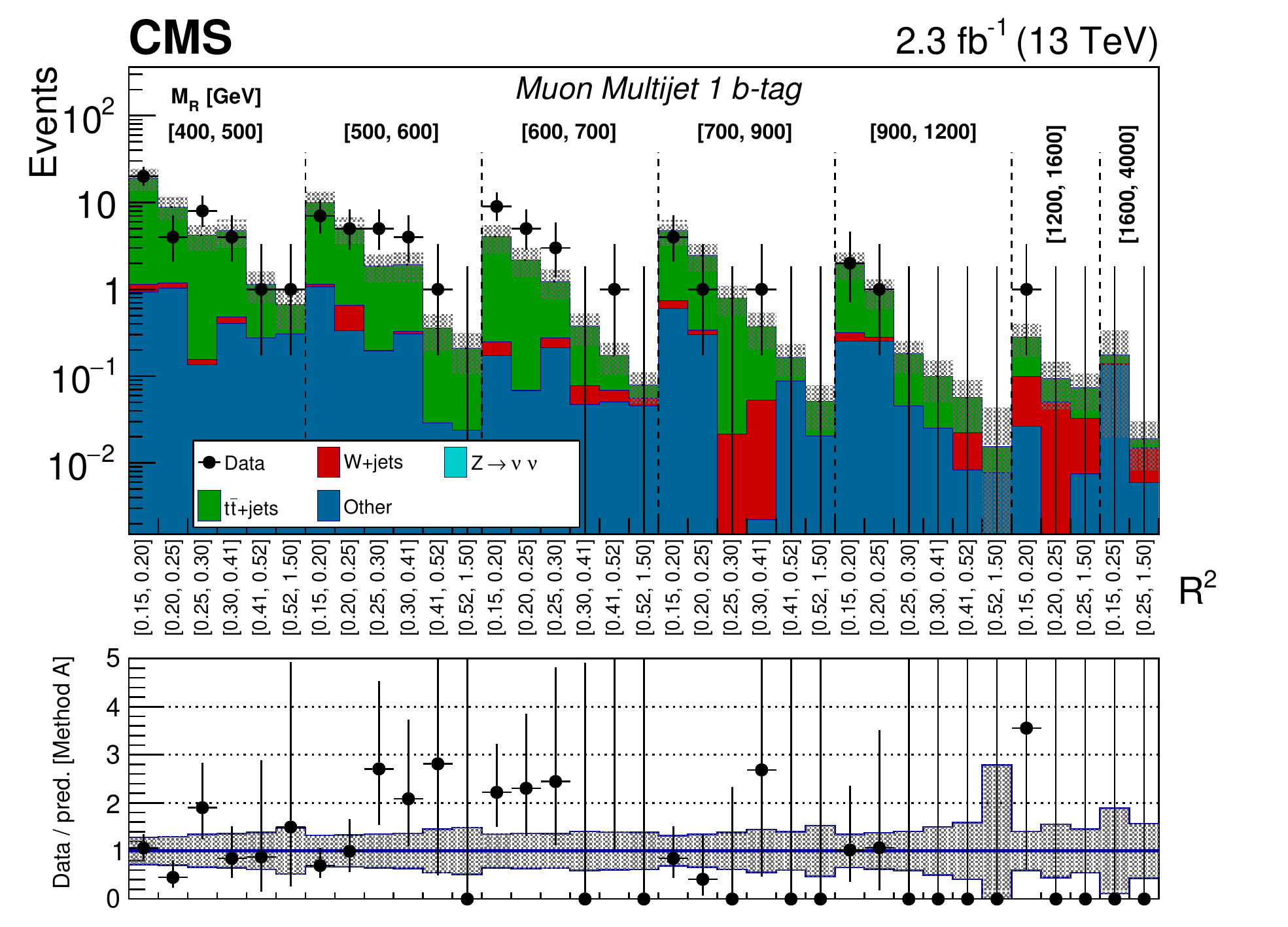}
\caption{ The $\MR$-$\Rtwo$ distribution observed in data is shown along with the background prediction
obtained from method A for the Muon Multijet event category in the 0 \PQb-tag (upper) and 1 \PQb-tag (lower) bins.
A detailed explanation of the panels is given in the caption of
Fig.~\ref{fig:ResultsMultiJet0btag1btag}.
}
\label{fig:ResultsMuMultiJet0btag1btag}
\end{figure}

\begin{figure}[!ptb] \centering
\includegraphics[width=0.50\textwidth]{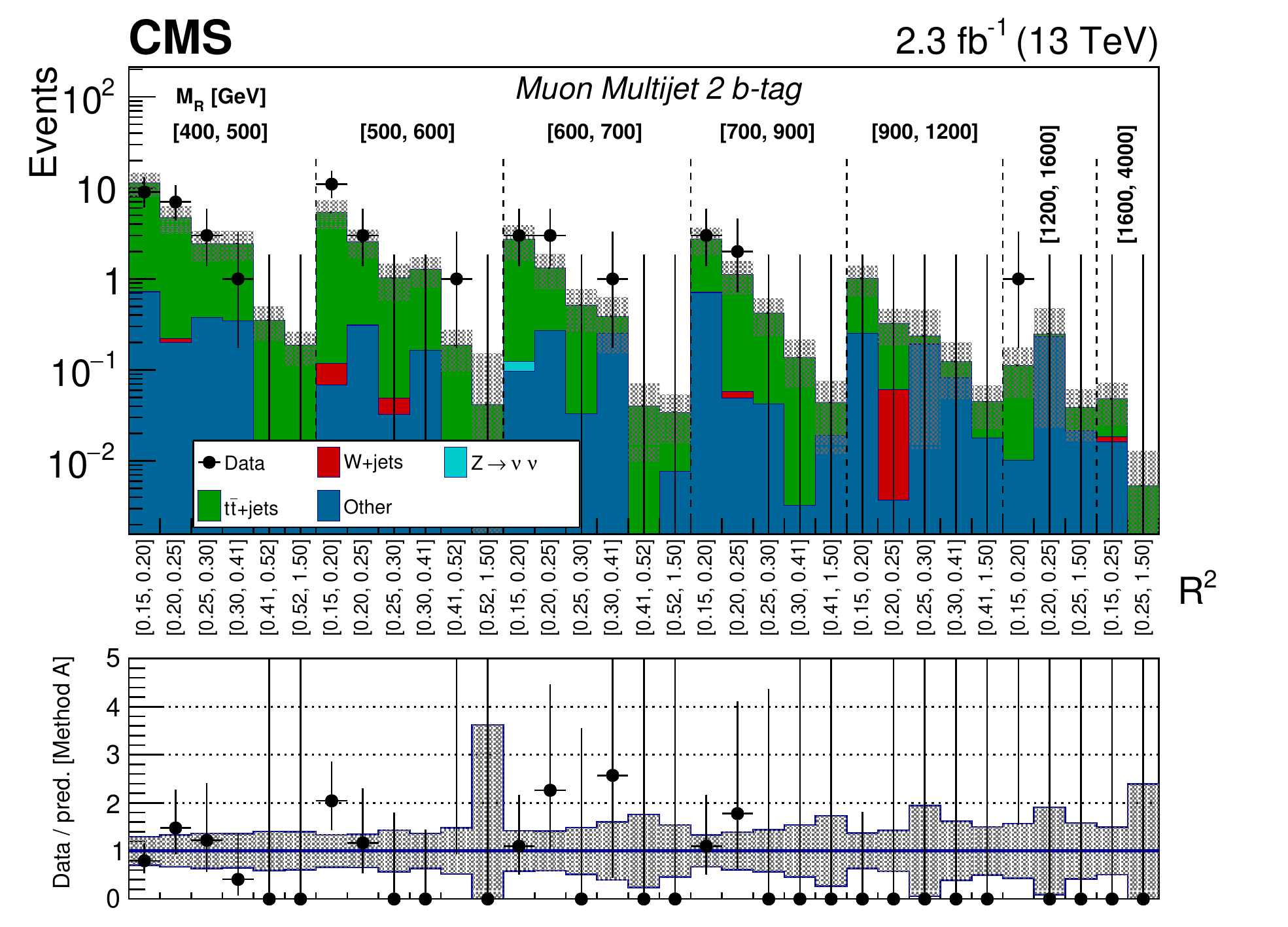}\\
\includegraphics[width=0.50\textwidth]{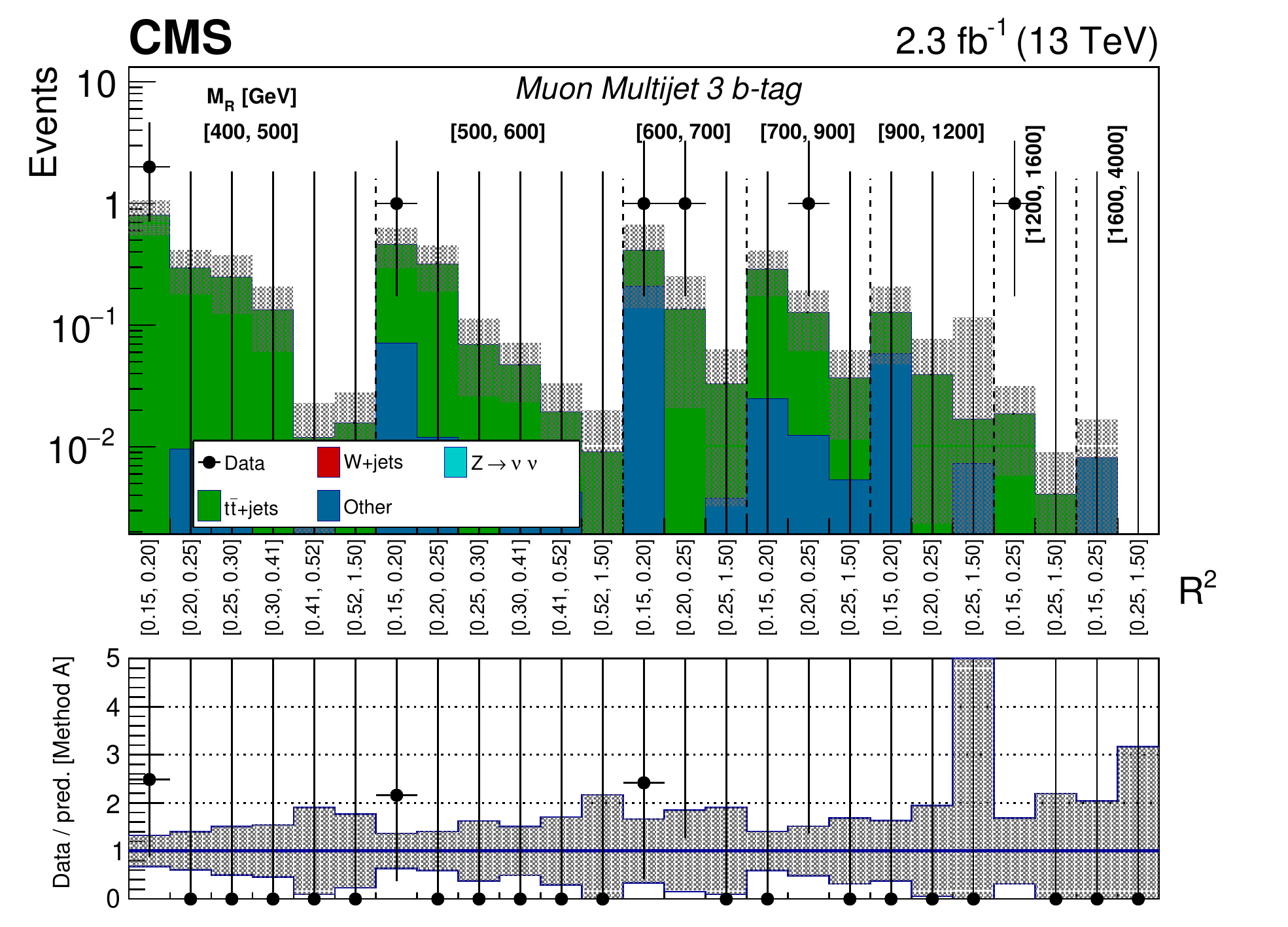}
\caption{ The $\MR$-$\Rtwo$ distribution observed in data is shown along with the background prediction
obtained from method A for the Muon Multijet event category in the 2 \PQb-tag (upper) and ${\geq}3$ \PQb-tag (lower) bins.
A detailed explanation of the panels is given in the caption of
Fig.~\ref{fig:ResultsMultiJet0btag1btag}.
}
\label{fig:ResultsMuMultiJet2btag3btag}
\end{figure}

\begin{figure}[!ptb] \centering
\includegraphics[width=0.50\textwidth]{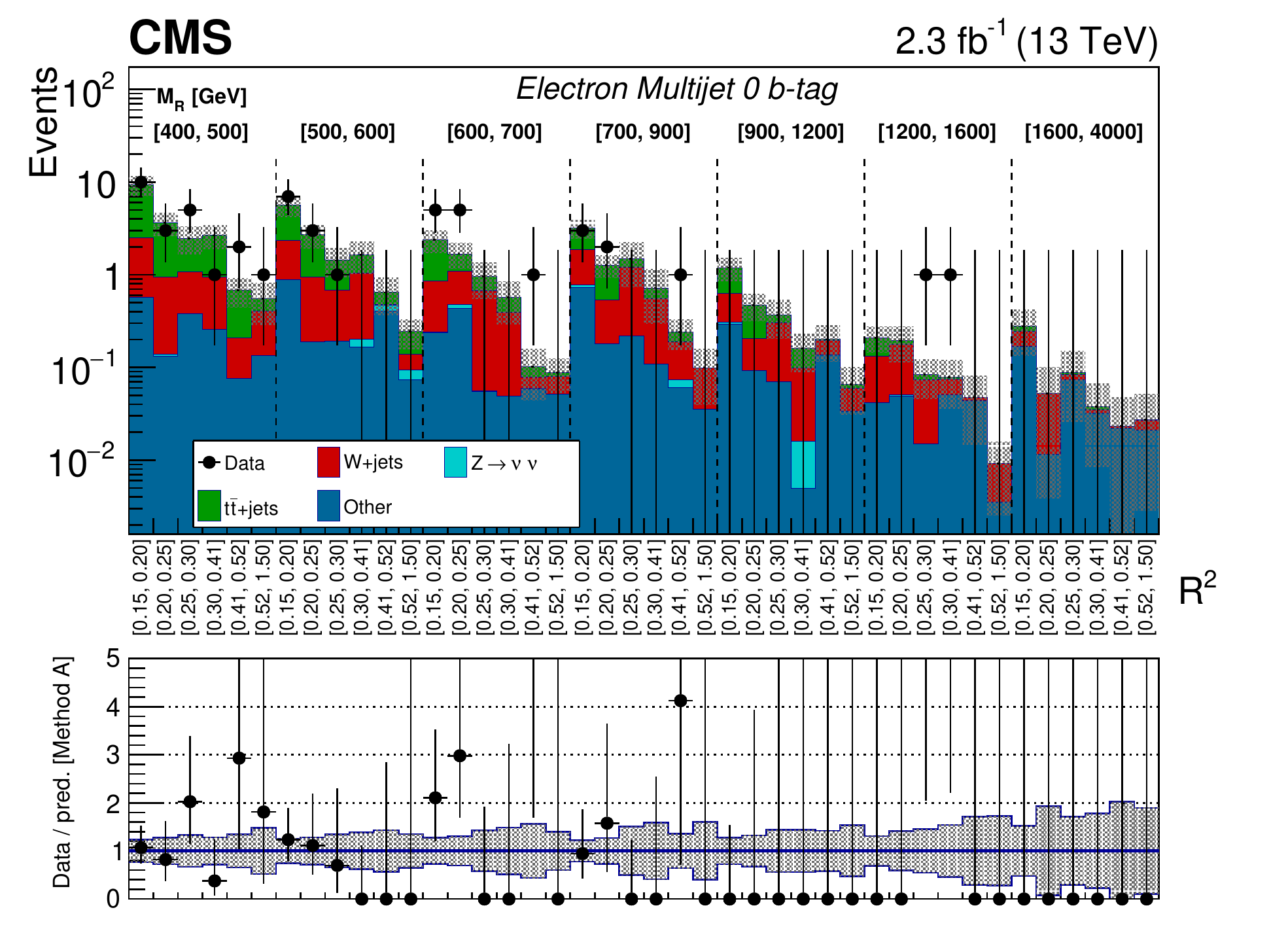}\\
\includegraphics[width=0.50\textwidth]{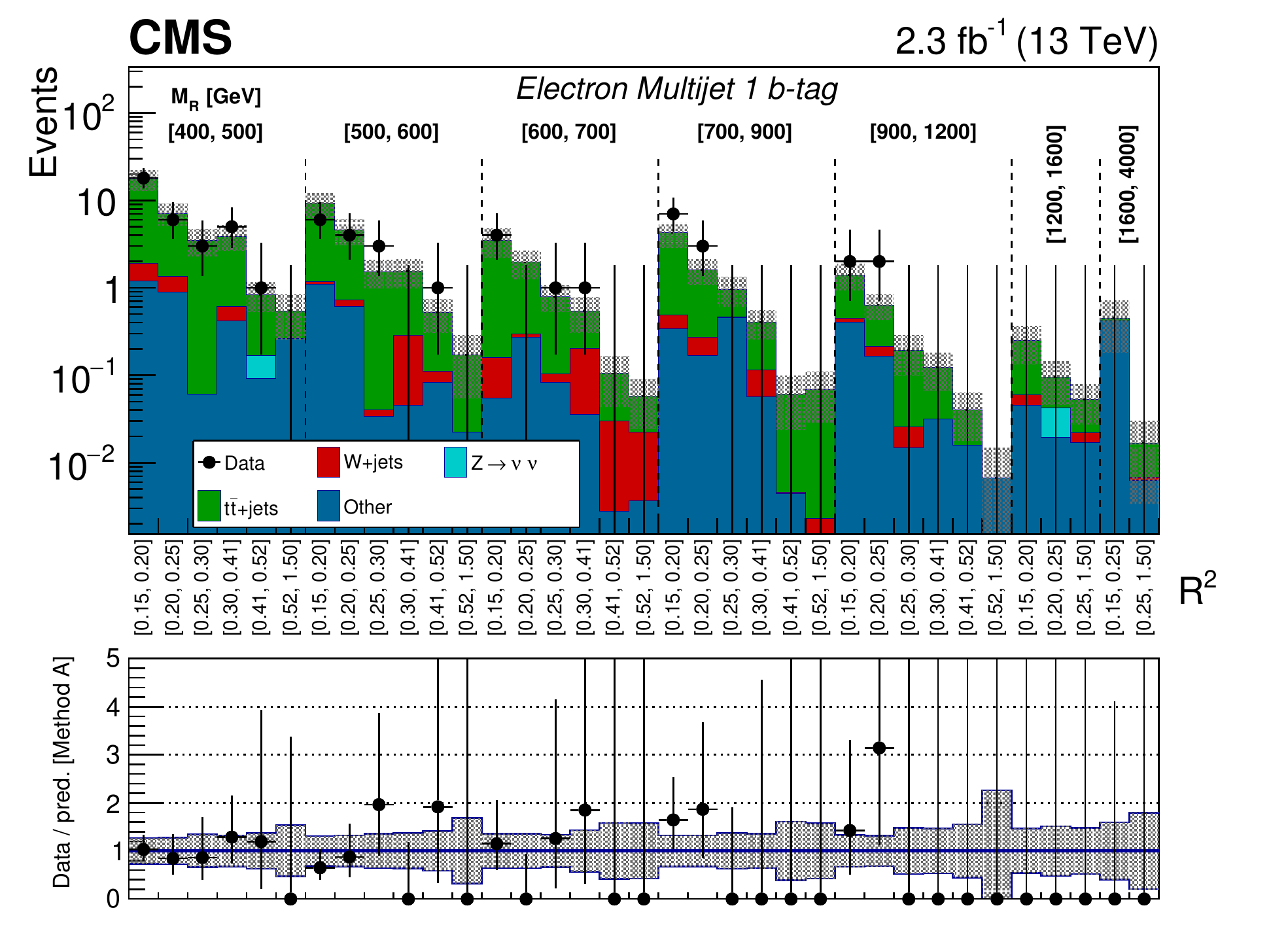}
\caption{ The $\MR$-$\Rtwo$ distribution observed in data is shown along with the background prediction
obtained from method A for the Electron Multijet event category in
the 0 \PQb-tag (upper) and 1 \PQb-tag (lower) bins. A detailed explanation of the panels is given in the caption of
Fig.~\ref{fig:ResultsMultiJet0btag1btag}.
}
\label{fig:ResultsEleMultiJet0btag1btag}
\end{figure}

\begin{figure}[!ptb] \centering
\includegraphics[width=0.50\textwidth]{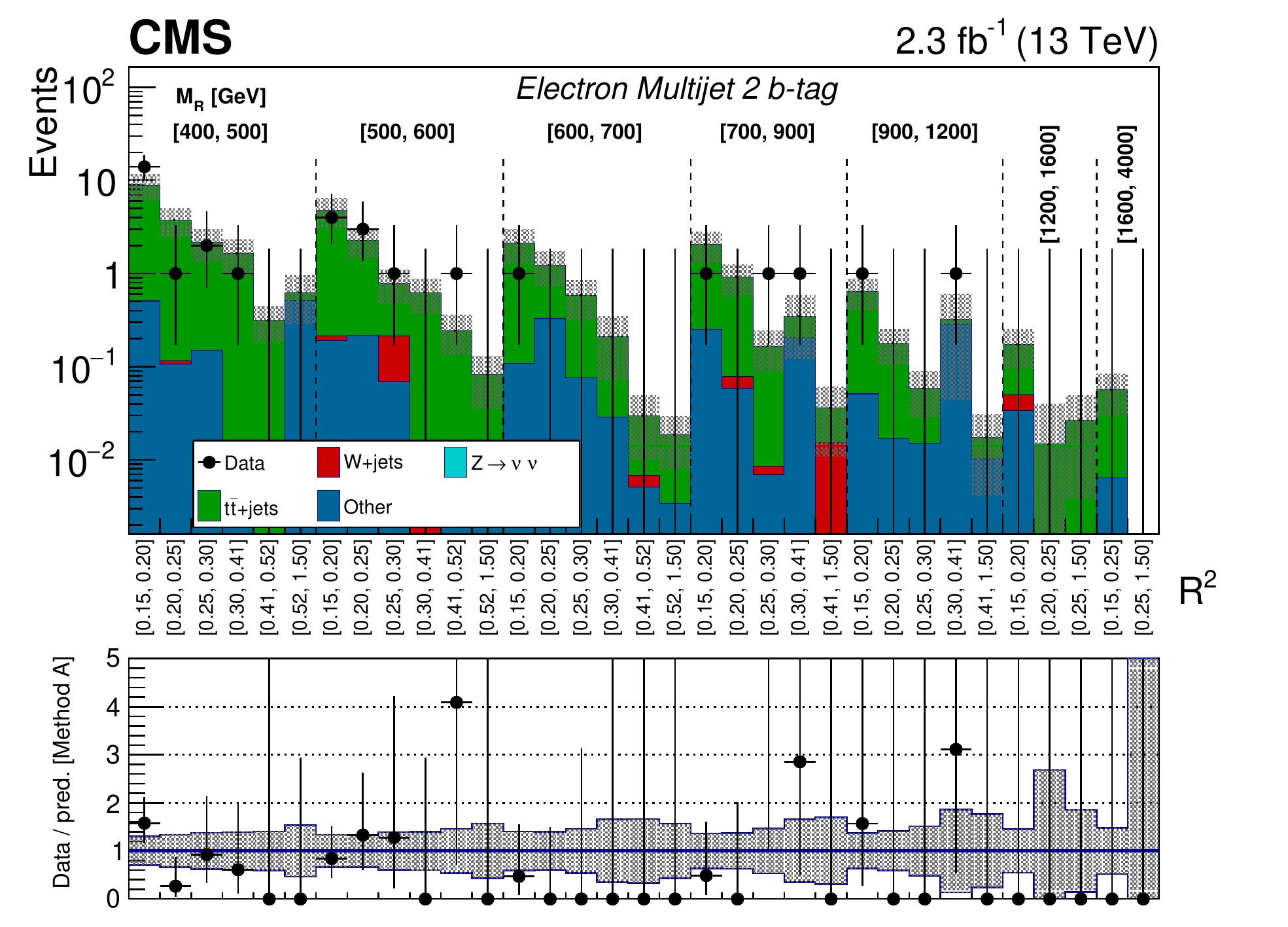}\\
\includegraphics[width=0.50\textwidth]{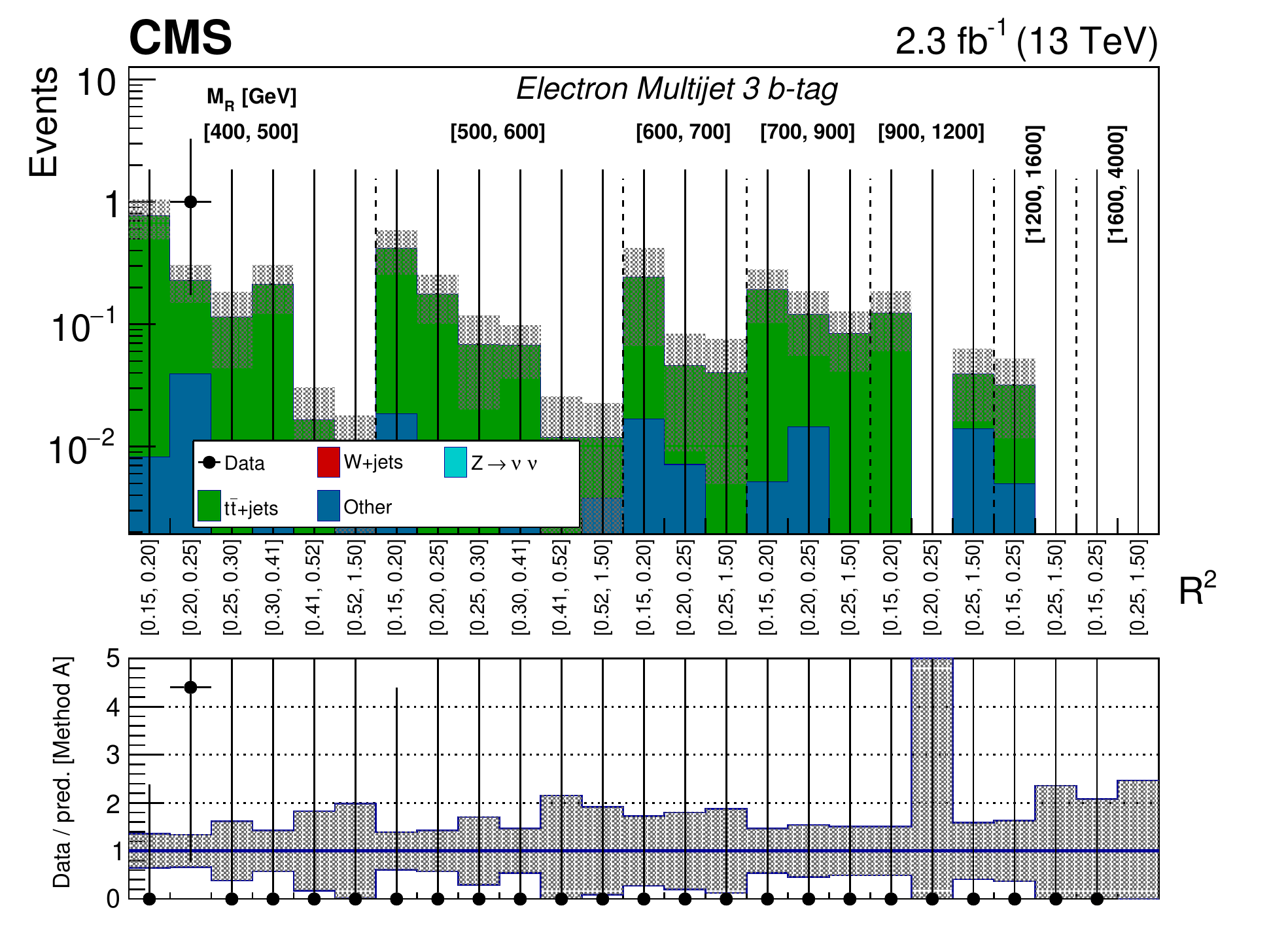}
\caption{ The $\MR$-$\Rtwo$ distribution observed in data is shown along with the background prediction
obtained from method A for the Electron Multijet event category in the 2 \PQb-tag (upper) and ${\geq}3$ \PQb-tag (lower) bins.
A detailed explanation of the panels is given in the caption of Fig.~\ref{fig:ResultsMultiJet0btag1btag}.
}
\label{fig:ResultsEleMultiJet2btag3btag}
\end{figure}

We present results of the search using method A as it provides slightly better sensitivity.
The two-dimensional $\MR$-$\Rtwo$ distributions for the search regions in the
Multijet, Electron Multijet, and Muon Multijet categories observed in data are shown in
Figures~\ref{fig:ResultsMultiJet0btag1btag}-\ref{fig:ResultsEleMultiJet2btag3btag},
along with the background prediction from method A.
We observe no statistically significant discrepancies and interpret the null search
result using method A by determining the 95\%
confidence level (CL) upper limits on the production cross sections of
the SUSY models presented in Section~\ref{sec:intro} using a global likelihood determined by combining the
likelihoods of the different search boxes and sidebands. Following the LHC $\CLs$
procedure~\cite{ATLAS:2011tau}, we use
the profile likelihood ratio test statistic and the asymptotic
formula to evaluate the 95\% CL observed and expected limits on the
SUSY production cross section $\sigma$.
Systematic uncertainties are taken into account by
incorporating nuisance parameters $\boldsymbol{\theta}$, representing different sources of
systematic uncertainty, into the likelihood function $\mathcal L(\sigma,\boldsymbol{\theta})$.
For each signal model the simulated SUSY events are used to estimate the effect of possible signal
contamination in the analysis control regions, and the method A background prediction is corrected
accordingly.
To determine a confidence interval for $\sigma$, we construct the profile likelihood ratio test
statistic $-2\ln[\mathcal L(\sigma,\boldsymbol{\hat\theta}_{\sigma})/\mathcal L(\hat\sigma,\boldsymbol{\hat\theta})]$ as a function of
$\sigma$, where $\boldsymbol{\hat\theta}_{\sigma}$ refers to the conditional maximum
likelihood estimators of $\boldsymbol\theta$ assuming a given value
$\sigma$, and $\hat\sigma$ and $\boldsymbol{\hat\theta}$ correspond to the
global maximum of the likelihood. Then for example, a 68\% confidence interval for $\sigma$
can be taken as the region for which the test statistic is less than 1. By
allowing each nuisance parameter to vary, the test statistic
curve is wider, reflecting the systematic uncertainty arising from
each source, and resulting in a larger confidence interval for $\sigma$.
First, we consider the scenario of gluino pair production
decaying to third-generation quarks. Gluino decays to the third-generation
are enhanced if the masses of the third-generation squarks are significantly
lighter than those of the first two generations, a scenario that is
strongly motivated in natural SUSY
models~\cite{naturalSUSY,Agashe:2014kda,DINE1990250,Cohen:1996vb}. Prompted by this, we consider the three decay
modes:
\begin{itemize}
\item $\PSg\to\bbbar\PSGcz$;
\item $\PSg\to\ttbar\PSGcz$;
\item
$\PSg\to\PQb\cPaqt\chip_1\to\PQb\cPaqt\PW^{\ast+}\PSGczDo$~or
charge conjugate,
\end{itemize}
where $\PW^{\ast}$ denotes a virtual $\PW$ boson. Due to a
technical limitation inherent in the event generator, we consider these
three decay modes for $\abs{m_{\PSg}-m_{\PSGczDo}} \geq 225\GeV$. For
$\abs{m_{\PSg}-m_{\PSGczDo}} < 225\GeV$, we only consider the $\PSg\to\bbbar\PSGcz$ decay mode.

The three-body gluino decays considered here capture
all of the possible final states within this natural SUSY context
including those of two-body gluino decays with intermediate top or bottom
squarks. Past studies have shown that LHC searches exhibit a similar sensitivity to
three-body and two-body gluino decays with a only a weak dependence on
the intermediate squark mass~\cite{Khachatryan:2016uwr}.

\begin{figure*}[!ptb] \centering
\includegraphics[width=0.45\textwidth]{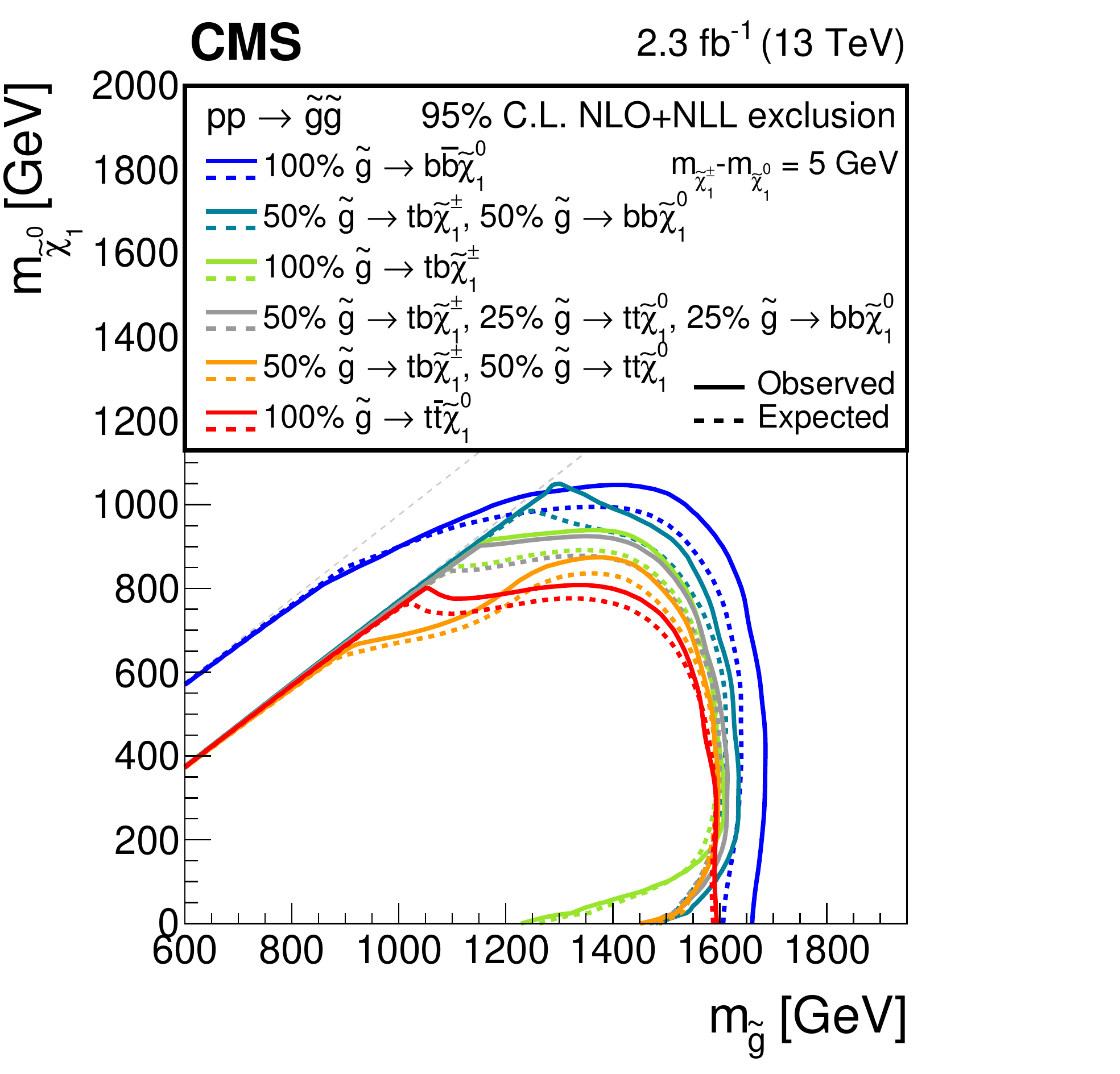}
\includegraphics[width=0.45\textwidth]{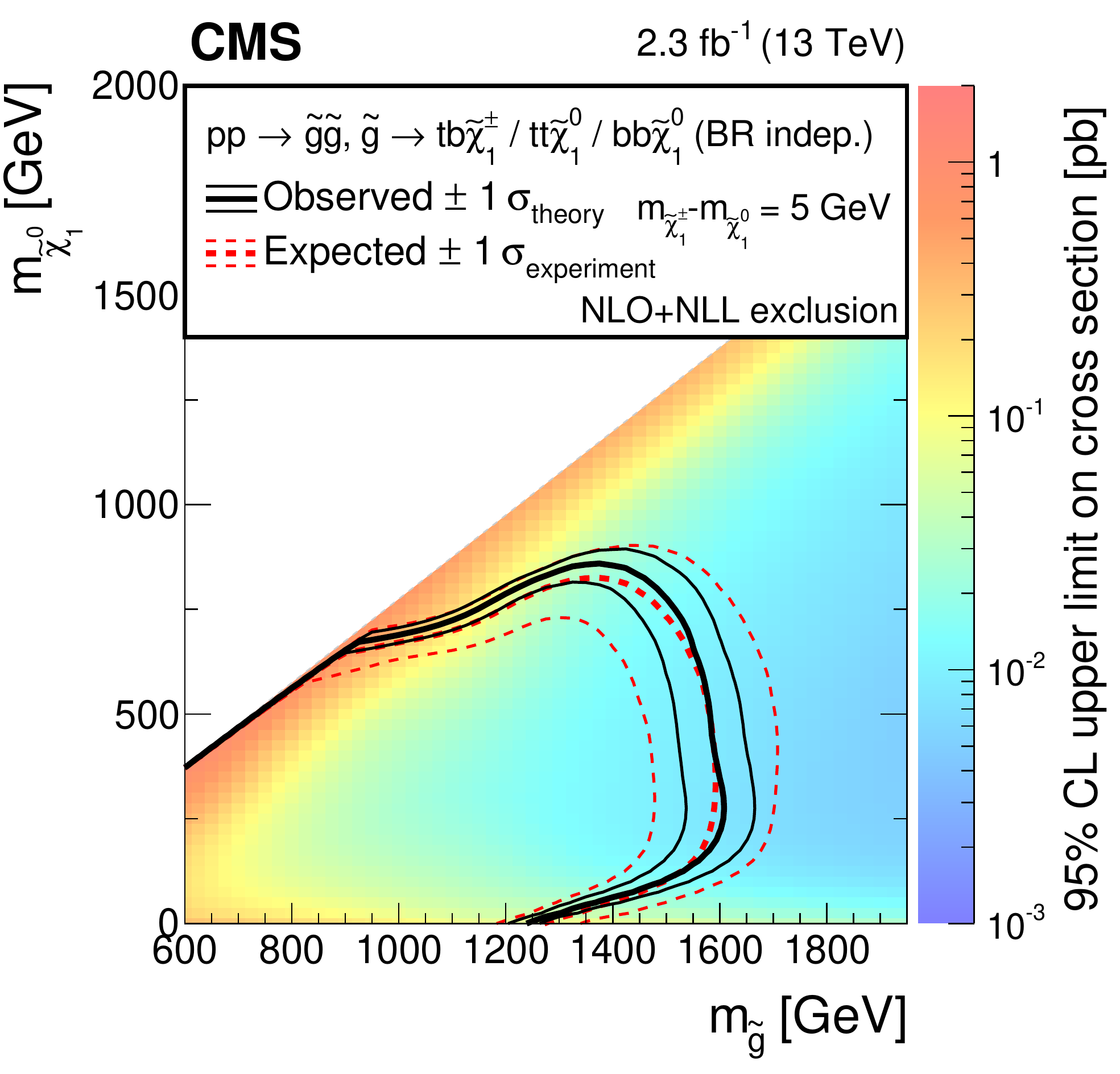}
\caption{ (Left) the expected and observed 95\% confidence level (CL) upper limits on the production
cross section for gluino pair production decaying to third-generation quarks under various
assumptions of the branching fractions. The two gray dashed diagonal
lines correspond to $\abs{m_{\PSg}-m_{\PSGczDo}} = 25\GeV$, which is
where the scan ends for the $\PSg\to\bbbar\PSGczDo$ decay
mode, and $\abs{m_{\PSg}-m_{\PSGczDo}} = 225\GeV$, which is where the scan
ends for the remaining modes due to a technical limitation inherent
in the event generator. For $\abs{m_{\PSg}-m_{\PSGczDo}} < 225\GeV$,
we only consider the $\PSg\to\bbbar\PSGczDo$ decay mode. (Right) the analogous upper limits on
the gluino pair production cross section valid for any values of the gluino
decay branching fractions.
}
\label{fig:GluinoToThirdGenLimits}
\end{figure*}

We perform a scan over all possible branching fractions to these three decay modes
and compute limits on the production cross section under each such scenario. The production cross section
limits for a few characteristic branching fraction scan points are shown on the left of
Fig.~\ref{fig:GluinoToThirdGenLimits} as a function of the gluino and neutralino masses. We find a range of excluded regions
for different branching fraction assumptions and generally observe the strongest limits for
the $\PSg\to\bbbar\PSGczDo$ decay mode over the full two-dimensional mass plane
and the weakest limits for the $\PSg\to\ttbar\PSGczDo$ decay
mode. For scenarios that include the intermediate decay
$\chipm_1\to\PW^{\ast \pm}\PSGczDo$ and small values of $m_{\PSGczDo}$ the sensitivity
is reduced because the LSP carries very little momentum in both the
NLSP rest frame and the laboratory frame, resulting in small values of
$\ETmiss$ and $\Rtwo$. By considering the limits obtained for all scanned branching fractions, we
calculate the exclusion limits valid for any assumption on the branching
fractions, presented on the right of Fig.~\ref{fig:GluinoToThirdGenLimits}. For
an LSP with mass of a few hundred \GeV, we exclude pair production of gluinos decaying
to third-generation quarks for mass below about 1.6\TeV. This result
represents a unique attempt to obtain a branching fraction independent limit on
gluino pair production at the LHC for the scenario in which gluino decays are dominated by
three-body decays to third-generation quarks and a neutralino LSP.

\begin{figure*}[!ptb] \centering
\includegraphics[width=0.45\textwidth]{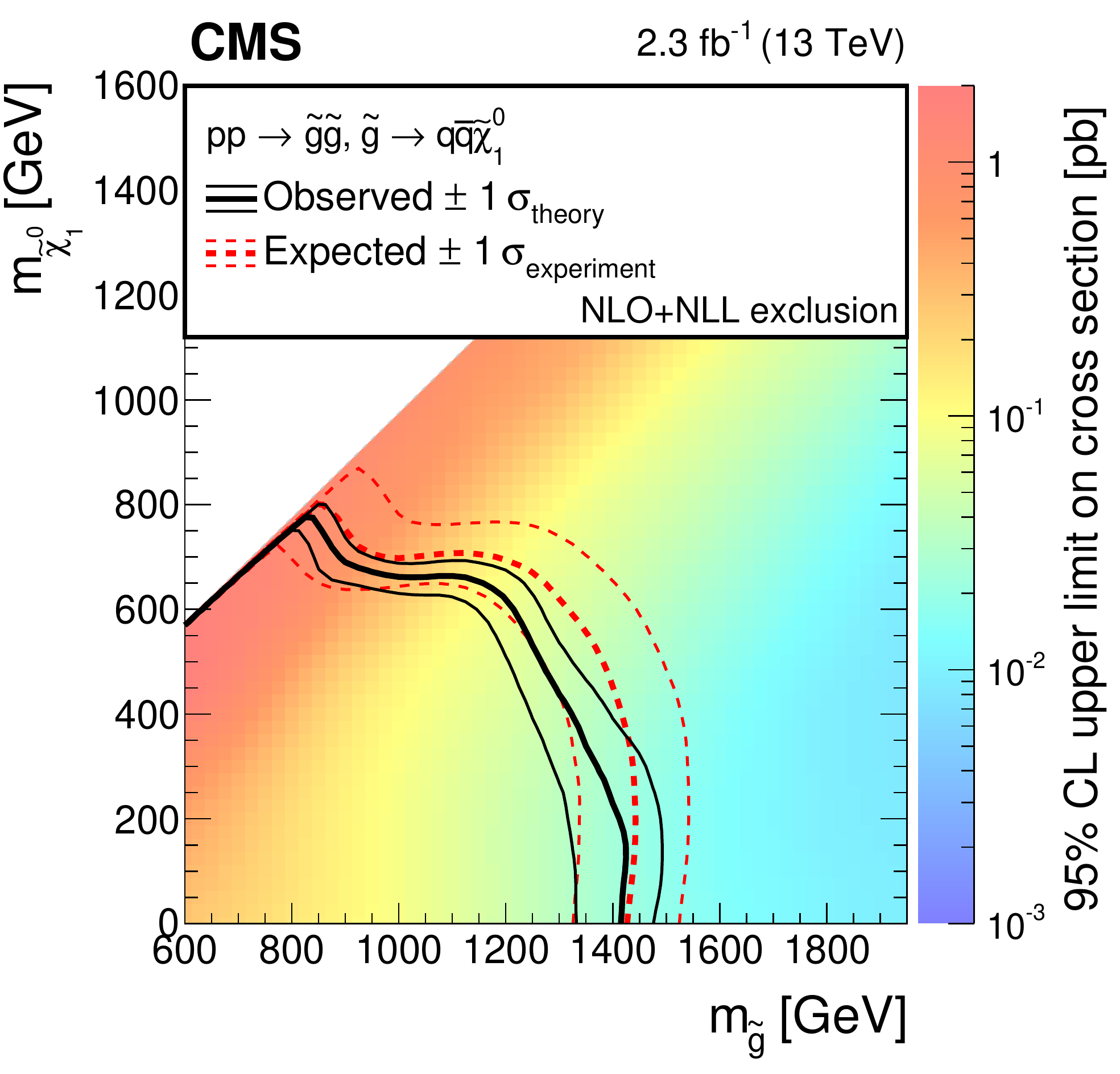}
\includegraphics[width=0.45\textwidth]{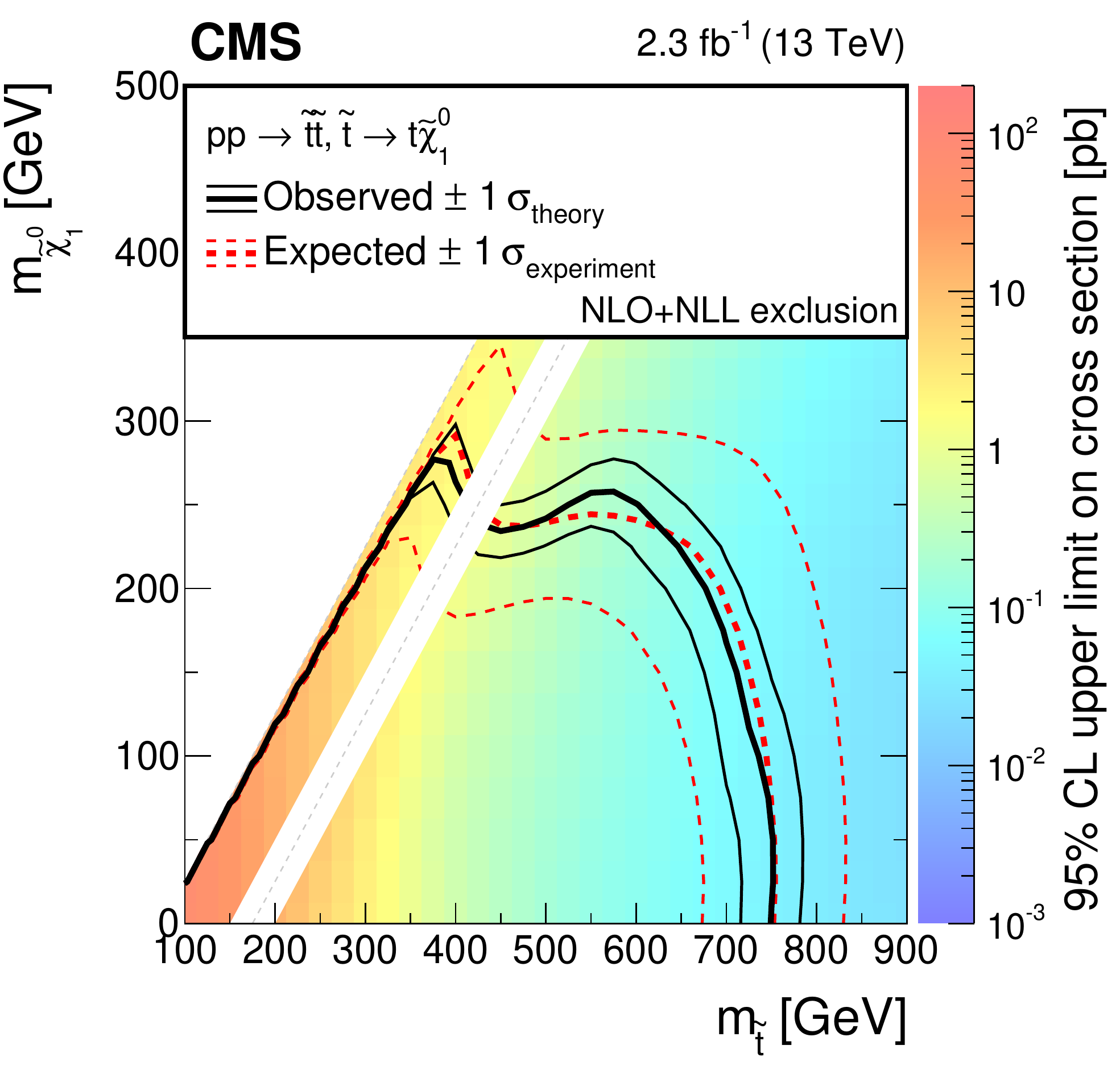}
\caption{ Expected and observed 95\% confidence level (CL) upper limits on the production cross section
for (left) gluino pair production decaying to two light-flavored quarks and the LSP
and (right) top squark pair production decaying to a top quark and the LSP.
The white diagonal band in the right plot corresponds to the region
$\abs{m_{\PSQt}-m_{\PQt}-m_{\PSGczDo}} < 25\GeV$, where the signal
efficiency is a strong function of $m_{\PSQt}-m_{\PSGczDo}$, and as a
result the precise determination of the cross section upper limit is uncertain
because of the finite granularity of the available MC samples
in this region of the ($m_{\PSQt}$, $m_{\PSGczDo}$)  plane.
}
\label{fig:limitT1qqqqT2tt}
\end{figure*}

In Figure~\ref{fig:limitT1qqqqT2tt}, we present additional interpretations for
simplified model scenarios of interest. On the left, we show the production cross section
limits on gluino pair production where the gluino decays to two light-flavored
quarks and the LSP, and on the right we show the production cross section limits on
top squark pair production where the top squark decays to a top quark and the LSP.
For a very light LSP, we exclude top squark production with mass below
750\GeV.
\section{Summary}
\label{sec:Summary}
We have presented an inclusive search for supersymmetry in events
with no more than one lepton, a large multiplicity of energetic jets, and
missing transverse energy. The search is sensitive to a broad
range of SUSY scenarios including pair production of gluinos and top
squarks. The event categorization in the number of leptons and
the number of \PQb-tagged jets enhances the search sensitivity for a variety of different SUSY
signal scenarios.
Two background estimation methods are presented, both
based on transfer factors between data control regions and the search
regions, but having very different systematic assumptions:
one relying on the simulation and associated corrections derived in
the control regions, and the other relying on the accuracy of an
assumed functional form for the shape of background distributions in the $\MR$ and $\Rtwo$ variables.
The two predictions agree within their uncertainties, thereby
demonstrating the robustness of the background modeling.

No significant deviations from the predicted standard model background are
observed in any of the search regions, and this result is interpreted
in the context of simplified models of gluino or top
squark pair production. For top squark decays to a top quark and an LSP with a mass of 100\GeV, we
exclude top squarks with masses below 750\GeV. Considering
separately the gluino decays to bottom quarks and the LSP or first- and
second-generation quarks and the LSP, gluino masses up to
1.65\TeV or 1.4\TeV are excluded, respectively.
Furthermore, this search goes beyond the existing simplified model paradigm by
interpreting results in a broader context inspired by natural SUSY,
with multiple gluino decay modes considered simultaneously.
By scanning over all possible branching fractions
for three-body gluino decays to third generation quarks, exclusion
limits are derived on gluino pair production that are valid for any
values of the gluino decay branching fractions.
For a chargino NLSP nearly degenerate in mass with the LSP and LSP
masses in the range between 200 and 600\GeV, we exclude gluinos
with mass below 1.55 to 1.6\TeV, regardless of their decays. This
result is a more generic constraint on gluino production than
previously reported at the LHC.
\clearpage
\begin{acknowledgments}
\hyphenation{Bundes-ministerium Forschungs-gemeinschaft Forschungs-zentren} We congratulate our colleagues in the CERN accelerator departments for the excellent performance of the LHC and thank the technical and administrative staffs at CERN and at other CMS institutes for their contributions to the success of the CMS effort. In addition, we gratefully acknowledge the computing centres and personnel of the Worldwide LHC Computing Grid for delivering so effectively the computing infrastructure essential to our analyses. Finally, we acknowledge the enduring support for the construction and operation of the LHC and the CMS detector provided by the following funding agencies: the Austrian Federal Ministry of Science, Research and Economy and the Austrian Science Fund; the Belgian Fonds de la Recherche Scientifique, and Fonds voor Wetenschappelijk Onderzoek; the Brazilian Funding Agencies (CNPq, CAPES, FAPERJ, and FAPESP); the Bulgarian Ministry of Education and Science; CERN; the Chinese Academy of Sciences, Ministry of Science and Technology, and National Natural Science Foundation of China; the Colombian Funding Agency (COLCIENCIAS); the Croatian Ministry of Science, Education and Sport, and the Croatian Science Foundation; the Research Promotion Foundation, Cyprus; the Ministry of Education and Research, Estonian Research Council via IUT23-4 and IUT23-6 and European Regional Development Fund, Estonia; the Academy of Finland, Finnish Ministry of Education and Culture, and Helsinki Institute of Physics; the Institut National de Physique Nucl\'eaire et de Physique des Particules~/~CNRS, and Commissariat \`a l'\'Energie Atomique et aux \'Energies Alternatives~/~CEA, France; the Bundesministerium f\"ur Bildung und Forschung, Deutsche Forschungsgemeinschaft, and Helmholtz-Gemeinschaft Deutscher Forschungszentren, Germany; the General Secretariat for Research and Technology, Greece; the National Scientific Research Foundation, and National Innovation Office, Hungary; the Department of Atomic Energy and the Department of Science and Technology, India; the Institute for Studies in Theoretical Physics and Mathematics, Iran; the Science Foundation, Ireland; the Istituto Nazionale di Fisica Nucleare, Italy; the Ministry of Science, ICT and Future Planning, and National Research Foundation (NRF), Republic of Korea; the Lithuanian Academy of Sciences; the Ministry of Education, and University of Malaya (Malaysia); the Mexican Funding Agencies (BUAP, CINVESTAV, CONACYT, LNS, SEP, and UASLP-FAI); the Ministry of Business, Innovation and Employment, New Zealand; the Pakistan Atomic Energy Commission; the Ministry of Science and Higher Education and the National Science Centre, Poland; the Funda\c{c}\~ao para a Ci\^encia e a Tecnologia, Portugal; JINR, Dubna; the Ministry of Education and Science of the Russian Federation, the Federal Agency of Atomic Energy of the Russian Federation, Russian Academy of Sciences, and the Russian Foundation for Basic Research; the Ministry of Education, Science and Technological Development of Serbia; the Secretar\'{\i}a de Estado de Investigaci\'on, Desarrollo e Innovaci\'on and Programa Consolider-Ingenio 2010, Spain; the Swiss Funding Agencies (ETH Board, ETH Zurich, PSI, SNF, UniZH, Canton Zurich, and SER); the Ministry of Science and Technology, Taipei; the Thailand Center of Excellence in Physics, the Institute for the Promotion of Teaching Science and Technology of Thailand, Special Task Force for Activating Research and the National Science and Technology Development Agency of Thailand; the Scientific and Technical Research Council of Turkey, and Turkish Atomic Energy Authority; the National Academy of Sciences of Ukraine, and State Fund for Fundamental Researches, Ukraine; the Science and Technology Facilities Council, UK; the US Department of Energy, and the US National Science Foundation.

Individuals have received support from the Marie-Curie programme and the European Research Council and EPLANET (European Union); the Leventis Foundation; the A. P. Sloan Foundation; the Alexander von Humboldt Foundation; the Belgian Federal Science Policy Office; the Fonds pour la Formation \`a la Recherche dans l'Industrie et dans l'Agriculture (FRIA-Belgium); the Agentschap voor Innovatie door Wetenschap en Technologie (IWT-Belgium); the Ministry of Education, Youth and Sports (MEYS) of the Czech Republic; the Council of Science and Industrial Research, India; the HOMING PLUS programme of the Foundation for Polish Science, cofinanced from European Union, Regional Development Fund; the Mobility Plus programme of the Ministry of Science and Higher Education (Poland); the OPUS programme of the National Science Center (Poland); the Thalis and Aristeia programmes cofinanced by EU-ESF and the Greek NSRF; the National Priorities Research Program by Qatar National Research Fund; the Programa Clar\'in-COFUND del Principado de Asturias; the Rachadapisek Sompot Fund for Postdoctoral Fellowship, Chulalongkorn University (Thailand); the Chulalongkorn Academic into Its 2nd Century Project Advancement Project (Thailand); and the Welch Foundation, contract C-1845.
\end{acknowledgments}
\bibliography{auto_generated}
\clearpage
\appendix
\section{Results of method B fit-based background prediction}
\label{app:FitResults}
\begin{figure*}[!ptb] \centering
\includegraphics[width=0.50\textwidth]{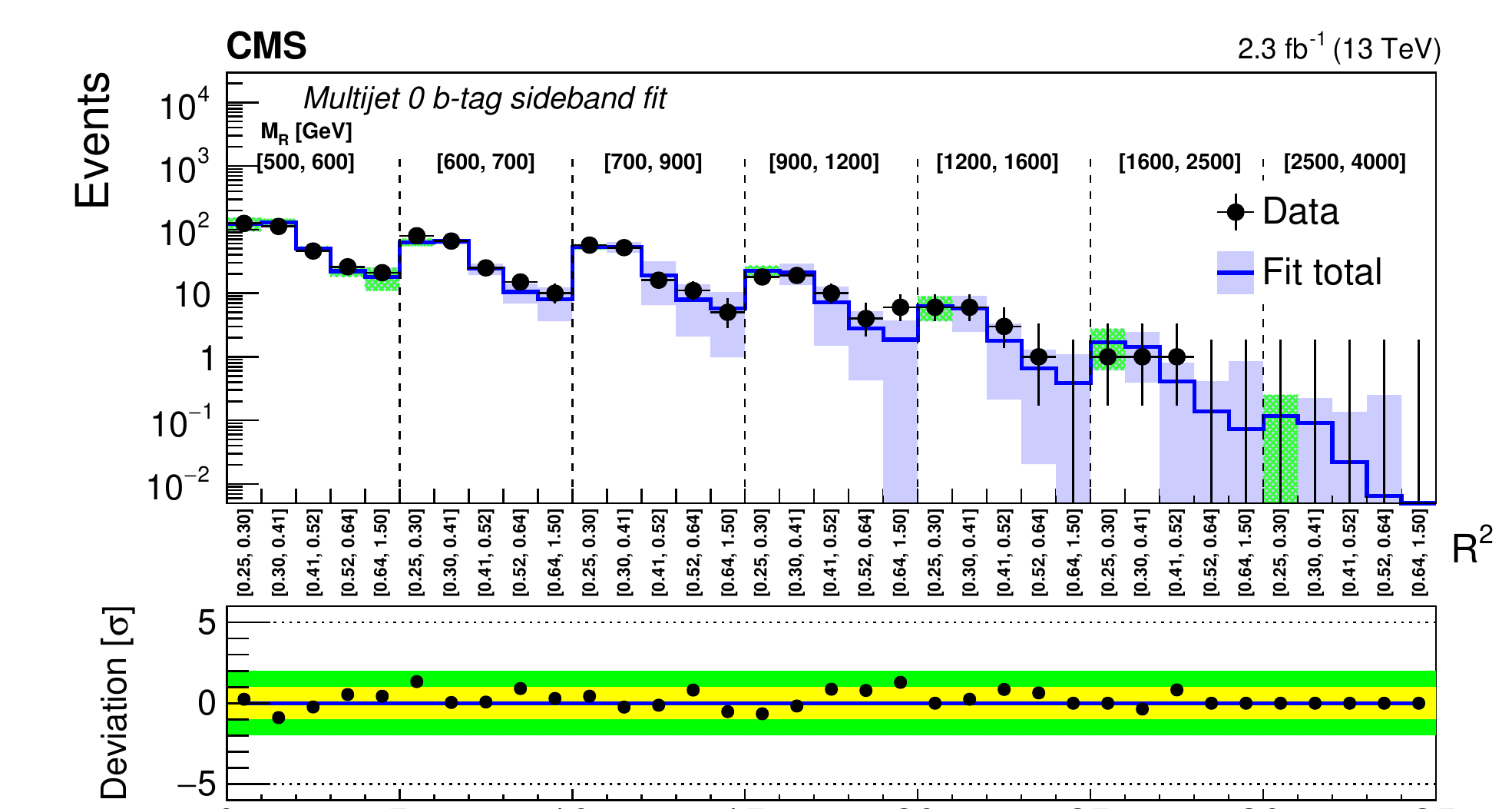}\\
\includegraphics[width=0.50\textwidth]{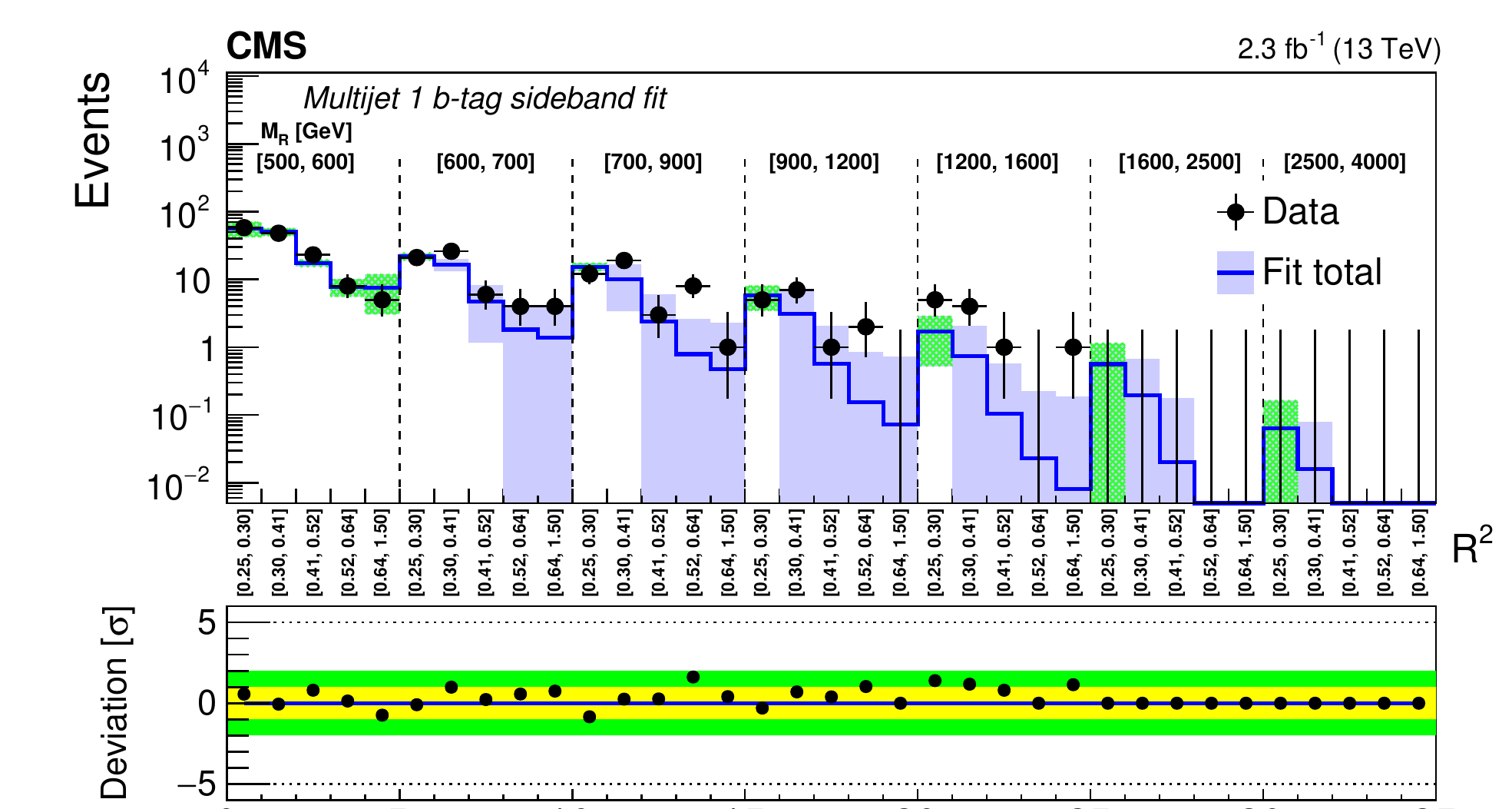}
\caption{Comparison of the predicted background with the observed data
in bins of $\MR$ and $\Rtwo$ variables in the Multijet category for
the 0 \PQb-tag (upper) and 1 \PQb-tag (lower) bins. A detailed explanation of the panels is given in the caption of
Fig.~\ref{fig:results_Multijet2btag3btag}. }
\label{fig:results_Multijet0btag1btag}
\end{figure*}

\begin{figure*}[!ptb] \centering
\includegraphics[width=0.50\textwidth]{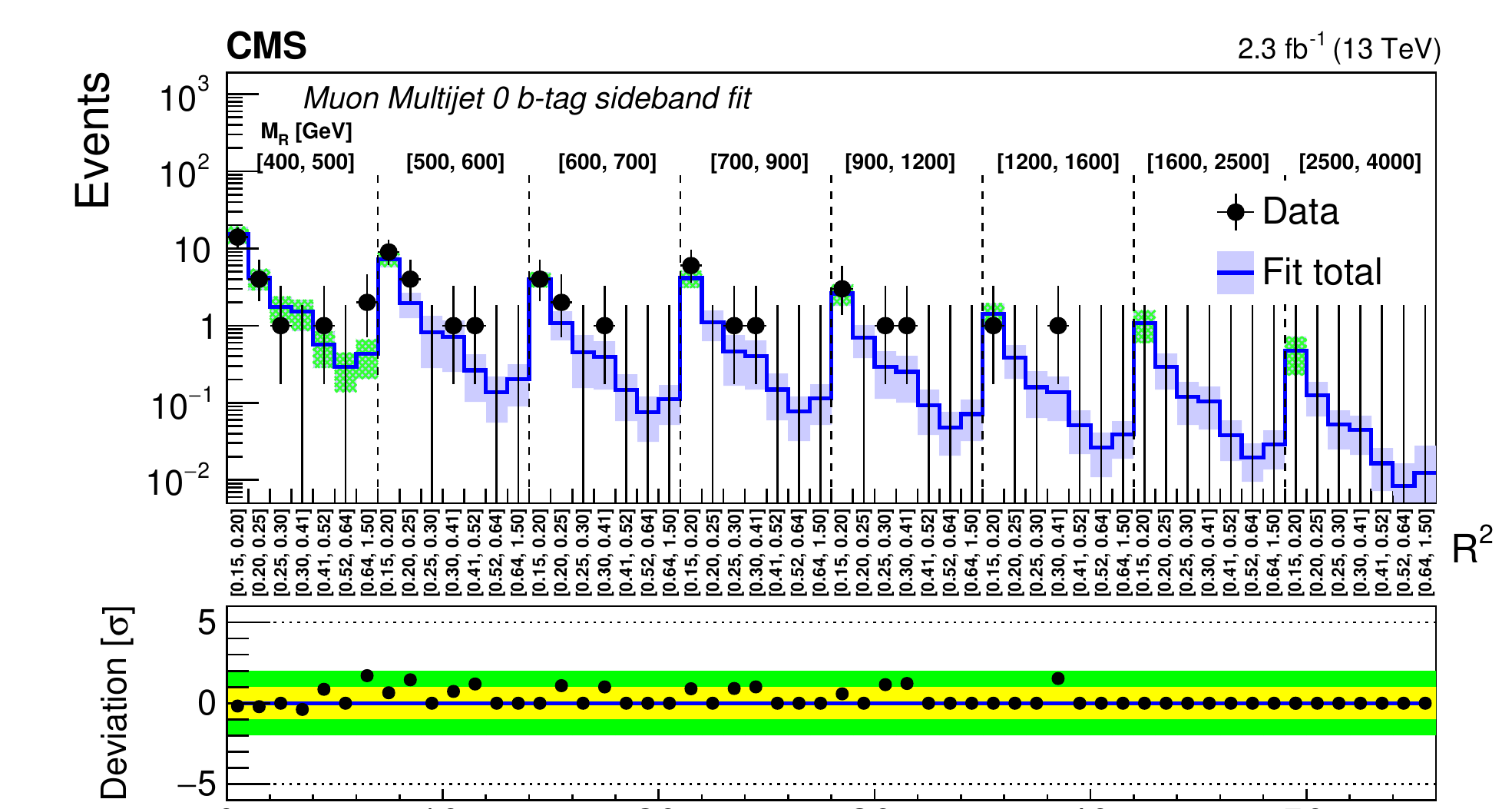} \\
\includegraphics[width=0.50\textwidth]{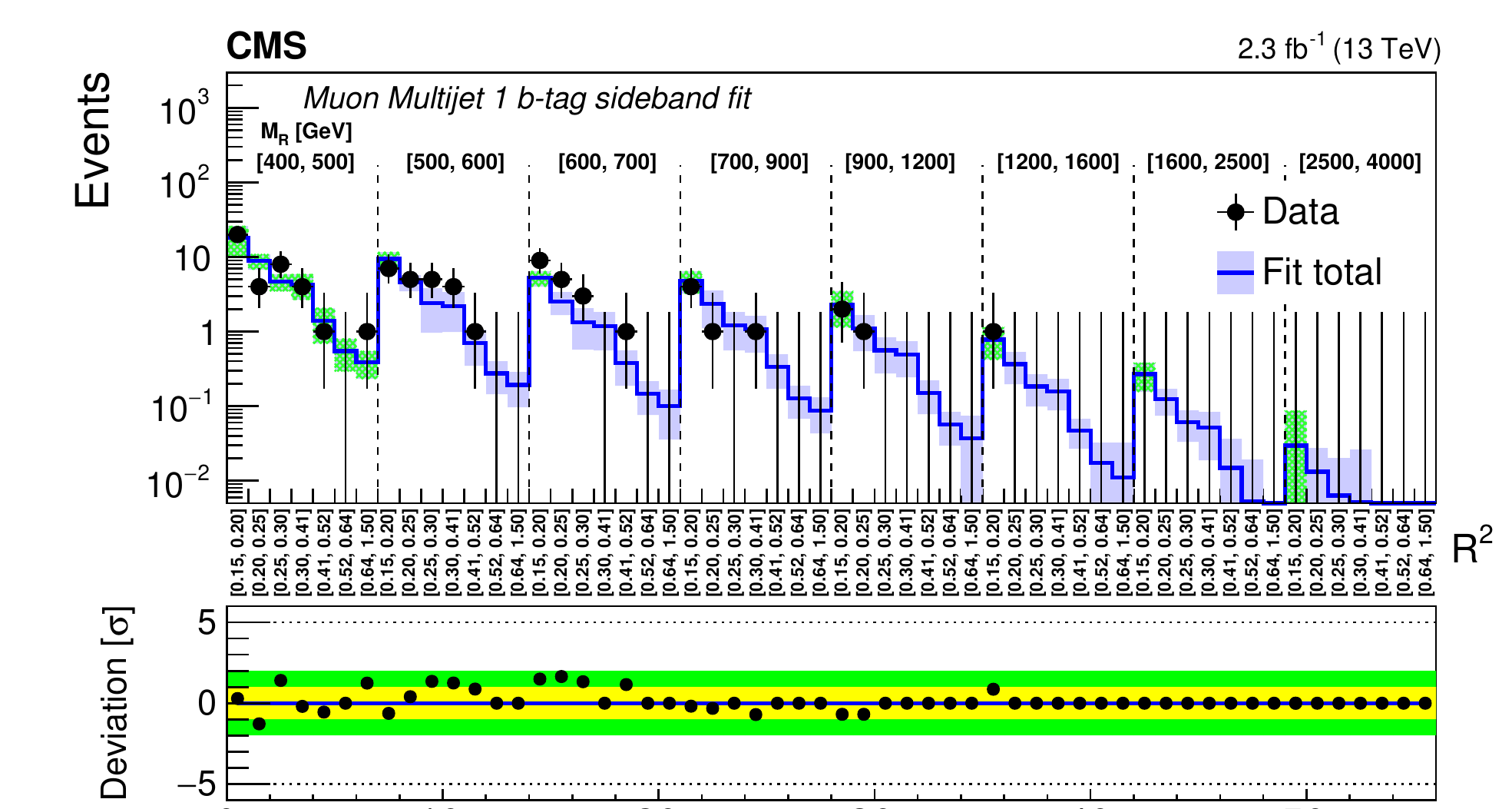} \\
\caption{Comparison of the predicted background with the observed data
in bins of $\MR$ and $\Rtwo$ variables in the Muon Multijet
category for the 0 \PQb-tag (upper) and 1 \PQb-tag (lower) bins. A detailed explanation of the panels is given in the caption of
Fig.~\ref{fig:results_Multijet2btag3btag}. }
\label{fig:results_MuMultijet0btag1btag}
\end{figure*}

\begin{figure*}[!ptb] \centering
\includegraphics[width=0.50\textwidth]{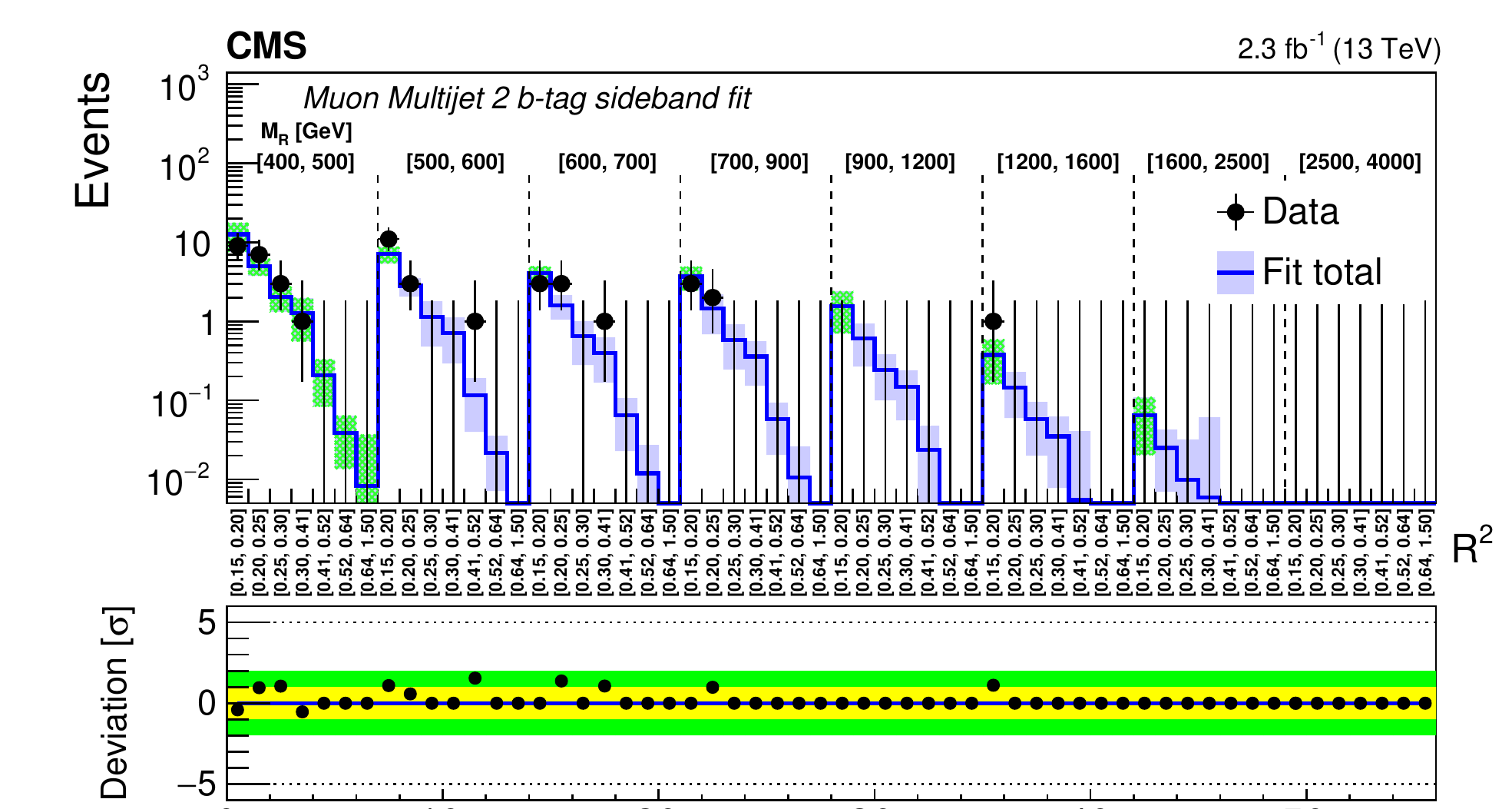} \\
\includegraphics[width=0.50\textwidth]{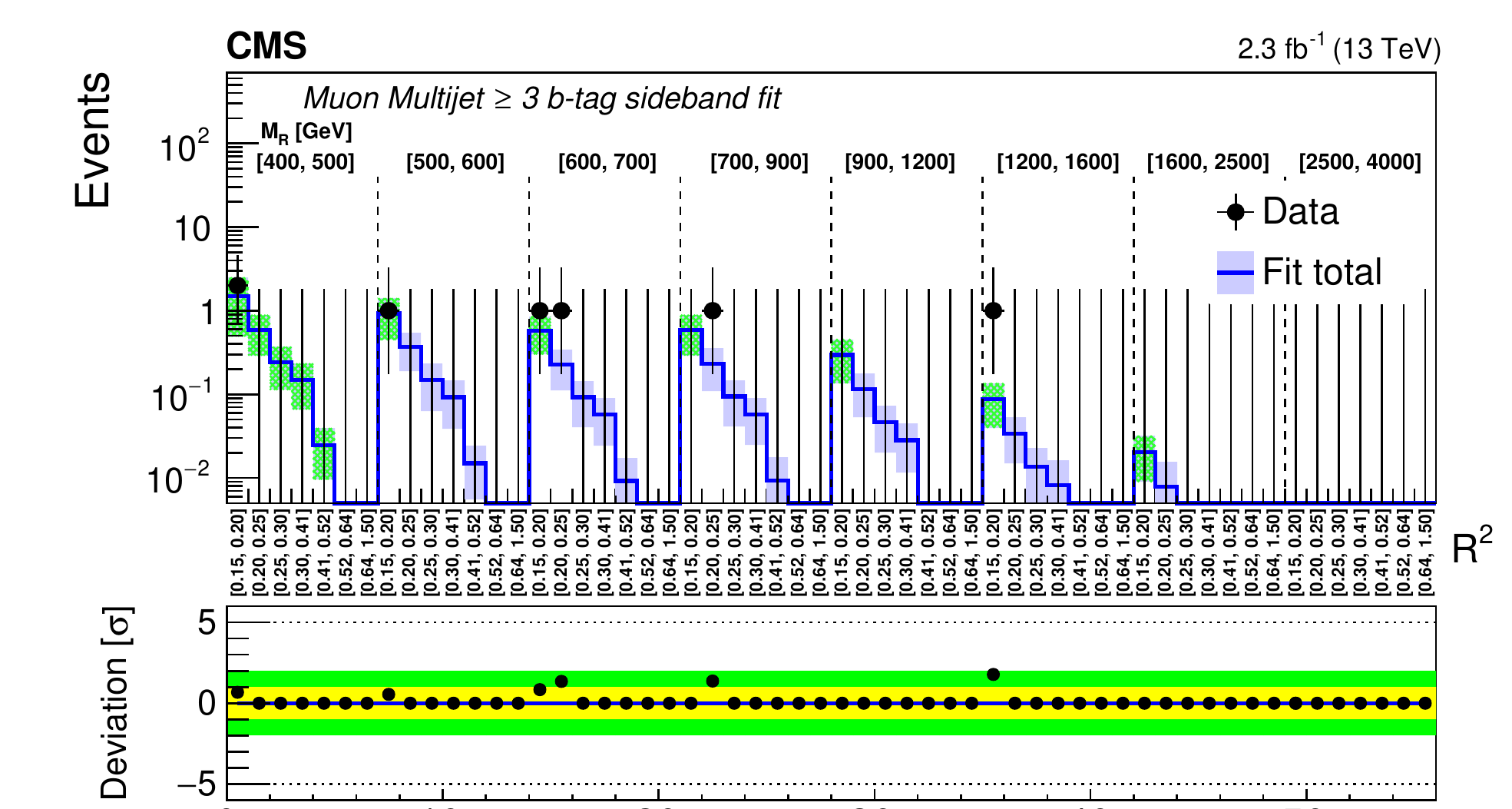}
\caption{Comparison of the predicted background with the observed data
in bins of $\MR$ and $\Rtwo$ variables in the Muon Multijet
category for the 2 \PQb-tag (upper) and ${\geq}3$ \PQb-tag (lower) bins. A detailed explanation of the panels is given in the caption of
Fig.~\ref{fig:results_Multijet2btag3btag}. }
\label{fig:results_MuMultijet2btag3btag}
\end{figure*}

\begin{figure*}[!ptb] \centering
\includegraphics[width=0.50\textwidth]{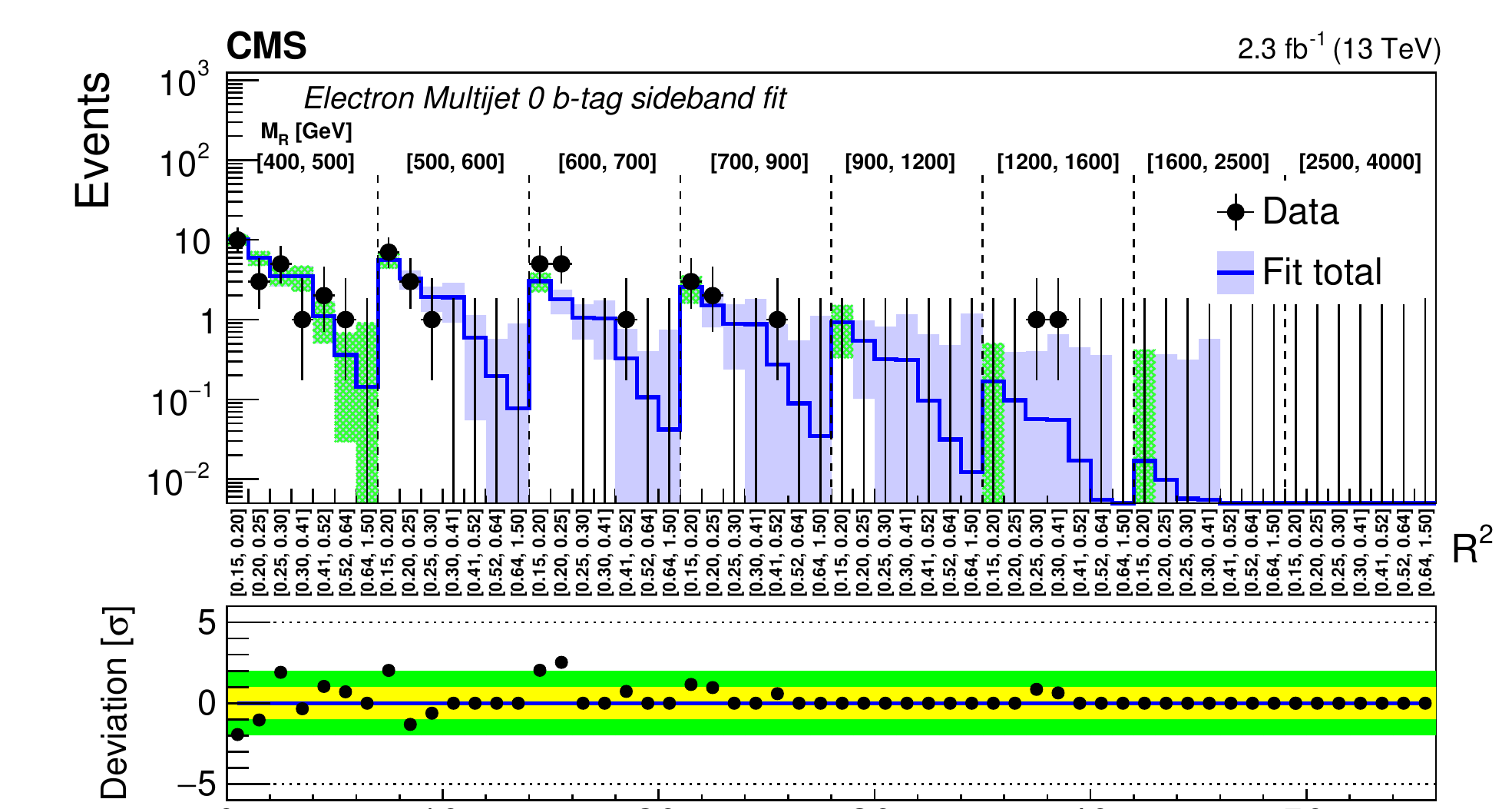} \\
\includegraphics[width=0.50\textwidth]{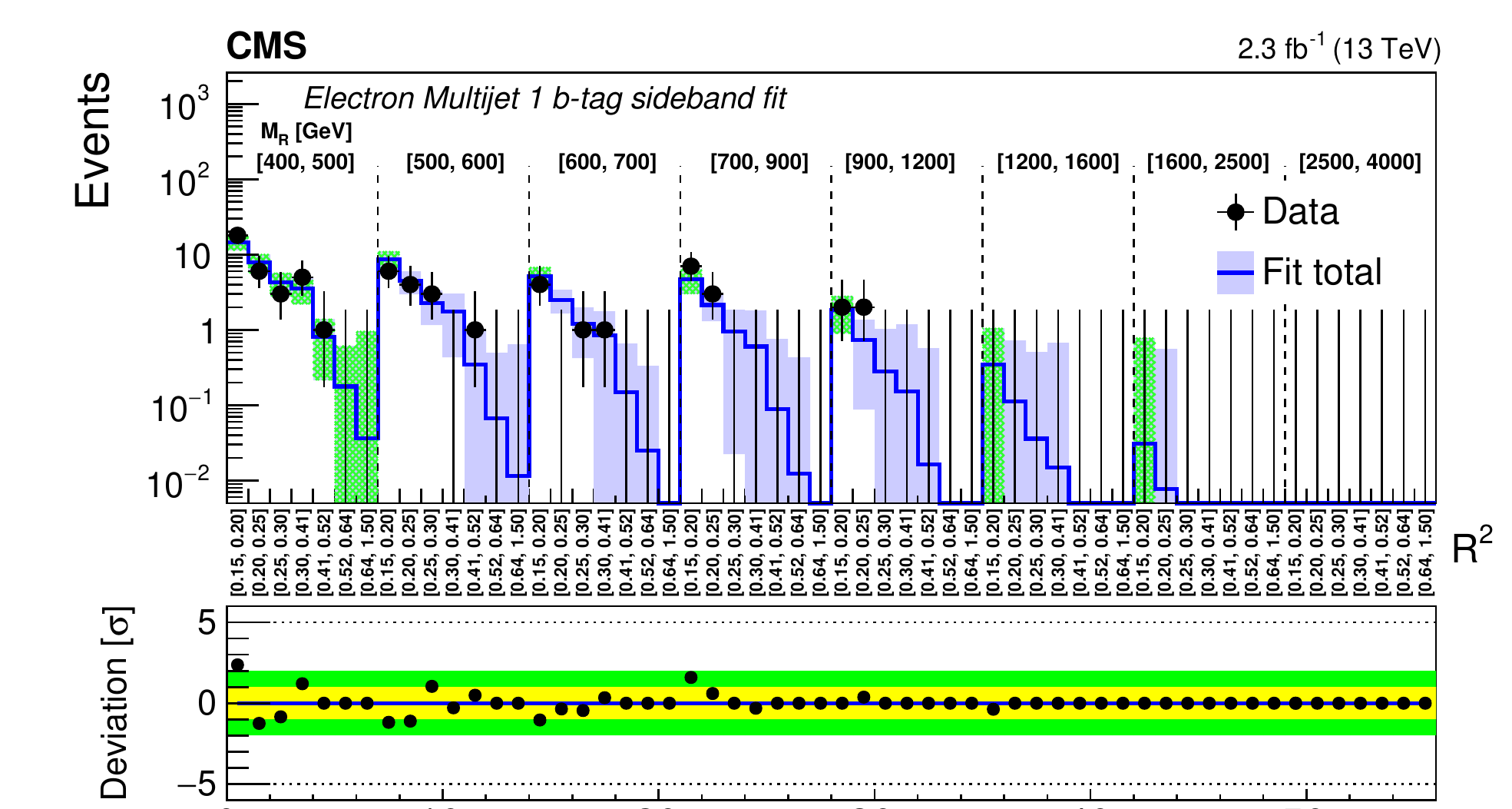}
\caption{Comparison of the predicted background with the observed data
in bins of $\MR$ and $\Rtwo$ variables in the Electron Multijet
category for the 0 \PQb-tag (upper) and 1 \PQb-tag (lower) bins. A detailed explanation of the panels is given in the caption of
Fig.~\ref{fig:results_Multijet2btag3btag}. }
\label{fig:results_EleMultijet0btag1btag}
\end{figure*}

\begin{figure*}[!ptb] \centering
\includegraphics[width=0.50\textwidth]{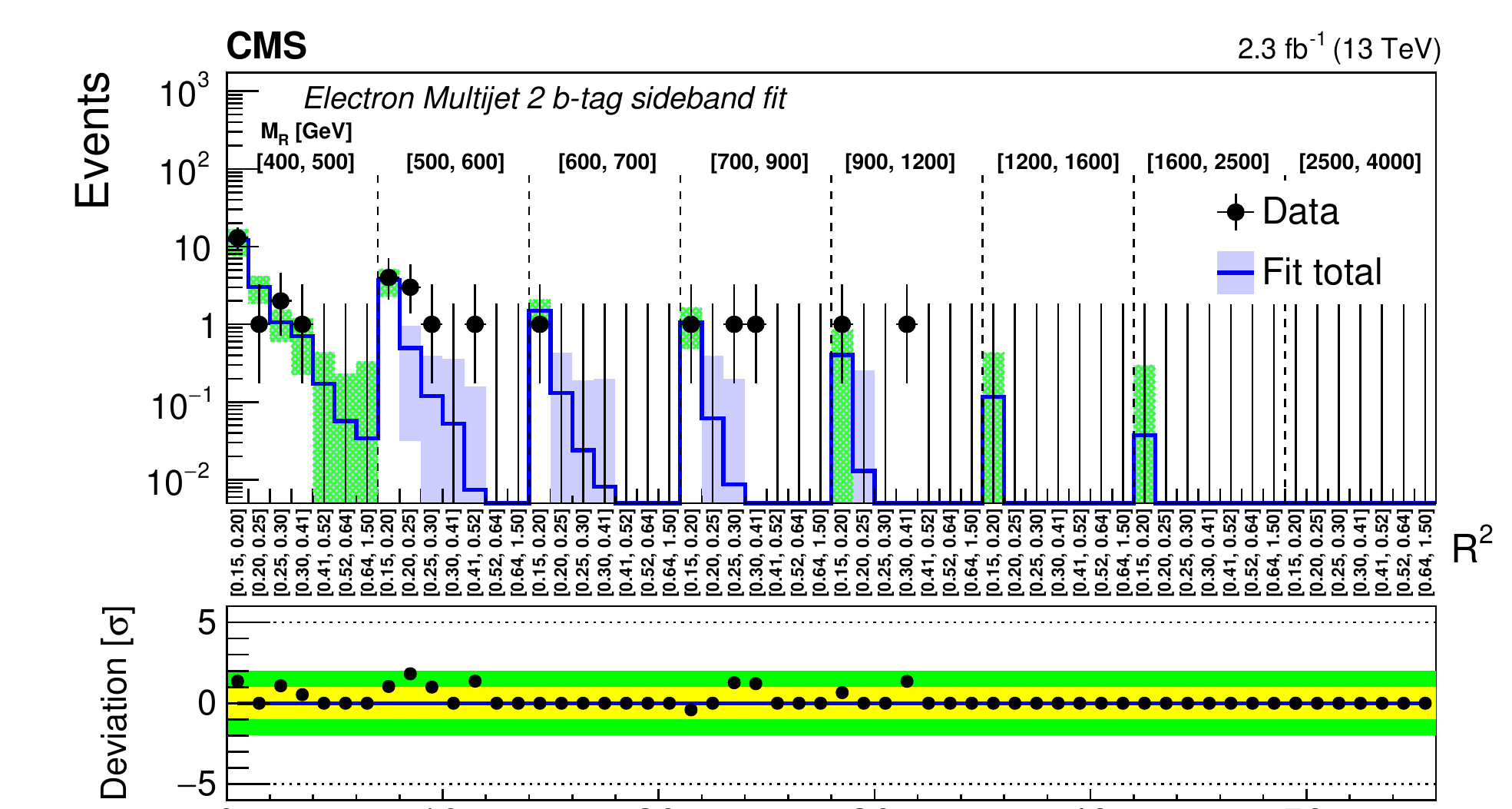} \\
\includegraphics[width=0.50\textwidth]{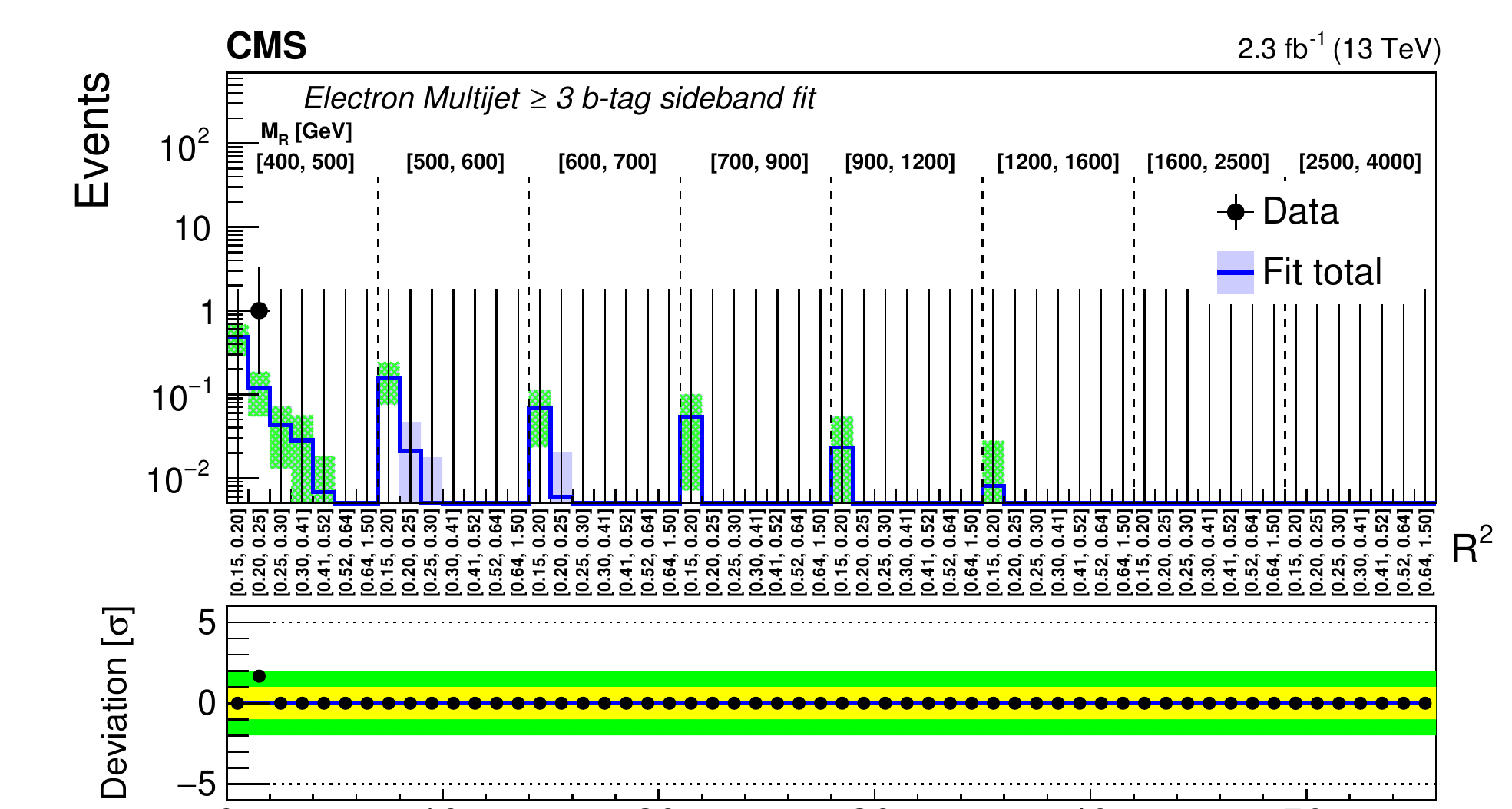}
\caption{Comparison of the predicted background with the observed data
in bins of $\MR$ and $\Rtwo$ variables in the Electron Multijet
category for the 2 \PQb-tag (upper) and ${\geq}3$ \PQb-tag (lower) bins. A detailed explanation of the panels is given in the caption of
Fig.~\ref{fig:results_Multijet2btag3btag}. }
\label{fig:results_EleMultijet2btag3btag}
\end{figure*}

In Section~\ref{sec:FitBkg}, we detail the fit-based background
prediction methodology and present the model-independent SUSY search
results in the 2 \PQb-tag and ${\geq}3$
\PQb-tag bins of the Multijet category in Fig.~\ref{fig:results_Multijet2btag3btag}.
In Figs.~\ref{fig:results_Multijet0btag1btag}-\ref{fig:results_EleMultijet2btag3btag}
in this Appendix, we present the results of the search for SUSY signal
events in the remaining categories, namely the 0 \PQb-tag and 1
\PQb-tag bins of the Multijet, the Muon Multijet, and Electron Multijet
categories. No statistically significant deviations from the expected
background predictions are observed in these categories in data.
\cleardoublepage \section{The CMS Collaboration \label{app:collab}}\begin{sloppypar}\hyphenpenalty=5000\widowpenalty=500\clubpenalty=5000\textbf{Yerevan Physics Institute,  Yerevan,  Armenia}\\*[0pt]
V.~Khachatryan, A.M.~Sirunyan, A.~Tumasyan
\vskip\cmsinstskip
\textbf{Institut f\"{u}r Hochenergiephysik,  Wien,  Austria}\\*[0pt]
W.~Adam, E.~Asilar, T.~Bergauer, J.~Brandstetter, E.~Brondolin, M.~Dragicevic, J.~Er\"{o}, M.~Flechl, M.~Friedl, R.~Fr\"{u}hwirth\cmsAuthorMark{1}, V.M.~Ghete, C.~Hartl, N.~H\"{o}rmann, J.~Hrubec, M.~Jeitler\cmsAuthorMark{1}, A.~K\"{o}nig, I.~Kr\"{a}tschmer, D.~Liko, T.~Matsushita, I.~Mikulec, D.~Rabady, N.~Rad, B.~Rahbaran, H.~Rohringer, J.~Schieck\cmsAuthorMark{1}, J.~Strauss, W.~Treberer-Treberspurg, W.~Waltenberger, C.-E.~Wulz\cmsAuthorMark{1}
% ...
\vskip\cmsinstskip
% ...
1:~~Also at Vienna University of Technology, Vienna, Austria\\
% ...

\end{sloppypar}
\end{document}